\documentclass[usenatbib]{mn2e}
\RequirePackage{snapshot}
\usepackage{hyperref}
\usepackage{amsmath}
\usepackage{times}
\usepackage{aas_macros}
\usepackage{graphicx}
\usepackage{xspace}
\usepackage{ctable}
\usepackage{datatool}
\usepackage[xindy, toc, hyperfirst=false, nolist, nostyles, sanitize={name=false,description=false,symbol=false}]{glossaries}
\glsdisablehyper

\newcommand{\plotdir}{./plots/}

\newcommand{\jblue}{\ensuremath{J_{lu}^b}\xspace}
\newcommand{\trad}{\ensuremath{T_\textrm{R}}\xspace}
\newcommand{\tinner}{\ensuremath{T_\textrm{i}}\xspace}
\newcommand{\vinner}{\ensuremath{v_\textrm{i}}\xspace}
\newcommand{\vouter}{\ensuremath{v_\textrm{o}}\xspace}
\newcommand{\rinner}{\ensuremath{r_\textrm{i}}\xspace}
\newcommand{\louter}{\ensuremath{L_\textrm{o}}\xspace}
\newcommand{\telectron}{\ensuremath{T_\textrm{e}}\xspace}
\newcommand{\nelectron}{\ensuremath{n_\textrm{e}}\xspace}

\newcommand{\tauelectron}{\ensuremath{\tau_\textrm{e}}\xspace}
\newcommand{\tausobolev}{\ensuremath{\tau_\textrm{S}}\xspace}

\newcommand{\jestimator}{\ensuremath{J_\textrm{est}}\xspace}
\newcommand{\nubarestimator}{\ensuremath{\bar{\nu}}\xspace}

\newcommand{\dline}{\ensuremath{d_\textrm{l}}\xspace}
\newcommand{\delectron}{\ensuremath{d_\textrm{e}}\xspace}
\newcommand{\dshell}{\ensuremath{d_\textrm{s}}\xspace}
\newcommand{\tauevent}{\ensuremath{\tau_\textrm{n}}\xspace}

\newcommand{\taucombined}{\ensuremath{\tau_\textrm{tot}}\xspace}

\newcommand{\simgt}%
        {\,\hbox{\lower0.6ex\hbox{$\sim$}\llap{\raise0.6ex\hbox{$>$}}}\,}
\newcommand{\simlt}%
        {\,\hbox{\lower0.6ex\hbox{$\sim$}\llap{\raise0.6ex\hbox{$<$}}}\,}

\makeglossaries

\newcommand{\lsun}{\hbox{$L_{\odot}$}\xspace}

\newglossaryentry{vrad}{name={radial velocity~}, text={radial velocity}, symbol={\ensuremath{v_\textrm{rad}}}, description={radial velocity}, sort=vrad}
\newglossaryentry{vrot}{name={stellar rotation~}, name={stellar rotation}, symbol={\ensuremath{v_\textrm{rot}}}, description={radial velocity}, sort=vrot}

\newcommand{\kms}{\ensuremath{\textrm{km}~\textrm{s}^{-1}}\xspace}

\newcommand{\xray}{X-ray\xspace}

\newcommand{\nififtysix}{$^{56}\textrm{Ni}$\xspace}

\makeatletter \newcommand{\ion}[2]{#1~\textsc{\@roman{#2}}\xspace} \makeatother

\newglossaryentry{angstrom}{name=\AA, description={unit of length $10^{-10}$\,m}, sort=angstrom}
\newglossaryentry{nir}{name=NIR,description={near infrared},first = {near infrared (NIR)}}
\newglossaryentry{psf}{name=PSF,description={Point Spread Function},first = {PSF}}
\newglossaryentry{fwhm}{name=FWHM,description={Full Width Half Maximum},first = {FWHM}}
\newglossaryentry{rms}{name=RMS,description={Root Mean Square},first = {RMS}}
\newglossaryentry{signalnoise}{name=S/N,description={signal to noise}}
\newglossaryentry{uv}{name=UV,description={ultra violet},first = {ultra violet (UV)}}
\newglossaryentry{halpha}{name=\ensuremath{\textrm{H}\alpha}, description={First line of the Balmer series at 6563\,\AA}, sort=halpha}
\newglossaryentry{mgb}{name={Mg \textsc{i} b}, description={Triplet at 5167\,\AA, 5173\,\AA and 5184\,\AA}}
\newglossaryentry{sobolevapprox}{name={Sobolev approximation}, description={Lines are approximation with an infinitley thin interaction region \citep[e.g. no broadening][]{1960mes..book.....S}}, first={Sobolev approximation }}
\newglossaryentry{radeq}{name={radiative equilibrium}, description={The net flux of energy between matter and radiation field is zero}}
\newglossaryentry{nebularapprox}{name={nebular approximation}, description={Assumes that the plasma condition are controlled by a central radiation source. The radiation field decreases with the distance to the source by geometrical dilution. See \citet{1978stat.book.....M} for details}}
\newglossaryentry{modnebularapprox}{name={modified nebular approximation}, description={In contrast to \gls{nebularapprox} where only geometrical dilution is taken into account, the modified nebular approximation also takes dilution by other radiative processes into account }, first={modified nebular approximation}, parent=nebularapprox}
\newglossaryentry{thompsonscat}{name={Thomson scattering}, description={Scattering of photons on low energy electrons}}
\newglossaryentry{lte}{name={LTE}, description={Local Thermodynamic Equilibrium}, first={local thermodynamic equilibrium (LTE)}}
\newglossaryentry{lsr}{name={LSR}, description={Local Standard of Rest}, first={\textit{local standard of rest} (LSR)}}
\newglossaryentry{montecarlo}{name={MC}, description={Monte Carlo }, first={\textit{Monte Carlo} (MC)}}

\newglossaryentry{sfit}{name=SFIT, text=\textsc{sfit}, description={spectral fitting program for hot stars \citep{2001A&A...376..497J}}, first={\textsc{sfit} \citep{2001A&A...376..497J}}}
\newglossaryentry{iraf}{name=IRAF, text=\textsc{iraf}, description={Image Reduction and Analysis Facility maintained by NOAO}, first={\textsc{iraf}\protect\footnote{IRAF: the Image Reduction and Analysis Facility is distributed by the National Optical Astronomy Observatory, which is operated by the Association of Universities for Research in Astronomy (AURA) under cooperative agreement with the National Science Foundation (NSF).}}}
\newglossaryentry{pyraf}{name=PyRAF, text=\textsc{PyRAF}, description={Python wrap of \gls{iraf} maintained by STSCI}, first=\textsc{PyRAF} \protect\footnote{PyRAF is a product of the Space Telescope Science Institute, which is operated by AURA for NASA.}}
\newglossaryentry{scipy}{name=SciPy, text=\textsc{Scipy}, description={Scientific Python \citep{Jones:2001fk}}}
\newglossaryentry{moog}{name=MOOG,text={\textsc{moog}}, description={spectral synthesis software \citep{1973ApJ...184..839S}}, first={\textsc{Moog} \citep{1973ApJ...184..839S}}}
\newglossaryentry{atlas9}{name=ATLAS9,description={grid of stellar atmospheres \citep{2004astro.ph..5087C}}, first={ATLAS9 \citep{2004astro.ph..5087C}}}
\newglossaryentry{vald}{name=VALD,description={Vienna Atomic Line Database \citep{2000BaltA...9..590K}}, first={Vienna Atomic Line Database \citep[VALD;][]{2000BaltA...9..590K}}}
\newglossaryentry{sextractor}{name=SExtractor, text=\textsc{SExtractor}, description={Source Extractor photometry program \citep{1996A&AS..117..393B}}, first={\textsc{SExtractor} \citep{1996A&AS..117..393B}}}
\newglossaryentry{idl}{name=IDL,text={\textsc{idl}}, description={Interactive Data Language}}
\newglossaryentry{makee}{name=MAKEE,text=\textsc{makee}, description={MAuna Kea Echelle Extraction by Tom Barlow available}}
\newglossaryentry{minuit}{name=MINUIT,text={\textsc{minuit}}, description={collection of numerical optimization tools \citep{James:1975dr}}}
\newglossaryentry{migrad}{name=MIGRAD,text={\textsc{migrad}}, description={numerical gradient optimization tools - part of \gls{minuit}}}
\newglossaryentry{dolphot}{name=DOLPHOT, text=DOLPHOT, description=photometry package for HST, first=DOLPHOT \citep{2000PASP..112.1383D}}
\newglossaryentry{chianti}{name=CHIANTI, text=\textsc{chianti}, description= CHIANTI Database 7.1, first =\textsc{chianti} 7.1 \citep{1997A&AS..125..149D,2012ApJ...744...99L}}
\newglossaryentry{chiantipy}{name=CHIANTIPY, text=\textsc{ChiantiPy}, description= \textsc{ChiantiPy}{0.5.2}, first =\textsc{ChiantiPy} 0.5.2 \protect\footnote{This work uses the version 0.5.2 with some corrections applied to the excitation calculation in collaboration with Ken Dere.}}
\newglossaryentry{synpp}{name=SYNPP, text=\textsc{syn++}, description= SYN++ software, first =SYN++ \citep{2011PASP..123..237T}}
\newglossaryentry{python_rt}{name=PYTHON, text=\textsc{python}, description= Python radiative transfer software, first = \textsc{python} \citep{2002ApJ...579..725L}}
\newglossaryentry{astropy}{name=ASTROPY, text=\textsc{astropy}, description=\textsc{astropy} framework, first = \textsc{astropy} \citep{2013A&A...558A..33A}}
\newglossaryentry{artis}{name=ARTIS, text=\textsc{artis}, description= ARTIS MC code, first = \textsc{artis} \citep{2009MNRAS.398.1809K}}
\newglossaryentry{sedona}{name=SEDONA, text=\textsc{sedona}, description= Sedona MC code, first = \textsc{sedona} \citep{2006ApJ...651..366K}}
\newglossaryentry{tardis}{name=TARDIS, text=\textsc{tardis}, description= TARDIS MC code}
\newglossaryentry{mlmc}{name=MLMC, text=ML93, description= Mazzali Lucy Monte Carlo, first ={Mazzali \& Lucy (1993, ML93) code}}

\newglossaryentry{2mass}{name=2MASS,description={Two Micron All Sky Survey \citep{2006AJ....131.1163S}}, first={Two Micron All Sky Survey \citep{2006AJ....131.1163S}}}
\newglossaryentry{nomad}{name=NOMAD,first={Naval Observatory Merged Astrometric Dataset \citep[NOMAD; ][]{2005yCat.1297....0Z}}, description={nomad catalogue}}
\newglossaryentry{wifes}{name=WIFES, text=\textsc{WiFeS}, first={\textsc{WiFeS} \citep{2007Ap&SS.310..255D}},  description={Wide Field Spectrograph - \gls{ifu} mounted on the 2.3\,m telescope at Siding Spring Observatory}}
\newglossaryentry{scp}{name=SCP,description={Supernova Cosmology Project, led by Saul Perlmutter}, first={Supernova Cosmology Project (SCP)}}
\newglossaryentry{hzsns}{name=HZSNS,description={High Z Supernova Search, led by Brian Schmidt}, first={High Z Supernova Search (HZSNS)}}
\newglossaryentry{vlt}{name=VLT,description={Very Large Telescope located on Cerro Paranal (Chile)}, first={Very Large Telescope (VLT)}}
\newglossaryentry{flames}{name=FLAMES,description={Multi-object, intermediate and high resolution spectrograph mounted on the  \gls{vlt}}}
\newglossaryentry{hires}{name=HIRES, description={High Resolution Echelle Spectrometer mounted on the Keck Telescope}, first={High Resolution Echelle Spectrometer \citep[HIRES;][]{1994SPIE.2198..362V}}}
\newglossaryentry{lris}{name=LRIS,description={Low Resolution Imaging Spectrometer mounted on the Keck Telescope}, first={Low-Resolution Imaging Spectrometer \citep[LRIS;][]{Oke95}}}

\newglossaryentry{essence}{name=ESSENCE,description={The `Equation of State: SupErNovae trace Cosmic Expansion' project \citep[ESSENCE;][]{2002AAS...201.7809G}}, first={`The Equation of State: SupErNovae trace Cosmic Expansion' \citep[ESSENCE;][]{2002AAS...201.7809G}}}
\newglossaryentry{ifu}{name=IFU,description={Optical instrument combining spectrographic and imaging capabilities, used to obtain spatially resolved spectra}, first={Integral Field Unit (IFU)}, firstplural={Integral Field Units (IFUs)}} 

\newglossaryentry{besancon}{name=Besan\c{c}on Model, description={Model of stellar population synthesis of the Galaxy, including kinematics.}}

\newglossaryentry{int}{name=INT,description={Isaac Newton 2.5\,m Telescope}, first={Isaac Newton 2.5\,m Telescope (INT)}}
\newglossaryentry{iau}{name=IAU,description={International Astronomical Union}, first={IAU}}
\newglossaryentry{chandra}{name=Chandra,description={Chandra \xray\ Observatory (space-based)}}
\newglossaryentry{hst}{name=HST,description={Hubble Space Telescope}}
\newglossaryentry{wfpc2}{name=WFPC2,description={Wide-Field Planetary Camera 2 mounted on the \gls{hst}}, first={Wide-Field Planetary Camera 2 (WFPC2)}}
\newglossaryentry{acs}{name=ACS,description={Advanced Camera for Surveys mounted on the \gls{hst}}, first={Advanced Camera for Surveys (ACS)}}
\newglossaryentry{snls}{name=SNLS,description={Supernova Legacy Survey \citep{2003AAS...203.8209P}}, first={Supernova Legacy Survey \citep[SNLS;][]{2003AAS...203.8209P}}}
\newglossaryentry{dass}{name=DASS, description={Digitized Astronomy Supernova Survey \citep{1975PASP...87..565C}}, first={Digitized Astronomy Supernova Survey \citep[DASS;][]{1975PASP...87..565C}}}
\newglossaryentry{bait}{name=BAIT, description={Berkley Automatic Imaging Telescope \citep{1993PASP..105.1164R}}, first={Berkley Automatic Imaging Telescope \citep[BAIT;][]{1993PASP..105.1164R}}}
\newglossaryentry{kait}{name=KAIT, description={Katzman Automatic Imaging Telescope \citep{2001ASPC..246..121F}}, first={Katzman Automatic Imaging Telescope \citep[KAIT;][]{2001ASPC..246..121F}}}
\newglossaryentry{loss}{name=LOSS, description={Lick Observatory Supernova Search  \citep{2000AIPC..522..103L}}, first={Lick Observatory Supernova Search \citep[LOSS;][]{2000AIPC..522..103L}}}
\newglossaryentry{ctss}{name=CTSS,description={Cal\'{a}n/Tololo Supernova Survey \citep{1993AJ....106.2392H}}, first={Cal\'{a}n/Tololo supernova survey \citep[CTSS;][]{1993AJ....106.2392H}}}
\newglossaryentry{ctio}{name= CTIO, description={Cerro Tololo Inter-American Observatory}, first={Cerro Tololo Inter-American Observatory (CTIO)}}
\newglossaryentry{ptf}{name=PTF, description={Palomar Transient Factory \citep{2009PASP..121.1395L}}, first={Palomar Transient Factory \citep[PTF;][]{2009PASP..121.1395L}}}
\newglossaryentry{batse}{name=BATSE, description={Burst and Transient Source Experiment mounted on the Compton Gamma Ray Observatory}, first={Burst and Transient Source Experiment (BATSE)}}
\newglossaryentry{bepposax}{name=BeppoSAX, description={\xray\ satellite named in honor of Giuseppe "Beppo" Occhialini}}
\newglossaryentry{rosat}{name=ROSAT, description={short for R\"{o}ntgensatellit}, first={ROSAT}}
\newglossaryentry{hete2}{name=HETE2, description={High Energy Transient Explorer}, first={High Energy Transient Explorer (HETE)}}
\newglossaryentry{gnirs}{name=GNIRS, description={Gemini Near InfraRed Spectrograph mounted on the Gemini North Telescope}}
\newglossaryentry{swift}{name=Swift, description={Swift Gamma-Ray Burst Mission}}
\newglossaryentry{vla}{name=VLA, description={Very Large Array radio telescope located in North America}, first={Very Large Array (VLA)}}
\newglossaryentry{evla}{name=EVLA, description={Extended Very Large Array radio telescope located in North America}, first={Extended Very Large Array (EVLA)}}
\newglossaryentry{sdss}{name=SDSS, description={Sloan Digital Sky Survey}}
\newglossaryentry{dss}{name=DSS, description={Digitized Sky Survey}}
\newglossaryentry{skymapper}{name=SkyMapper, description={SkyMapper telescope \citep{2007PASA...24....1K}}, first={SkyMapper \citep{2007PASA...24....1K}}}
\newglossaryentry{panstarrs}{name=PanSTARRS, description={Panoramic Survey Telescope \& Rapid Response System \citep{2010SPIE.7733E..12K}}, first={Panoramic Survey Telescope \& Rapid Response System \citep[PanSTARRS;][]{2010SPIE.7733E..12K}}}
\newglossaryentry{lsst}{name=LSST, description={Large Synoptic Survey Telescope}, first={Large Synoptic Survey Telescope \citep[LSST;][]{2006AAS...209.8604P}}}
\newglossaryentry{ppmxl}{name=PPMXL, description={PPMXL Catalog of Positions and Proper Motions on the ICRS \citep{2010AJ....139.2440R}}}
\newglossaryentry{gaia}{name=GAIA, description={Global Astrometric Interferometer for Astrophysics \citep{2001A&A...369..339P}}, first={Global Astrometric Interferometer for Astrophysics \citep[GAIA;][]{2001A&A...369..339P}}}
\newglossaryentry{ligo}{name=LIGO, description={Laser Interferometer Gravitational Wave Observatory}, first={Laser Interferometer Gravitational Wave Observatory \citep[LIGO;][]{1992Sci...256..325A}}}
\newglossaryentry{aligo}{name=Advanced LIGO, description={Advanced LIGO}, sort=ligo2}
\newglossaryentry{lisa}{name=LISA, description={Laser Interferometer Space Antenna \citep{1994ESAJ...18..219J}}, first={Laser Interferometer Space Antenna \citep[LISA;][]{1994ESAJ...18..219J}}}
\newglossaryentry{ccd}{name=CCD,description={Charged Coupled Device}, first={charged coupled device (CCD)}, firstplural={charged coupled devices (CCDs)}}
\newglossaryentry{nist}{name=NIST, description={National Institute of Standards}, first=NIST \citep{kramida2012nist}}
\newglossaryentry{kuruczlinelist}{name=Kurucz line list, first={Kurucz line list \citep{1995KurCD..23.....K}}, description={kurucz}}
\newcommand{\sn}[2]{SN~#1#2}

\newglossaryentry{sn}{name=supernova, text={SN}, plural={SNe}, description={exploding star}, nonumberlist=true, first={supernova (SN)}, firstplural={supernovae (SNe)}}
\newglossaryentry{snia}{name=Type~Ia (SN~Ia), text={SN~Ia}, description={Thermonuclear explosion of a white dwarf - spectra show no hydrogen but a strong silicon line},first={Type~Ia supernova (SN~Ia)}, firstplural={Type Ia supernovae (SNe~Ia)}, plural={SNe~Ia}, parent=sn, nonumberlist=true}
\newcommand{\sneia}{\glspl*{snia}\xspace}
\newcommand{\snia}{\gls*{snia}\xspace}

\newglossaryentry{branchnormal}{name={branch-normal}, text=\textit{Branch-normal}, description={Large homogeneous class of Type Ia Supernovae, defined in \citet{1993AJ....106.2383B}}, first={\textit{Branch-normal} SNe Ia \citep{1993AJ....106.2383B}}, parent=snia} 
\newglossaryentry{91t}{name={91T-like}, description={Luminous class of Type Ia supernovae similar to \sn{1991}{T} \citep{1992AJ....103.1632P}} , first={91T-like}, parent=snia} 
\newglossaryentry{91bg}{name={91bg-like}, description={Faint class of Type Ia supernovae similar to \sn{1991}{bg} \citep{1992AJ....104.1543F}}, first={91bg-like}, parent=snia} 
\newglossaryentry{02cx}{name={02cx-like}, description={Peculiar class of Type Ia supernovae similar to \sn{2002}{cx} \citep{2003PASP..115..453L}}, first={02cx-like \sneia\ \citep{2003PASP..115..453L}}, parent=snia} 

\newglossaryentry{snibc}{name=Type~Ib/c, text={SN~Ib/c}, description={Collapse of the core of a massive star -  spectrum shows no hydrogen and no silicon line},first={Type~Ib/c supernova (SN~Ib/c)}, firstplural={Type~Ib/c supernovae (SNe~Ib/c)}, plural={SNe~Ib/c}, parent=sn}

\newglossaryentry{snib}{name=Type~Ib, text={SN~Ib}, description={Spectrum shows no hydrogen and no silicon, but helium line},first={Type Ib supernova (SN~Ib)}, firstplural={Type~Ib supernovae (SNe~Ib)}, plural={SNe~Ib}, parent=snibc}

\newglossaryentry{snic}{name=Type~Ic, text={SN~Ic}, description={Spectrum shows no hydrogen, no silicon and no helium line},first={Type~Ic supernova (SN~Ic)}, firstplural={Type~Ic supernovae (SNe~Ic)}, plural={SNe~Ic}, parent=snibc}

\newglossaryentry{snii}{name=Type~II, text={SN~II}, description={Collapse of the core of a massive star - spectrum shows strong hydrogen line},first={Type~II supernova (SN~II)}, firstplural={Type~II supernovae (SNe~II)}, plural={SNe~II}, parent=sn}

\newglossaryentry{sniib}{name=Type~IIb, text={SN~IIb}, description={Spectrum shows hydrogen and helium lines},first={Type~IIb supernova (SN~IIb)}, firstplural={Type~IIb supernovae (SNe~IIb)}, plural={SNe~IIb}, see=snib, parent=snii}

\newglossaryentry{sniip}{name=Type~II~Plateau (Type IIP), text={SN~IIP}, description={Lightcurve shows plateau},first={Type~IIP supernova (SN~IIP)}, firstplural={Type~II Plateau supernovae \citep[SNe~IIP;][]{1979A&A....72..287B}}, plural={SNe~IIP}, parent=snii}

\newglossaryentry{sniil}{name=SN~II~Linear, text={SN~IIL}, description={Lightcurve shows no plateau, but linear decline},first={Type~IIL supernova (SN~IIL)}, firstplural={Type~II~Linear supernovae \citep[SNe~IIL;][]{1990MNRAS.244..269S}}, plural={SNe~IIL}, parent=snii}

\newglossaryentry{sniin}{name=Type II narrow-lined (Type IIn), description={Spectrum shows narrow lines},first={Type~II~narrow-lined supernova (SN IIn)}, firstplural={Type~IIn supernovae (SNe~IIn)}, plural={SNe~IIn}, parent=snii}

\newglossaryentry{snr}{name=Remnant (SNR), text=SNR, description={Remnant left visible post-explosion}, first={supernova remnant (SNR)}, firstplural={supernova remnants (SNRs)}, parent=sn}

\newglossaryentry{dtd}{name=DTD,description={delay time distribution - expected supernova rate over time after a brief outburst of starformation},first={delay time distribution (DTD)}, firstplural={delay time distributions (DTDs)}, plural=DTDs}

\newglossaryentry{hvg}{name=HVG,description={high velocity gradient - Type Ia supernovae with a fast evolution of photospheric velocity},first={high velocity group (HVG)}, firstplural={high velocity groups (HVGs)}, plural=HVGs, parent=snia}

\newglossaryentry{lvg}{name=LVG,description={low velocity gradient - Type Ia supernovae with a slow evolution of photospheric velocity},first={low velocity group (LVG)}, firstplural={low velocity groups (LVGs)}, plural=LVGs, parent=snia}

\newglossaryentry{wd}{name=white dwarf (WD), text=WD, description={White Dwarf - extremely dense stellar remnant}, first={white dwarf (WD)}}
\newglossaryentry{onemgwd}{name= Oxygen/Neon (ONe), text={ONe-WD},description={Oxygen/Neon White Dwarf}, first={oxygen/neon White Dwarf (ONe-WD)}, parent=wd}
\newglossaryentry{cowd}{name=carbon/oxygen (CO), text={CO-WD}, description={carbon/oxygen white dwarf}, first={carbon/oxygen white dwarf (CO-WD)}, firstplural = {carbon/oxygen white dwarfs (CO-WDs)}, parent=wd}

\newglossaryentry{sds}{name=SD-Scenario,description={single-degenerate scenario (single white dwarf accreting from non-degenerate companion)}, first={single-degenerate scenario (SD-scenario)}}

\newglossaryentry{dds}{name=DD-Scenario, description={double degenerate scenario (merging of two white dwarfs)}, first={double-degenerate scenario (DD-scenario)}}

\newglossaryentry{sss}{name=SSS, text={supersoft \xray\ source}, description={supersoft \xray\ source - believed to be emitted by nuclear fusion on a white dwarf's surface}}

\newglossaryentry{amcvn}{name=AM CVn, description={AM Canum Venaticorum star (white dwarf accreting hydrogen poor matter from a companion star; see \citet{2005ASPC..330...27N})}}

\newglossaryentry{rlof}{name=RLOF, description={Roche Lobe Overflow (see \citet{1971ARA&A...9..183P} for a more detailed description)}, first={Roche-lobe overflow (RLOF)}}

\newglossaryentry{mchan}{name={Chandrasekhar mass~}, text={Chandrasekhar~mass}, symbol={\ensuremath{M_\textrm{Chan}}}, plural={Chandrasekhar~masses}, description={Mass when the core of a star collapses due to insufficient degeneracy pressure - for a white dwarf $\approx1.38\,M_\odot$ see \citet{1931ApJ....74...81C}}, first={Chandrasekhar~mass \citep[$M_\textrm{Chan}=1.38\,M_\odot$;][]{1931ApJ....74...81C}}, sort=mchan}

\newglossaryentry{w7}{name={W7 model},description={W7 model \citep{1984ApJ...286..644N}},first = {W7 model \citep{1984ApJ...286..644N}}}

\newglossaryentry{ew}{name=Equivalent Width, text={EW}, description={width of a rectangle that has the same area as a spectral line when taken to zero flux}, first={equivalent width (EW)}, firstplural={equivalent widths (EWs)}}
\newglossaryentry{agb}{name=AGB,description={Asymptotic Giant Branch}}
\newglossaryentry{cmb}{name=CMB,description={Cosmic Microwave Background}}
\newglossaryentry{csm}{name=CSM,description={Circumstellar Medium}, first={circumstellar medium (CSM)}}
\newglossaryentry{csi}{name=CSI,description={Circumstellar Interaction}, first={circumstellar interaction (CSI)}}
\newglossaryentry{ism}{name=ISM,description={Interstellar Medium}, first={interstellar medium (ISM)}}
\newglossaryentry{ige}{name=IGE,description={Iron Group Element}, first={iron group element (IGE)}, firstplural={iron group elements (IGEs)}}
\newglossaryentry{epm}{name=EPM,description={Expanding Photosphere Method \citep{1974ApJ...193...27K}}, first={Expanding Photosphere Method (EPM)}}
\newglossaryentry{aic}{name=AIC,description={Accretion Induced Collapse}, first={accretion induced collapse (AIC)}}
\newglossaryentry{ime}{name=IME,description={Intermediate Mass Element}, first={intermediate mass element (IME)}, firstplural={intermediate mass elements (IMEs)}}
\newglossaryentry{h0}{name=\ensuremath{H_0},description={Hubbles constant}}
\newglossaryentry{nse}{name=NSE,description={Nuclear Statistical Equilibrium}, first={nuclear statistical equilibrium (NSE)}}
\newglossaryentry{cdm}{name=CDM,description={Cold Dark Matter}}
\newglossaryentry{grb}{name=GRB,description={Gamma Ray Burst}, first={Gamma Ray Burst (GRB)}, firstplural={Gamma Ray Bursts (GRBs)}}
\newglossaryentry{donor}{name=donor,description={non-degenerate companion in the \gls{sds}}}
\newglossaryentry{mainsequence}{name=main sequence,description={main sequence star}}
\newglossaryentry{redgiant}{name=red giant,description={red giant star}}
\newglossaryentry{mlcs}{name=MLCS,description={Multicolor Light Curve Shape method \citep[MLCS;][]{1996ApJ...473...88R}}, first={Multicolor Light-Curve Shape method \citep[MLCS;][]{1996ApJ...473...88R}}}
\newglossaryentry{rsoph}{name=RS~Ophiuci ,description={white dwarf accreting from a red giant - assumed progenitor of the \gls{sds}}, sort=rsoph}
\newglossaryentry{usco}{name=U~Scorpii,description={white dwarf accreting from a main sequence star - assumed progenitor of the \gls{sds}}, sort=usco}
\newglossaryentry{rcw86}{name=RCW~86,description={supernova remnant sometimes associated with \sn{185}{}}, sort=rcw86}
\newglossaryentry{casa}{name=Cas~A,description={Cassiopeia A supernova remnant - probably a \gls{snib} event}}
\newglossaryentry{cepheid}{name=Cepheid,description={very luminous variable star with a strong luminosity period relationship}}
\newglossaryentry{urca}{name=Urca, text=\textit{Urca}, description={process predominatly contributing to cooling in stars. The \textit{Urca} process consists of alternating electron-capture and $\beta^{-}$ decay of two nuclei pairs.},sort=urca} 
\newglossaryentry{alphacen}{name=Alpha Centauri,description={one of the brightest stars in the night sky and a close binary}}
\newglossaryentry{pcygni}{name={P Cygni}, text={P Cygni},description={a hypergiant luminous blue variable with strong winds. Often referred to as a description for their line profiles showing a emission peak at the rest wavelength of the line and a blue-shifted absorption trough.}}

\newglossaryentry{teff}{name={effective temperature~}, text={effective temperature}, symbol={\ensuremath{T_\textrm{eff}}}, description={Temperature of a blackbody emitting the same total energy}, sort=teff}

\newglossaryentry{logg}{name={surface gravity~}, text={surface gravity}, symbol={\ensuremath{\textrm{log}\,g}}, description={gravity at the surface of a star}, sort=logg}
\newglossaryentry{feh}{name={metallicity~}, text={metallicity}, symbol=\textrm{[Fe/H]},description={iron abundance relative to the sun}, sort=feh}

\newglossaryentry{texp}{name={time since explosion~}, text={time since explosion}, text={time since explosion}, symbol={\ensuremath{t_{\rm exp}}},description={time since explosion (measured in days)}, sort=texp, first={time since explosion (\ensuremath{t_{\rm exp}})}}
\newcommand{\texp}{\glssymbol*{texp}}
\newglossaryentry{lmc}{name=LMC,description={Large Magellanic Cloud}, first={Large Magellanic Cloud (LMC)}, sort=lmc}
\newglossaryentry{smc}{name=SMC,description={Small Magellanic Cloud}, sort=smc}
\newglossaryentry{z}{name=\ensuremath{z},description={redshift}, sort=z}

\renewcommand{\sn}[2]{\object{SN~#1#2}}

\begin{document}

\title{A spectral synthesis code for rapid modelling of supernovae}

\author[Kerzendorf \& Sim]{Wolfgang~E.~Kerzendorf$^1$, Stuart~A.~Sim$^2$\\
$^1$Department of Astronomy and Astrophysics, University of Toronto, 50 Saint George Street, Toronto, ON M5S 3H4, Canada\\
$^2$Astrophysics Research Centre, School of Mathematics and Physics, Queen's University Belfast, Belfast BT7 1NN, UK}

\maketitle
\begin{abstract}
We present \gls{tardis} - an open-source code for rapid spectral modelling of supernovae (SNe).
Our goal is to develop a tool that 
is sufficiently fast to allow exploration of the complex parameter spaces of models for SN ejecta. 
This can be used to analyse the growing number of high-quality SN spectra being obtained by transient surveys.
The code uses \textit{Monte Carlo} methods to obtain a self-consistent description of the plasma state and to compute a synthetic spectrum. 
It has a modular design to 
facilitate the implementation of a range of physical approximations that can be compared to asses both accuracy and computational expediency. 
This will allow users to choose a level of sophistication appropriate for their application. Here, we describe the operation of the code and make comparisons with alternative radiative transfer codes of differing levels of complexity (\textsc{syn++},  \textsc{python}, and \textsc{artis}). We then explore the consequence of adopting simple prescriptions for the calculation of atomic excitation, focussing on four species of relevance to Type Ia supernova spectra -- \ion{Si}{2}, \ion{S}{2}, \ion{Mg}{2}, and \ion{Ca}{2}. We also investigate the influence of three methods for treating line interactions on our synthetic spectra and the need for accurate radiative rate estimates in our scheme.
\end{abstract}

\begin{keywords}
radiative transfer -- methods: numerical --  supernovae: general 
\end{keywords}

\nocite{2002ApJ...579..725L}

\section{Introduction}
\label{sec:intro}
\glsresetall

\nocite{1993A&A...279..447M}

The goal of \gls{sn} modelling  is typically to infer the composition, mass and explosion energy of an event with the aim of understanding the explosion mechanisms and progenitor systems. Studies often rely on radiative transfer / spectrum synthesis codes to interpret the complex spectra originating in rapidly expanding and often metal-rich ejecta. 
A variety of different codes exist and are used for different purposes: these range from very simple approaches designed purely for line identification [run times of second e.g. \gls{synpp}] via computationally inexpensive one-dimensional codes [runtimes of minutes, e.g. the \gls{mlmc}] to vastly more complex codes that implement sophisticated micro-physics, time-dependence and/or multi-dimensionality \citep[e.g.][]{1996A&A...312..525P, 1996MNRAS.283..297B, 2001ApJ...557..266L, 2003ASPC..288..185H, 2005A&A...437..667D, 2006ApJ...651..366K,2009MNRAS.398.1809K, 2013MNRAS.429.2127B, 2006A&A...459..229S}. Such sophisticated approaches, however, come at the expense of execution times running to many hours (even days) on massively parallel machines, making them impractical for efficient exploration of parameter space or rapid characterization of new data sets. 
Thus, for the foreseeable future, the need for the simpler approaches will remain. Nevertheless, given the quality of modern data sets, it is important to test and understand the implications of any simplifying assumptions that are made.

Our goal is to provide analysis tools that strike an optimal balance between accuracy and computational expense, ideally in a manner that can be adjusted to suit the demands of (and resources available to) a variety of studies. To this end, we have undertaken the development of a modular \gls{sn} radiative transfer code (\gls{tardis}). As input, the code takes a one-dimensional model for the supernova eject (which can include arbitrary density and/or abundance stratification, if required) together with a luminosity and time since explosion. From these, ionization and excitation states are iteratively estimated and a synthetic spectrum is calculated.
The code serves two main purposes. First, it is designed to have an execution time of minutes, which allows for rapid exploration of parameter space with the possibility of fitting and interpretation of observational data. Secondly, it serves as a platform to trial different approximations for important microphysics (e.g. line interaction, level populations, etc.). \gls{tardis} allows for swift implementation and testing of new physics that -- if successful -- might be used in the more complete but complex codes like \gls{artis} and \gls{sedona}, which are founded on the same principles as \gls{tardis}. Even if the newly implemented physics is computationally too costly to use in large parameter-space studies, it can be still be used to quantify the uncertainty associated with using faster, but less accurate, approaches (i.e. Uncertainty Quantification).

In Section~\ref{sec:method}, we present the numerical methods and a sub-set of the modes of operation currently implemented in the code. Sources of atomic data are described in Section~\ref{sec:datasources}. Sections~\ref{sec:numerics} and \ref{sec:code_comparisons} present results of numerical convergence tests and detailed comparisons of our synthetic spectra with those from a selection of existing radiative transfer codes of varying degrees of complexity. 

In this paper we explore the effect of approximations used by two different physics sub-modules, focussing on their importance in the modelling of \sneia. 

Studies of SNe~Ia using 1D codes that combine realistic treatments of ionization and excitation \citep[e.g.][]{1996MNRAS.283..297B, 1996A&A...312..525P,2013MNRAS.429.2127B} have already shown that the effects of departures from \gls{lte} can be rather important. Consequently, for our goal of developing a code that is both reasonably fast and accurate, we must attempt to identify suitable approaches and quantify the impact of approximations made in the interests of computational expediency. 
Thus, we first (in Section~\ref{sec:results_nlte})
examine the influence of adopting a simple analytic treatment for the populations of excited atomic/ionic levels by comparing to results using level populations obtained by numerically solving a set of equations of statistical equilibrium.

Secondly, in Section~\ref{sec:results_line_interaction}, we compare results obtained using different levels of sophistication in the treatment of line scattering and fluorescence. It has been clearly established that non-resonance line scattering has a key role in shaping the spectra of metal-rich SN ejecta \citep{2000A&A...363..705M, 2000ApJ...530..757P, 1996MNRAS.283..297B, 2005A&A...437..667D,2013MNRAS.429.2127B}. In this work, we will compare  two versions of the ``macro atom'' scheme \citep{2005A&A...429...19L, 2009MNRAS.398.1809K} to a resonance scattering scheme and quantify their influence on synthetic spectra.

In Section~\ref{sec:conclusions}, we summarise our findings and outline future plans for the further development of this project.

\section{Method}
\label{sec:method}

The \gls{tardis} code is a new implementation of indivisible energy-packet \gls{montecarlo} methods (\citealt{1985ApJ...288..679A,1993ApJ...405..738L}, ML93, \citealt{1999A&A...345..211L,2002A&A...384..725L,2003A&A...403..261L}) that have previously been used to model radiation transport in Type Ia supernovae \glsdisp{snia} (SNe Ia; e.g. ML93; \citealt{2006ApJ...651..366K,2007MNRAS.375..154S,2009MNRAS.398.1809K}). 
The algorithm operates by seeding a set of \gls{montecarlo} quanta representing photon bundles \citep[$r$-packets in the nomenclature of ][]{2002A&A...384..725L} and following their propagation through a model for the \gls{sn} ejecta, as described below.

\gls{tardis} is based on the same methods used by \gls{artis}, but the scope of \gls{tardis} is much more limited so that runtimes are orders of magnitude shorter. 
In particular,
having in mind the goal of rapid fitting of observations (i.e. using \gls{tardis} to compute synthetic spectra that can be fed into a fitting algorithm and explore model parameter space), 
\gls{tardis} currently neglects multi-dimensionality and time-dependence. Thus, a \gls{tardis} calculation provides a single snapshot spectrum while an \gls{artis} run provides a time series of spectra from which synthetic light curves can be constructed.\footnote{We note, however, that \gls{tardis} can be used to compute a time series of synthetic spectra for a model via a sequence of calculations in which the input parameters specifying luminosity and time since explosion are varied.}
In addition, \gls{tardis} adopts an inner boundary approximation, which greatly simplifies the calculations by avoiding the need to describe the diffusion of radiation at very high optical depths.

\gls{tardis} already includes a number of different options and modules that make it possible to compare the effects of particular assumptions on both the outputs and the computational expense. The current options will each be introduced in the sections below but, for reference, we provide a summary in Table~\ref{tab:modes}.
So far, the development of \gls{tardis} has focused on \snia applications. However, we stress that our goal is to continue development and that the code has been structured to make it easy to implement additional physics and improve, alter or lift many of the existing approximations (see below).
Further modules extending the code for application to \glspl{snii} are being developed (Klauser, Kromer et al., in prep.).

\gls{tardis} is written in Python with C extensions and is available as open source (BSD 3 clause license) at \url{http://pypi.python.org/pypi/tardis-sn}. For all random numbers \gls{tardis} uses the Mersenne Twister random number generator \citep{Matsumoto:1998:MTE:272991.272995}.
In Appendix A we show an example input file and provide basic operational details for using \gls{tardis}. However, we refer potential users to the 
technical manual (available at \url{http://tardis.rtfd.org}) for further information on running the code.

\begin{table*}
\caption{Summary of the \gls{tardis} modes of operation used in this study.}
\begin{tabular}{ll}
Mode name & Mode of operation \\ \hline
\multicolumn{2}{l}{Ionization modes:} \\
\texttt{lte}& Ionization fractions from the Saha-Boltzmann equation and $\trad$ (Equation~\ref{eq:lte_ionization}) \\
\texttt{nebular} & Ionization fractions from the nebular approximation  (Equation~\ref{eq:modified_nebular})\\ 
\\
\multicolumn{2}{l}{Excitation modes:} \\
\texttt{lte} & Level populations from Boltzmann equation and $\trad$ (Equation~\ref{eq:lte_excitation})\\
\texttt{dilute-lte} & Excited level populations reduced by $W$ (Equation~\ref{eq:dilute_lte_excitation})\\
\texttt{nlte} & Population ratios determined from statistical equilibrium (Section~\ref{sec:plasma_state})\\ 
\\
\multicolumn{2}{l}{Radiative rate modes:}\\
\texttt{dilute-blackbody} & Radiative rates obtained using $J_{lu}^b = W B_{\nu_{lu}}(\trad)$\\
\texttt{detailed} & Radiative rates obtained using \gls{montecarlo} estimators for $J_{lu}^b$ (Section~\ref{sec:montecarlo_estimators})\\ 
\\
\multicolumn{2}{l}{Line interaction modes:}\\
\texttt{scatter} & All line interactions treated as resonance scattering events\\
\texttt{downbranch} & Macro atom scheme with all internal transition probabilities set to zero (Section~\ref{sec:rad_matter_interaction})\\
\texttt{macroatom} & Full macro atom scheme (Section~\ref{sec:rad_matter_interaction}) \\\hline
\end{tabular}
\label{tab:modes}
\end{table*}

\subsection{Model Setup}
In the interest of computational expedience, \gls{tardis} approximates the \gls{sn} ejecta as spherically symmetric such that all physical properties depend only on the radial coordinate $r$. The computational domain is defined by inner and outer radial boundaries. The code currently assumes that material inside the domain is in radiative equilibrium and neglects any non-radiative energy sources (i.e. it is assumed that energy injection by \nififtysix within the computation volume is negligible). 
This approximation limits the applicability of \gls{tardis} to epochs at which 
 the effective photosphere is external to the volume in which the majority of the luminosity is generated (for \sneia, this is most valid at early times and becomes an increasingly poor approximation at later epochs). 

The computational domain is discretised into multiple cells (spherical shells). For each of these, the density and elemental abundances must be specified (the ``input model''). The input model is setup at runtime and can involve any density profile and set of stratified abundances. Simple models with uniform abundances and standard density profiles are directly created by the code (see Appendix~\ref{sec:using_tardis}) while models with stratified abundances or arbitrary density profiles are setup by providing simple input files, as described in the manual.
During the simulation, various quantities are computed on a cell-by-cell basis, most importantly the parameters of a simple radiation-field model (see Section~\ref{sec:rad_model}), which are used to estimate ionization fractions and level populations (see Section~\ref{sec:plasma_state}). 

We assume that the ejecta are in homologous expansion, which becomes an adequate approximation for \sneia\ within $\sim~100$ seconds after explosion \citep{2005A&A...432..969R}. 

\subsection{Radiation Field Model}
\label{sec:rad_model}

In general, we assume that both ionization and excitation are primarily controlled by the radiation field and we follow \citet{1985ApJ...288..679A}, \citet{1993ApJ...405..738L} and ML93 in adopting approximations based on a simple model for the frequency ($\nu$) dependent mean intensity

\begin{equation}
J_{\nu} = W B_{\nu} (\trad) 
\end{equation}
where, $B_{\nu}$ is the Planck function. The radiation temperature ($\trad$) and dilution factor ($W$) are parameters that are iteratively derived from \gls{montecarlo} estimators in each grid cell (see Section~\ref{sec:montecarlo_estimators}).

\subsection{Plasma State}
\label{sec:plasma_state}

To compute opacities and handle radiation-matter interactions, it is necessary to specify
the number densities for the states of the atomic/ionic species included. 
In principal, these could be obtained by a full solution of the complete system of equations of statistical equilibrium \citep[as done by][]{1996A&A...312..525P,1996MNRAS.283..297B,2013MNRAS.429.2127B} with rate coefficients for radiative processes derived from properties of the \gls{montecarlo} radiation field \citep{2002A&A...384..725L,2003A&A...403..261L}. This will be implemented in future versions of \gls{tardis}. However, since the computational cost of such an approach will be significant, it is also valuable to consider simpler approximations that can be used for rapid calculations to 
explore parameter space.
We pursue the approximate approach here, closely following ML93 who have demonstrated the utility of such a philosophy \citep[see also][]{1996A&A...312..525P}.

Currently, \gls{tardis} determines the ionization balance for each of the model cells based on the density, elemental abundance, radiation temperature ($\trad$), electron temperature ($\telectron$)  and the dilution factor ($W$). In the calculations presented in this work, we will adopt either a standard Saha-Boltzmann equation (\texttt{lte} ionization mode), 
\begin{equation}
\frac{N_{i, j+1} n_e}{N_{i, j}} = \Phi_{i,j,\trad} \equiv \frac{2 Z_{i,j+1}(\trad)}{Z_{i,{j}}(\trad)} \left(\frac{2\pi m_e k \trad}{h^2} \right) ^{3/2} e^{-\frac{\chi_{i,j}}{k\,\trad}}
\label{eq:lte_ionization}
\end{equation}
or a modified nebular approximation (ML93), referred to as \texttt{nebular} ionization mode
\begin{equation}
\frac{N_{i, j+1} n_e}{N_{i, j}} = W [\delta\zeta_{i,j} + W(1-\zeta_{i,j})] \left( \frac{\telectron}{\trad} \right)^{1/2} \Phi_{i,j,\trad}
\label{eq:modified_nebular}
\end{equation}
Here, $n_e$ is the free electron number density, $N_{i,j}$ is the ion number density, $Z_{i,j}$ the partition function and $\chi_{i,j}$ the ionization potential for ion $j$ of element $i$. $\zeta_{i,j}$ is the fraction of recombinations that go directly to the ground state in recombination to ion $j$ of element $i$. $\delta$ is a correction factor introduced to approximately account for the dominance of locally created radiation at short wavelengths (see \gls{mlmc}). We follow ML93 in setting $T_{e} = 0.9 T_{R}$.

With ionization ratios determined using one of these approximations, the code solves for a complete set of ion number densities (and the electron density) by enforcing the appropriate total elemental number densities and insisting on charge conservation in the usual manner.

\gls{tardis} can use a variety of approximations for calculating the level populations within each ion. As a simplest case, we adopt the Boltzmann excitation formula,
\begin{equation}
n_{i, j, k} = \frac{g_{i,j,k}}{Z_{i, j}} N_{i, j} \exp (-\epsilon_{i,j,k} /  k \trad),
\label{eq:lte_excitation}
\end{equation}
where $n_{i, j, k}$ is the number density of level $k$ of ionization state $j$ of element $i$, which has excitation energy $\epsilon_{i,j,k}$ relative to the ion ground state. We will refer to this as \texttt{lte} excitation mode.

As a slightly more sophisticated approach, \gls{tardis} includes a crude NLTE approximation for level populations in which we continue to apply Equation~\ref{eq:lte_excitation} to all metastable levels but adopt

\begin{equation}
n_{i, j, k} = W \frac{g_{i,j,k}}{Z_{i, j}} N_{i, j} \exp{(-\epsilon_{i,j,k}/k \trad)}
\label{eq:dilute_lte_excitation}
\end{equation}
for all other excited state \citep[see Equation 4 in ][]{1999A&A...345..211L}. We refer to this approach as the \texttt{dilute-lte} excitation mode in \gls{tardis}.

Finally, \gls{tardis} is also able to obtain NLTE level populations by solving a set of statistical equilibrium equations for chosen species (\texttt{nlte} excitation mode).  
Here we formulate rates between pairs of levels (upper level $u$ and lower level $l$)\footnote{For clarity, we drop the explicit reference to atomic number and ionization number here.} in an ion as

\begin{align}
{\cal R}_{ul} =& \beta_{lu} A_{ul} n_l + \beta_{lu} B_{ul} n_l J_{lu}^b + C_{ul} n_u n_e\label{eq:nlte_rul}\\
{\cal R}_{lu} =& \beta_{lu} B_{lu} n_l J_{lu}^b + C_{lu} n_l n_e\label{eq:nlte_rlu},
\end{align}
where $A, B$ are the usual Einstein coefficients for radiative transitions, $C$ is the rate coefficient for electron collisions and 
$\beta_{lu}=\frac{1}{\tau_{lu}}[1-\exp(-\tau_{lu})]$ is the Sobolev escape probability \citep[see ][ and Section~\ref{sec:rad_matter_interaction}]{2002A&A...384..725L}. 
$J_{lu}^b$ is the mean intensity at the extreme violet wing of the bound-bound transition between levels $l$ and $u$ \citep[see][]{2003A&A...403..261L}. 
\gls{tardis} can currently estimate \jblue in one of two ways.
The first option (\texttt{dilute-blackbody}) sets $\jblue = W B_{\nu_{lu}}(\trad)$ (i.e. imposes the radiation-field model of Section~\ref{sec:rad_model}). Alternatively, we have also implemented a \texttt{detailed} setting that calculated \jblue-values using estimators (see Section~\ref{sec:montecarlo_estimators}).

The net rate of change for a level population is
\begin{equation}
\frac{dn_k}{dt} = \sum_{i \ne k} R_{ik} - \sum_{i \ne k} R_{ki} \; .
\end{equation}
In \texttt{nlte} excitation mode, \gls{tardis} solves for level populations satisfying $\frac{d n_{k}}{dt} = 0$ for all $k$ (with  $\sum_k n_{k} = N$, the total ion population). 
In practise,  the code is initialized with LTE level populations and initial guesses for $\beta_{lu}$ and $\jblue$. It then solves the statistical equilibrium equations using LU decomposition 
\citep[LAPACK;][]{laug}. After each \gls{montecarlo} iteration (see Section~\ref{sec:montecarlo_iteration}) the statistical equilibrium equations are re-solved using the last level populations to compute values for $\beta_{lu}$ and results from the \gls{montecarlo} step to estimate $\jblue$. This process eventually converges to a stable set of level populations (see Section~\ref{sec:montecarlo_iteration}).

Note that our current implementation includes only bound-bound rates in the set of statistical equilibrium equations -- bound-free processes are neglected (but will be implemented as part of future ionization modules). Consequently, our \texttt{nlte} excitation mode will not yield accurate populations for levels that are significantly (de)-populated via bound-free processes. 
Nevertheless, our \texttt{nlte} excitation mode is certainly an improvement over LTE (or dilute-LTE) populations and provides a convenient means by which to quantify the plausible scale of errors introduced by the simple analytic formulae (Section~\ref{sec:results_nlte}).

\subsection{Initialization of Monte Carlo quanta}

For each \gls{montecarlo} simulation of the radiation field, we begin by initializing a population of $r$-packets at the inner boundary of the computational domain. Each $r$-packet is an indivisible quantum of radiative energy with an associated photon frequency ($\nu$). We choose to initialize all our $r$-packets with equal co-moving frame energy $E$.

Currently it is assumed that the radiation field injected through the inner boundary has a black body frequency distribution at temperature $\tinner$ such that the luminosity launched at the inner boundary is

\begin{equation}
L_{i} = 4 \pi \rinner^2 \sigma \tinner^4
\end{equation}
where $\rinner$ is the radius of the inner boundary.
Consequently, the co-moving frame frequencies of the packets are selected by randomly sampling the Planck function (for $\tinner$) and the packet energies are chosen to match $L_{i}$: 
i.e., in a simulation with $N$ $r$-packets, the co-moving frame energy of each packet is initialised to

\begin{equation}
E = \frac{4 \pi \rinner^{2} \sigma \tinner^4}{N} \Delta t
\end{equation}

\noindent where $\Delta t$ is the (arbitrarily chosen) time interval represented by the simulation.
The procedure adopted to choose $\tinner$ is described in Section~\ref{sec:montecarlo_iteration}.
The $r$-packets are assigned initial directions of propagation specified by $\mu = \sqrt{z}, z \in (0, 1]$, as appropriate for zero limb darkening. 
\footnote{We specify direction of propagation by $\mu = \cos \theta$, where $\theta$ is the angle between the packet flight path and the radial direction.}

\subsection{Radiation-matter interactions}
\label{sec:rad_matter_interaction}

In our indivisible energy packet scheme, opacity does not destroy \gls{montecarlo} quanta but can change both the photon frequency and direction of propagation associated with a packet. The co-moving frame energy of the packet is always conserved during interaction. This ensures that radiative equilibrium is strictly enforced throughout the simulation.

Currently, \gls{tardis} treats only two classes of radiation-matter interaction -- Thomson scattering by free electrons and bound-bound interactions with atoms/ions. 
For Thomson scattering the opacity 
encountered by an $r$-packet in a path length $s$ in direction $\mu$ is given by 

\begin{equation} 
\tauelectron=\kappa_e s 
\end{equation}

\noindent where $\kappa_e = \sigma_\textrm{T} \nelectron D_{\mu}$ is the observer frame opacity coefficient. Here, $\sigma_\textrm{T}$ is the Thomson cross section and $\nelectron$ is the number density of free electrons. We include the first order Doppler factor $D_{\mu} = (1 - \mu v / c)$ 
to preserve $O(v/c)$ accuracy in transforming between observer and co-moving frames \citep[following][]{2005A&A...429...19L}. 
\gls{tardis} treats Thomson scattering as a coherent scattering process: it causes packets to change their direction but not their co-moving frame frequency or energy. Following a Thomson scattering event, the new direction is drawn from an isotropic distribution, $\mu = -1 + 2z, z \in [0, 1]$.

Bound-bound transitions are the dominant (and also most complex) opacity source in \snia ejecta \citep{2000ApJ...530..757P}. \gls{tardis} treats bound-bound opacity in the Sobolev approximation \citep[see e.g.][]{1999isw..book.....L}, which is appropriate for media with large velocity gradients.
In the case of homologous expansion, the Sobolev optical depth of the transition between lower state $l$ and upper state $u$ is given by  

\begin{equation}
\tausobolev =  \frac{\pi e^2}{m_e c}\, f \lambda_{lu} t_{\rm exp} n_l\, \left(1 - \frac{g_l n_u}{g_u n_l}\right) \, ,
\end{equation}
where $f$ is the absorption oscillator strength of the transition, $\lambda_{lu}$ is the wavelength and $t_\textrm{exp}$ the time since explosion. The last term (in brackets) corrects for stimulated emission.

During the \gls{montecarlo} simulations (see Section~\ref{sec:montecarlo_propagation}), $r$-packets can be absorbed by bound-bound transitions. Currently,
three approaches are implements in \gls{tardis} to 
described the re-emission of an absorbed $r$-packet. In the simplest approach (mode \texttt{scatter}), all bound-bound interactions are treated as resonance scattering events [cf. \citealt{1985ApJ...294..619B,1993ApJ...405..738L}; Long \& Knigge (2002, LK02)]. Consequently, absorbed $r$-packets are simply re-emitted with the same co-moving frame frequency but a new direction of propagation (determined by randomly sampling an isotropic distribution, as above).

Our most sophisticated approach to line interactions is an implementation of the macro atom scheme (mode \texttt{macroatom})  devised by \citet{2002A&A...384..725L,2003A&A...403..261L} and used by \citet{2009MNRAS.398.1809K}. In the \texttt{macroatom} mode, bound-bound absorption of an $r$-packet activates a macro atom to the upper level of the absorbing transition. 
The macro atom algorithm is then used to simulate the re-emission of the absorbed energy, in accordance with the assumptions of radiative and statistical equilibrium.
Currently, our macro atom implementation includes only bound-bound radiative transitions although we stress that the method can be readily extended to account for collisional and/or bound-free processes if suitable atomic data are provided.
When a macro atom is activated, it can undergo a sequence of internal state transitions until ultimately deactivating and returning an $r$-packet to the \gls{montecarlo} simulation \citep[see][]{2002A&A...384..725L,2003A&A...403..261L}.

Our third method of handling bound-bound interaction is a simplified version of macro atom, referred to as \texttt{downbranch}. This version is effectively the same as the \texttt{macroatom} scheme except that the probabilities for all internal transitions are set to zero. Thus macro atoms are forced to deactivate directly from the state originally activated. This approach avoids some of the computational cost of the full macro atom machinery while retaining a simplified treatment of fluorescent processes, akin to that introduced by \citet{1999A&A...345..211L}.

\subsection{Propagation of Monte Carlo quanta}
\label{sec:montecarlo_propagation}

In the current implementation, there are three processes that can act to terminate the flight path of an $r$-packet: reaching a grid zone boundary, undergoing electron scattering or being absorbed by a bound-bound transition. Choosing which class of event occurs is determined via a simple \gls{montecarlo} experiment. First, an optical depth ($\tauevent$) that the $r$-packet may propagate without being absorbed is randomly selected in accordance with the $\exp(-\tauevent)$ attenuation law:
$\tauevent = \ln{z}, z \in (0, 1]$. Next we identify the closest redward line transition to the co-moving frame frequency of the $r$-packet. Then we calculate the distance (\dline) the packet must travel to Doppler-shift into Sobolev resonance with that line. We also calculate \delectron, defined by $\tauevent = \kappa_e \delectron$.
Finally, we calculate the distance \dshell the $r$-packet would need to propagate to reach the boundary of the current shell. 

To determine which class of event will terminate the flight path, we compare the three distances 
\dline, \delectron and \dshell. If \dline is the shortest, then bound-bound absorption is possible. To test whether line absorption occurs, we compare $\taucombined = \tausobolev + \sigma_\textrm{T} \nelectron \dline$ to $\tauevent$. If $\taucombined > \tauevent$ an interaction with the line will occur: we propagate the $r$-packet to the position of Sobolev resonance and then process the line interaction event in accordance with the procedures described in Section~\ref{sec:rad_matter_interaction} for bound-bound absorption. Alternatively, if $\taucombined < \tauevent$, we reduce $\tauevent \rightarrow \tauevent - \tausobolev$ and recompute $\delectron$. We also recompute $\dline$, now considering the next redward line transition. With new values of $\delectron$ and $\dline$ we again compare the three distances (\dline, \delectron, \dshell) and continue this process until either a line absorption event is triggered or until \dline is no longer the shortest distance.

If \delectron is shortest, an electron scattering event terminates the $r$-packet flight path.
The packet is propagated through distance $\delectron$ and sent to the algorithm dealing with electron scattering events (see Section~\ref{sec:rad_matter_interaction}).

Finally, if \dshell is shortest, the $r$-packet is propagated to the appropriate grid cell boundary and flagged as having successfully crossed that boundary. If the cell boundary is internal to the computational domain, the propagation of the packet through the new shell continues as before (noting that all material properties relevant to computing the opacity may have changed). When an $r$-packet reaches either of the boundaries of the computational domain (outer or inner), its flight path terminates and its final properties are recorded. $r$-packets reaching the outer boundary are assumed to the escape freely (such that they can contribute to the observable spectrum). Packets crossing the inner boundary are assumed to be reabsorbed by the inner ejecta and thus lost.

\subsection{Monte Carlo estimators}
\label{sec:montecarlo_estimators}
During \gls{montecarlo} simulations, \gls{tardis} uses the $r$-packet trajectories to collect estimators for radiation field properties, as required for calculation of the ionization and excitation conditions.
Specifically, we record a pair of estimators (following ML93; LK02),
\begin{equation}
\jestimator = \frac{1}{4\pi \Delta t V} \sum{E l D_{\mu}}
\end{equation}

\noindent and
\begin{equation}
\nubarestimator = \frac{1}{4\pi \Delta t V} \sum{E \nu l D_{\mu}}
\end{equation}
 for each grid cell. The summation is made over all $r$-packet trajectories inside the cell. For each trajectory, $E$ and $\nu$ are the packet energy and frequency (in the co-moving frame) and $l$ is the length of the trajectory (measured in the observer frame).
After each \gls{montecarlo} calculation, we use these estimators to obtain new values for the parameters of our radiation field model (see Section~\ref{sec:rad_model}):

\begin{equation}
\trad = \frac{h}{k_{B}}  \frac{\pi^4}{360 \zeta(5)} \frac{\nubarestimator}{\jestimator}
\end{equation}

\noindent and

\begin{equation}
W = \frac{\pi \jestimator}{\sigma_{SB} \trad^4}\:\: .
\end{equation}

\noindent In {\texttt{detailed}} mode, we also record estimators for the mean intensity at the violet wing of each line transition $J_{lu}^b$ in each cell \citep{1999A&A...345..211L}:

\begin{equation}
J_{lu}^b = \frac{1}{4\pi \Delta t V} \frac{t_{exp}}{c} \sum \frac{E}{\nu} D_{\mu}
\label{eq:jblue}
\end{equation}
where the summation is now over all $r$-packets that pass through Sobolev resonance with the $l \rightarrow u$ transition in the cell.

\subsection{Iteration cycle}
\label{sec:montecarlo_iteration}

\gls{tardis} performs a sequence of \gls{montecarlo} simulation during which the values of $\tinner$, $\trad$ and $W$ (and, in \texttt{detailed} mode, $J_{lu}^b$) are iteratively improved until convergence is reached.

To initialize a calculation, we set $\trad$ (in all shells) to a sensible guess, typically $10000$~K. The dilution factor is initialized to follow geometric dilution, $W=\frac{1}{2}[1 - (1- (\rinner/r)^2)^{1/2}]$, where $r$ is the radius at the centre of each shell and $\rinner = \vinner t_{exp}$ is the radius of the inner boundary. In \texttt{detailed} mode, we initialize $\jblue = W B_{\nu_{lu}}(\trad)$.
Although $\tinner$ can also be treated as a simple input parameter, we anticipate the goal of fitting observations and so generally adopt the luminosity at the {\it outer} boundary ($L_o$) as a simulation parameter. As a first guess we then adopt

\begin{equation} 
\tinner = \left( {\frac{\louter}{ 4 \pi \rinner^2 \sigma}} \right)^{1/4} \; .
\end{equation}

A \gls{montecarlo} simulation is then performed (see Section~\ref{sec:montecarlo_propagation}) and the resulting estimators (see Section~\ref{sec:montecarlo_estimators}) are used to update the plasma properties (see Section~\ref{sec:plasma_state}) in each shell. 
We compare the total energy emerging through the outer boundary to the requested value of $\louter$ and modify $\tinner$ (and therefore the luminosity at the inner boundary) to obtain better agreement (i.e., if the emergent luminosity is lower than requested, $\tinner$ is increased and vice versa).
Using these updated quantities, a new \gls{montecarlo} calculation is carried out and used to update the model properties again. This process is repeated until the plasma state has converged to sufficient accuracy (or a chosen maximum number of iterations is reached); convergence properties are described for example calculations in Section~\ref{sec:numerics}.

\subsection{Synthetic spectrum}
\label{sec:synthetic_spectrum}

Synthetic spectra can be obtained directly from \gls{montecarlo} radiative transfer calculations by binning the quanta that emerge through the outer boundary in frequency-space.
However, 
\citet{1999A&A...345..211L} and \citet{2000A&A...363..705M}, showed that higher-quality synthetic spectra can be extracted from the simulations using only slightly more sophisticated algorithms 
and various such approaches have been used in \gls{montecarlo} codes (ML93; LK02; \citealt{2008MNRAS.385.1681S}).

In \gls{tardis}, we have implemented a simple approach to compute the synthetic spectrum, very similar to that adopted by LK02 and \cite{2010MNRAS.404.1369S}. This method is fully consistent with the macro atom scheme \citep{2002A&A...384..725L, 2003A&A...403..261L}. The spectrum is calculated during one final \gls{montecarlo} simulation that is performed once the iterative sequence of simulations discussed in Section~\ref{sec:montecarlo_iteration} has converged. During this last simulation, whenever an $r$-packet is about to be launched on a new trajectory (either because it is about to be injected through the inner boundary, or because an interaction occurred inside the computational domain), the main simulation is suspended and a set of $N_{v}$ test packets, which we refer to as {\it virtual} ($v$-) packets, are created. These $v$-packets have properties identical to those of the $r$-packet that was about to be launched, except that they each have a different propagation direction and their energies are assigned based on the probability distribution associated with the creation of the original $r$-packet (see below). Since the only purpose of $v$-packets is to estimate a contribution to the emergent spectrum, we do not spawn $v$-packets on trajectories that ultimately intersect the inner boundary. Thus, for $v$-packets created at radius $r$, we assign directions ($\mu_v$) in the interval $\mu_{min} < \mu_v < 1$, where $\mu_{min} = -\sqrt{(1 - (\rinner/r)^2)}$. For  $v$-packets generated following a physical interaction inside the domain, the co-moving frame energy assigned ($E_{v}$) is simply

\begin{equation}
E_{v} = E \frac{1 - \mu_{min}}{2 N_{v}}
\end{equation}
where $E$ is the co-moving frame energy of the parent $r$-packet.
For $v$-packets created when an $r$-packet is about to be injected through the inner boundary, the energies assigned are modified to account for the adopted angular distribution of the incoming radiation field
\begin{equation}
E_{v} = E \frac{2 \mu_v}{N_{v}}
\end{equation}
Once created, each  $v$-packet is propagated though the simulation domain. Unlike $r$-packets,  $v$-packets do not undergo physical interactions with the medium (they are never scattered), but the total optical depth along the trajectory of the$v$-packet ($\tau_v$) is recorded. Each $v$-packet makes a contribution to the photon frequency-bin of the synthetic luminosity (ergs s$^{-1}$ Hz$^{-1}$) spectrum in which its observer frame frequency lies

\begin{equation}
L_{v} = \frac{E_v}{\Delta t \Delta \nu} \exp({- \tau_v})
\end{equation}
where $\Delta \nu$ is the width of the spectral frequency bin. The spectrum computed from the $v$-packets agrees with that obtained by direct binning of emerging $r$-packets but typically has considerably less \gls{montecarlo} noise. The reduction in \gls{montecarlo} noise and the computational overhead introduced by the $v$-packet step are determined by the choice of $N_{v}$. For the calculations in this paper we typically use $N_{v} = 3$ which reduced the \gls{montecarlo} noise by a factor of $\approx 3$.

\section{Datasources}
\label{sec:datasources}

Our radiative transfer calculations require input atomic data for the calculation of the ionization/excitation state of the plasma (ionization potentials and atomic models), bound-bound opacities and radiative rate coefficients (lines lists and oscillator strengths) and, in our \texttt{nlte} excitation mode, electron collision rate coefficients (thermally averaged collision strengths). 

For the simulations presented below we use an atomic database that includes all elements with $Z\le30$.
Atomic masses were taken from \citet{wieser2011atomic} and ionization threshold energies from \gls{nist}. 
The modified nebular approximation (see Section~\ref{sec:plasma_state} equation~\ref{eq:modified_nebular}) requires that the fraction of recombinations to the ground state ($\zeta$-factor) be specified. 
For this, values were extracted from the data compiled by LK02 for a
 fixed set of temperatures (2000 -- 50000~K; 2000~K grid spacing). During code execution, linear interpolation between these values is used to obtain $\zeta$ as a function of temperature.

We have drawn atomic/ionic energy levels and bound-bound (line) transition data from two sources and constructed two atomic databases. In the first case, we populate the lines and levels database from the \gls{kuruczlinelist} omitting levels above the ionization threshold and lines with an oscillator strength $\log{gf} < -3$. This dataset contains no electron collision rate data. For ions not present in the \gls{kuruczlinelist}, we use atomic models consisting of only the ground state (statistical weights from \gls{nist}) and no line transitions.
We use this dataset for several of the test calculations presented in Section \ref{sec:code_comparisons}.

Secondly, 
we constructed a data set in which the \gls{kuruczlinelist} data are replaced with data from \gls{chianti} for select species of relevance to SNe~Ia: \ion{Si}{2}, \ion{S}{2} \ion{Mg}{2} and \ion{Ca}{2}. \gls{chianti} provides atomic models and data for both radiative and electron collision bound-bound transitions. For ions taken from \gls{chianti}, no cuts were applied to $\log{gf}$ in selecting radiative data. 
Electron collision rates were calculated using the \gls{chianti} atomic dataset for temperatures between 2000~K and 48000~K (in steps of 2000~K) and interpolated during code execution. 
This data set is used for most of the the calculations presented in Sections~\ref{sec:numerics}, \ref{sec:code_comparisons} and \ref{sec:results}.

The two atomic data sets used here have been constructed with the goals of this study in mind (code testing and simple differential tests of modelling assumptions). However, \gls{tardis} is capable of handling much larger atomic datasets and can easily be extended to include additional atomic processes.

\section{Convergence tests}
\label{sec:numerics}

\begin{figure}
   \centering
   \includegraphics[width=\columnwidth]{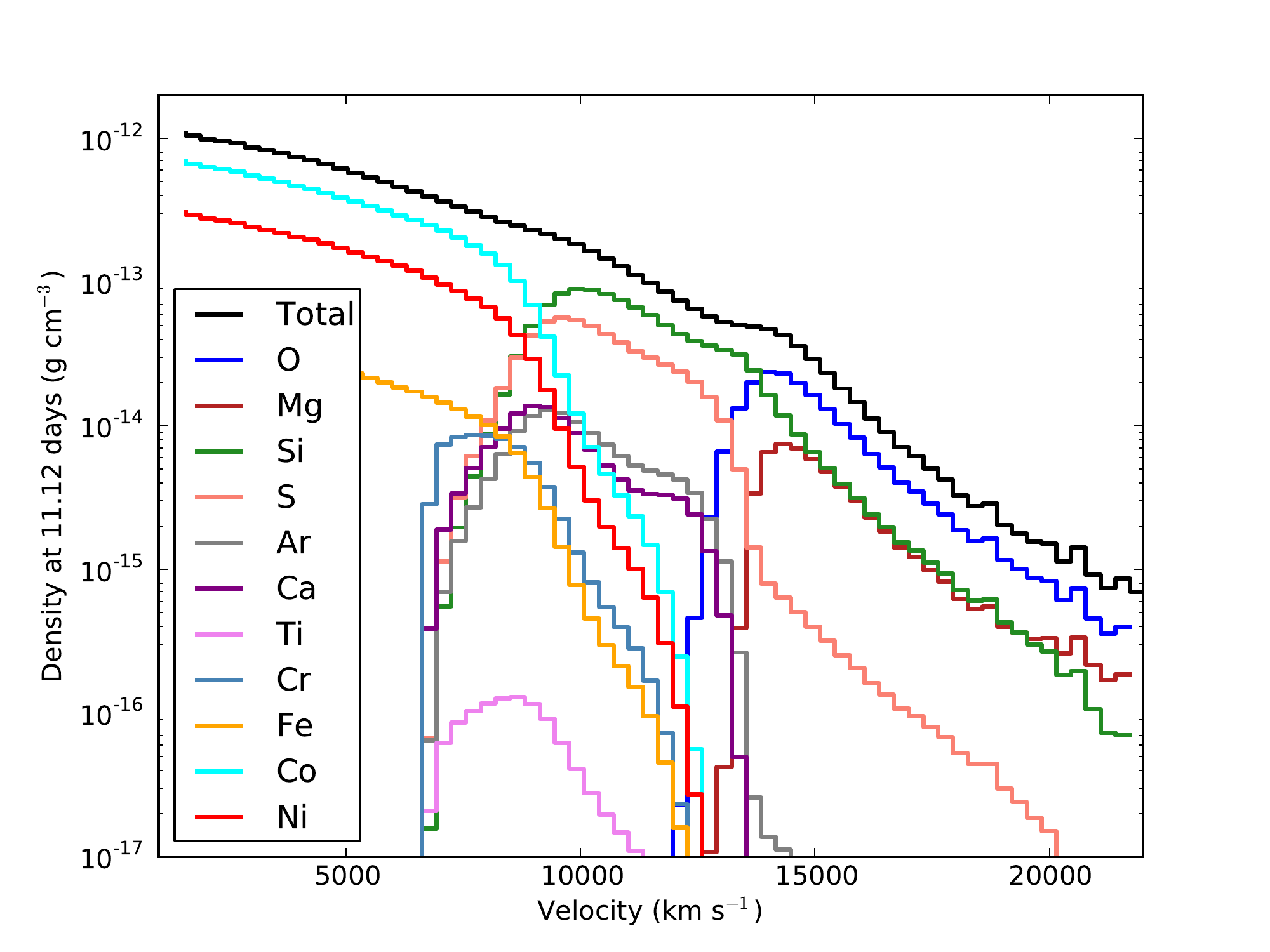} 
   \caption{Densities for the 1.06~M$_{\odot}$ CO WD detonation model of \citet{2010ApJ...714L..52S} at 11.1~days after explosion. The black line shows the total mass density and the coloured lines show the densities of important elements in the ejecta.}
   \label{fig:artis_model}
\end{figure}

As described in Section~\ref{sec:montecarlo_iteration}, \gls{tardis} performs an iterative sequence of simulations. Here we describe tests verifying the convergence properties of this iteration procedure.
For these tests, we have chosen to use a model that is based on the output of a 1D \snia\ explosion simulation -- this allows us to study the operation of the code applied in the most complex regime available (model with arbitrary density profile and stratified abundances including a large number of elements). \footnote{We note, however, that \gls{tardis} does not require such a complex input model (see Section~\ref{sec:method} and Appendix A). In many applications, simpler models will be more efficient for exploration of parameter space and attempting to fit observations.}
Specifically, we
adopt the model for a detonation of a 1.06~M$_{\odot}$ white dwarf (WD) \citep{2010ApJ...714L..52S} 
that was computed for a uniform initial composition of $^{12}$C and $^{16}$O (50:50 mix with no $^{22}$Ne).
The density and composition for this model during the homologous phase are illustrated (at $t = 11.1$~days)  in Fig.~{\ref{fig:artis_model}}. The model has strongly layered ejecta, with \glspl{ime} dominating the composition above $\sim 9000$~km~s$^{-1}$ and iron-group elements dominating below.
The complete model from Sim et al. (2010) consists of 70 shells. However, since inner and outer boundaries are imposed by \gls{tardis}, typically only around half of the shells are included in the \gls{tardis} calculations.
 For the test calculations we adopt $t_{exp} = 11.1$~days, $v_{i} = 11000$~km~s$^{-1}$, $\vouter = 22000$~km~s$^{-1}$ and $\log_{10} \louter / L_{\odot} = 9.34$ (the luminosity obtained at this epoch from \gls{artis} simulations -- see Section~\ref{sec:artis_compare}). 

We have tested convergence by carrying out a sequence of 30 iterations with $2\times 10^5$ \gls{montecarlo} quanta per iteration. To test all aspects of the current implementation, we show results from calculations with the most sophisticated set of modes of operation: \texttt{macroatom} mode for line interactions, \texttt{detailed} mode for calculation of radiative rates and \texttt{nebular} ionization mode. We use \texttt{nlte} excitation mode for \ion{Si}{2} and \texttt{dilute-lte} excitation for all other ions.

As described in Section~\ref{sec:montecarlo_iteration}, the code iteratively tries to match the requested SN luminosity ($\louter$) by changing the temperature of the black-body spectrum emitted from the inner boundary. This in turn changes the radiation field in the cells above (parametrized by \trad and $W$), which subsequently influences the ionization and excitation of the plasma. 
To speed convergence, we found it useful to adopt a scheme in 
which \tinner is only changed after every third iteration: this allows \trad and $W$ to respond to changes in \tinner, helping the code make a more informed choice when modifying \tinner in an attempt to match $\louter$.

\begin{figure*}
   \centering
   \includegraphics[width=0.48\textwidth]{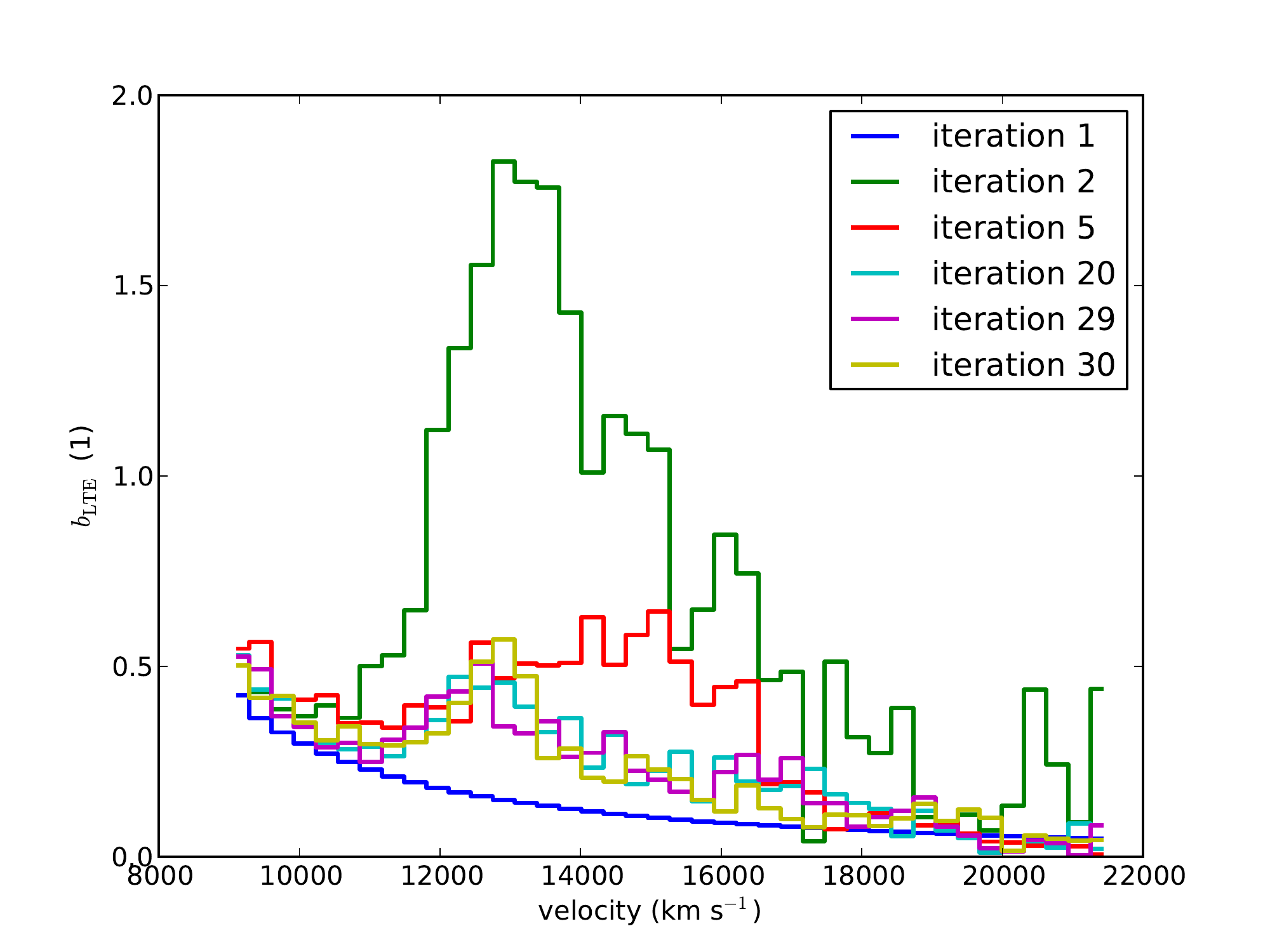} 
   \includegraphics[width=0.48\textwidth]{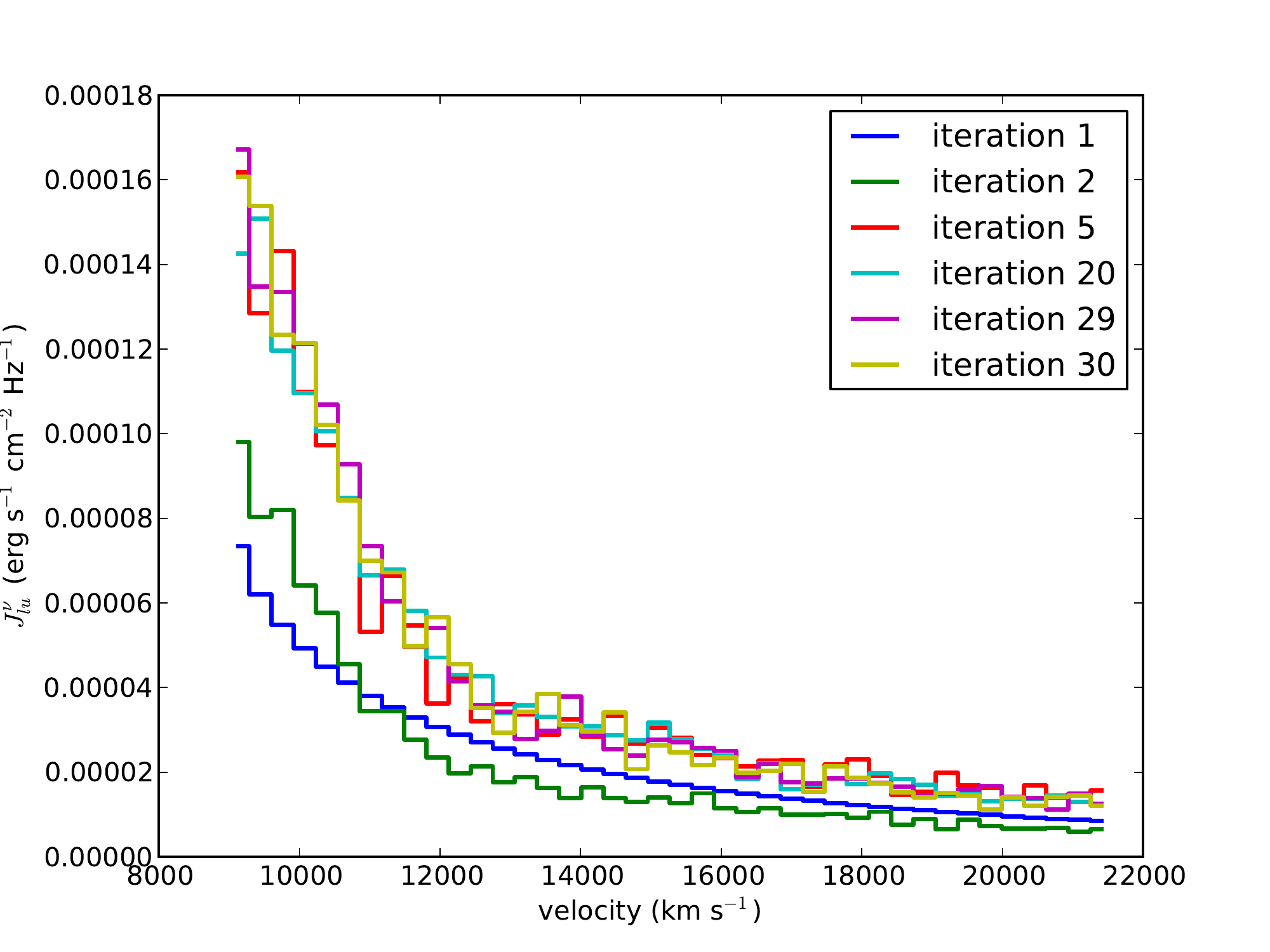} \\
   \caption{The left panel shows a convergence test for the
     departure coefficient $b_\textrm{LTE}$ (see
     Section~\ref{sec:compare_chianti}) of \ion{Si}{2} $3\textrm{s}^2
     4\textrm{s}\ ^2\textrm{S}_{1/2}$. For selected transitions, the rightpanel shows the
     convergence of \jblue calculated using \gls{montecarlo}
     estimators. The calculation used $2 \times 10^5$ \gls{montecarlo} quanta.}
   
   \label{fig:tardis_convergence_nlte}
\end{figure*}

\begin{figure*}
   \centering
   \includegraphics[width=0.48\textwidth]{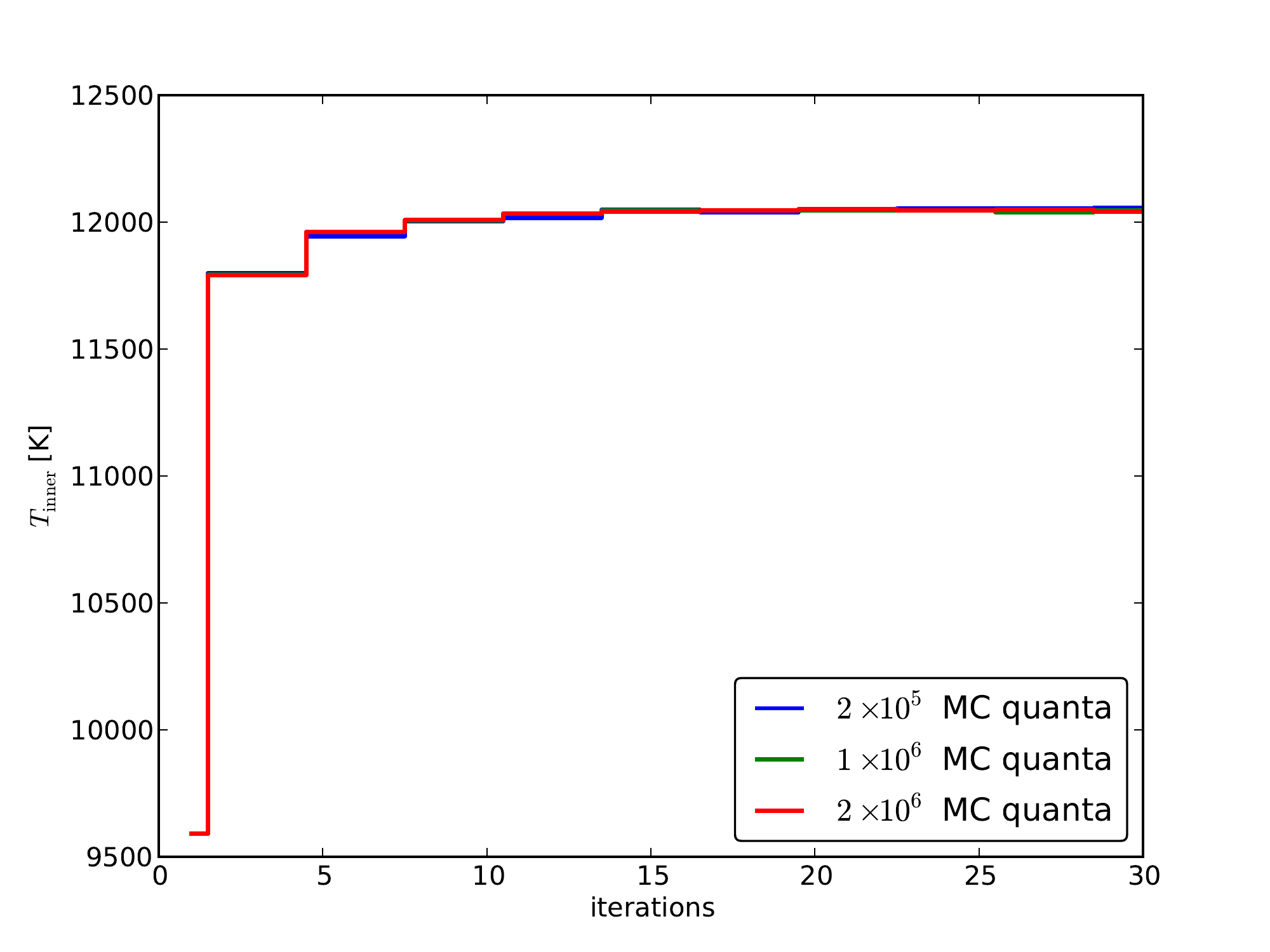} 
   \includegraphics[width=0.48\textwidth]{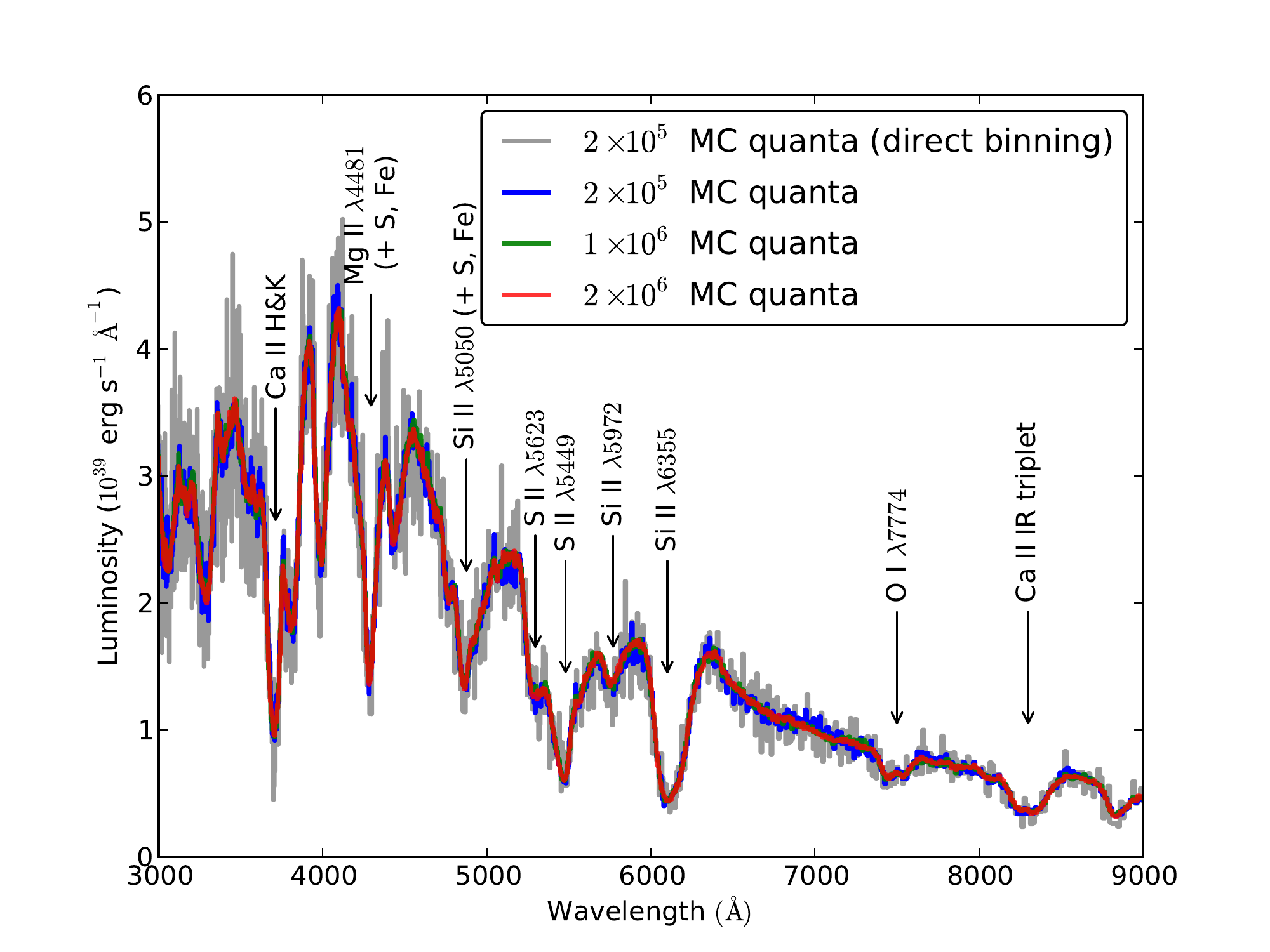} \\
   \caption{Results of \tinner convergence tests with different
     numbers of \gls{montecarlo} quanta are shown on the left. The final spectra are shown on the right.
For the $2 \times 10^5$ calculation we show both the spectrum obtained
by directly binning the emergent \gls{montecarlo} quanta and that
obtained using virtual packets (described in
Section~\ref{sec:synthetic_spectrum}). For all other cases we show
only the virtual packet spectrum. The positions of several prominent
features (that will be discussed in later sections) are marked (and
major blends are indicated, where appropriate).}
   \label{fig:tardis_convergence_mcquanta}
\end{figure*}

Fig.~\ref{fig:tardis_convergence_nlte} (upper panels) shows the rapid and uniform convergence for both \trad and $W$ for our test calculation.  By iteration 20 both quantities have ceased to evolve with iteration number. This is driven by the quick convergence of \tinner (see Fig.~\ref{fig:tardis_convergence_mcquanta}). 

\begin{figure*}
   \centering
   \includegraphics[width=0.48\textwidth]{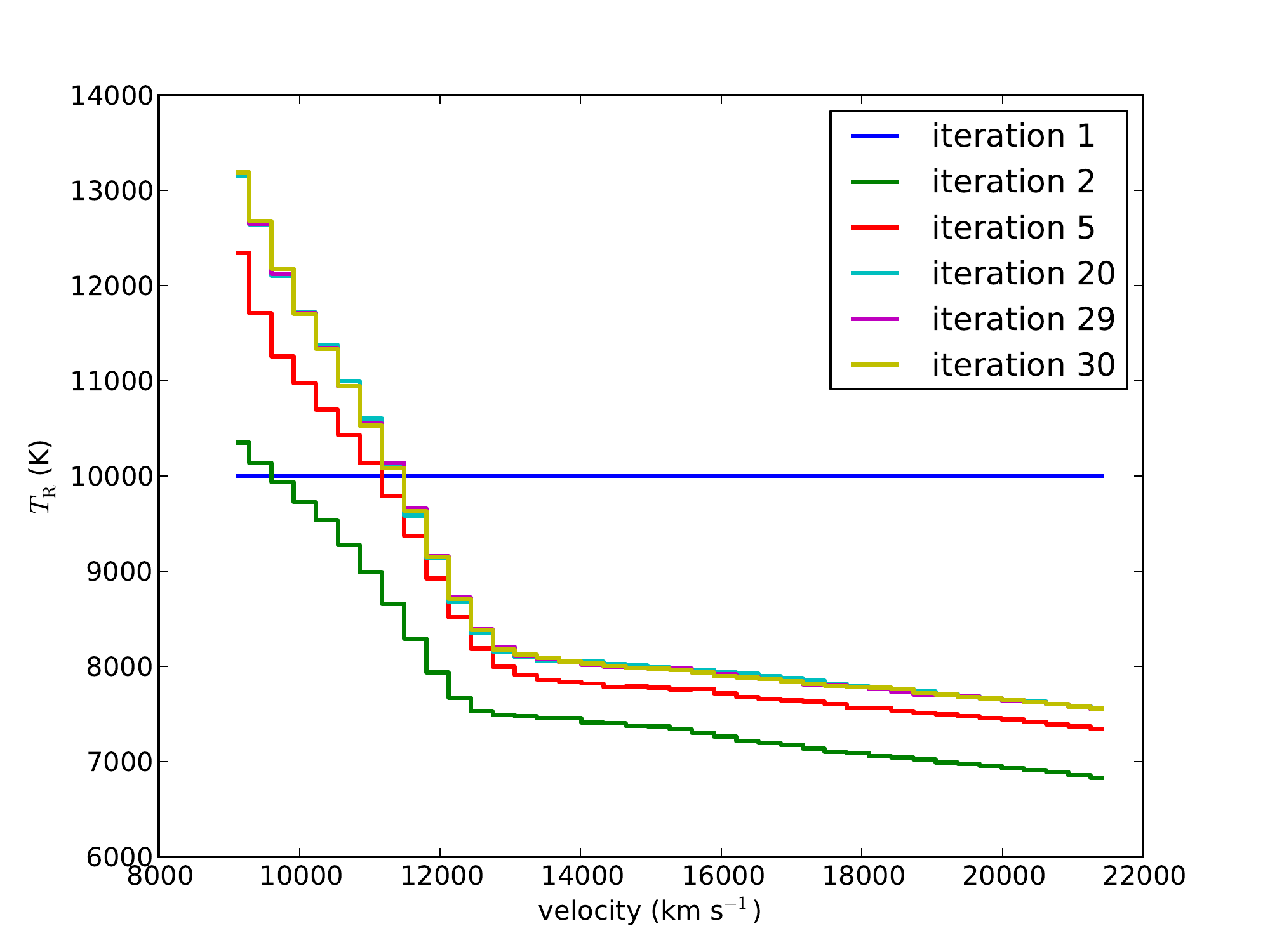} 
   \includegraphics[width=0.48\textwidth]{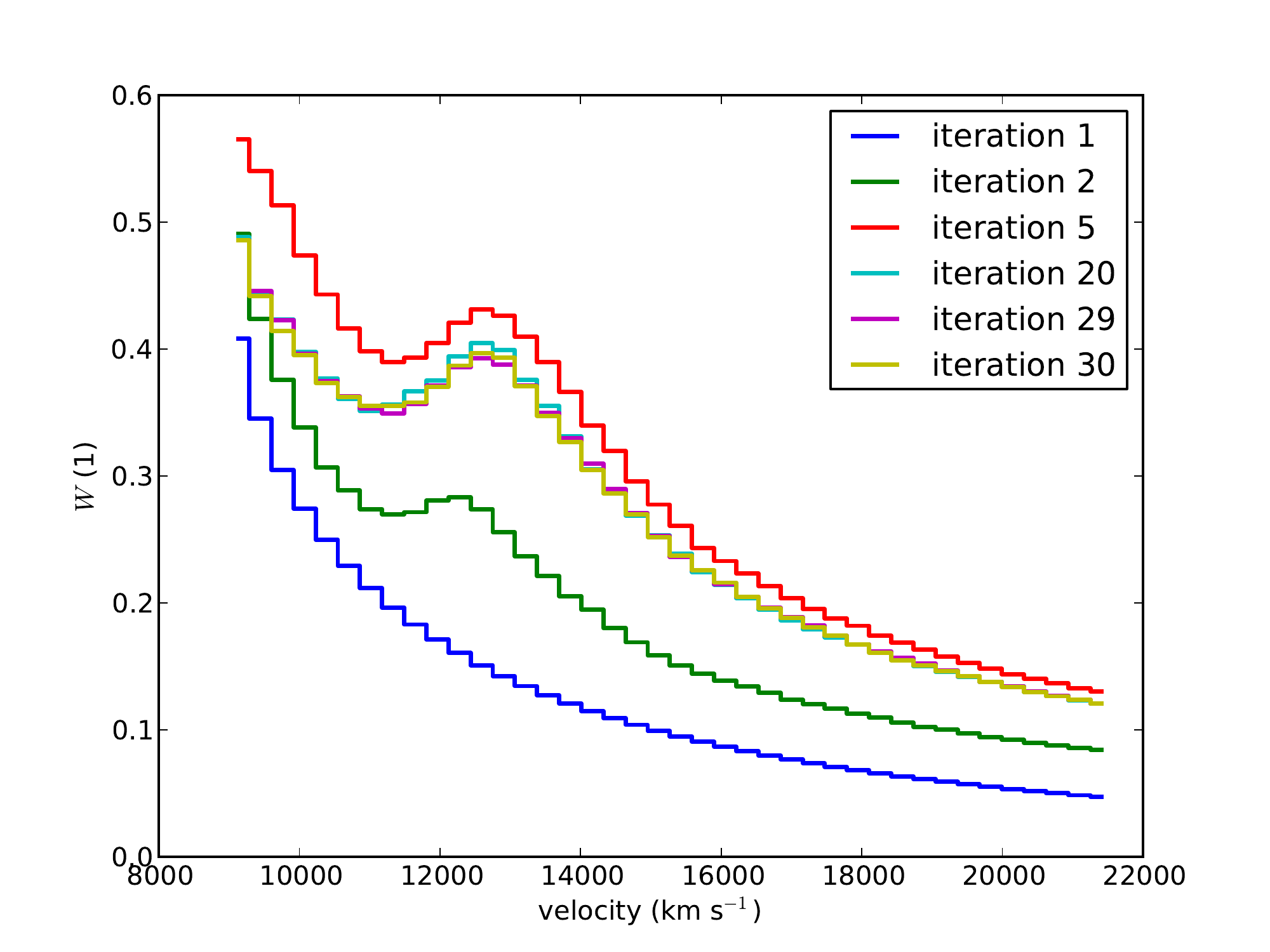}
   \caption{Convergence tests of \trad (left panel) and $W$
     (right panel) for a \gls{tardis} calculation with $2 \times 10^5$ \gls{montecarlo} quanta. }
   \label{fig:tardis_convergence_t_w}
\end{figure*}

We have tested the convergence properties of level populations for species treated with our 
\texttt{nlte} excitation mode by examining the evolution with iteration number of departure coefficients defined by

\begin{equation}
b_\textrm{LTE}=\frac{n_i/n_0}{n_i^\textrm{LTE}/n_0^{\textrm{LTE}}},
\end{equation}
where $n_i$ is the level population of the selected level and $n_0$ is that of the species ground state. 
This is illustrated for the
$3\textrm{s}^2 4\textrm{s}\ ^2\textrm{S}_{1/2}$  state of \ion{Si}{2} (see Fig.~\ref{fig:grotrian_si2}), in 
Fig.~\ref{fig:tardis_convergence_nlte}. 
Since the \texttt{nlte} excitation  level populations are strongly affected by the radiative rates, we also show the convergence of \jblue for \ion{Si}{2} $\lambda 6347$ in Fig.~\ref{fig:tardis_convergence_nlte}.
Although considerably more affected by \gls{montecarlo} noise than \trad or $W$, both \jblue and the departure coefficient have converged to a stable pattern after roughly 20 iterations.
We explore the influence of \gls{montecarlo} noise in the estimators in Section~\ref{sec:results_jblue}.


We have tested the sensitivity of our results to the number of \gls{montecarlo} quanta by repeating the calculation described above using $10^6$ and $2 \times 10^6$ \gls{montecarlo} quanta. The convergence of these runs was essentially identical to those of our calculation with $2 \times 10^5$ quanta and led to no changes in the output spectra (see Fig.~\ref{fig:tardis_convergence_mcquanta}). In addition, we show that the spectra extracted using \textit{virtual} packets (see Section~\ref{sec:synthetic_spectrum}) are much less noisy than those obtained by directly extracting a spectrum by binning emergent quanta.


Finally, to test the influence of grid resolution, we carried out additional runs in which we increased the number of shells used by factors of two, five and ten (using $2 \times 10^5$ \gls{montecarlo} quanta in each case). We interpolated the velocities linearly (the entire model follows a strictly homologous velocity law), but did not interpolate the densities or abundances (we aim to test the influence of resolution on our computed radiation field properties, not the input model). Despite the finer sampling, the converged properties and the emergent spectrum for these tests were unaffected indicating that the resolution adopted in our standard calculation was adequate.

\section{Code Comparisons}
\label{sec:code_comparisons}

We have made comparisons between \gls{tardis} calculations and those of several other codes. We first focus on the calculation of the emergent spectrum (for fixed plasma conditions) by comparison with \gls{synpp} (Section~\ref{sec:compare_synpp}). We then use the  \gls{chiantipy}  package to test our solver for NLTE level populations in Section~\ref{sec:compare_chianti}.

In Sections~\ref{sec:python_compare} and \ref{sec:artis_compare} we make more sophisticated tests by comparing \gls{tardis} synthetic spectra to those of two alternative radiative transfer codes, {\sc python} \glsdisp{python_rt} and \gls{artis}. 
Like \gls{tardis}, these codes use \gls{montecarlo} methods and a Sobolev treatment of line opacity. However, they have different treatments of NLTE ionization, bound-free opacity and radiation sources (as detailed below).
Comparing their results allows us both to identify potential issues (with any of the codes) and to quantify the effects of the different assumptions made by the codes. 

\subsection{Comparison to {\sc syn++}}
\label{sec:compare_synpp}

Similar to \gls{tardis}, \gls{synpp} adopts a spherically symmetric SN model in homologous expansion. \gls{synpp} handles line opacity in the Sobolev approximation and assumes all lines can be treated in the resonance scattering limit. In contrast to \gls{tardis}, \gls{synpp} does not accept density and abundances as input parameters, but rather the opacity for a reference line associated with each ion species. \gls{synpp} then calculates the opacities of other lines using the Boltzmann excitation formula at a user-specified temperature. Thus, comparison with \gls{synpp} does not test the calculation of the plasma state (ionization/excitation) in \gls{tardis}. However, it does provide a direct test of the accuracy to which the spectrum is computed for fixed plasma conditions.

\begin{figure}
   \centering
   \includegraphics[width=\columnwidth]{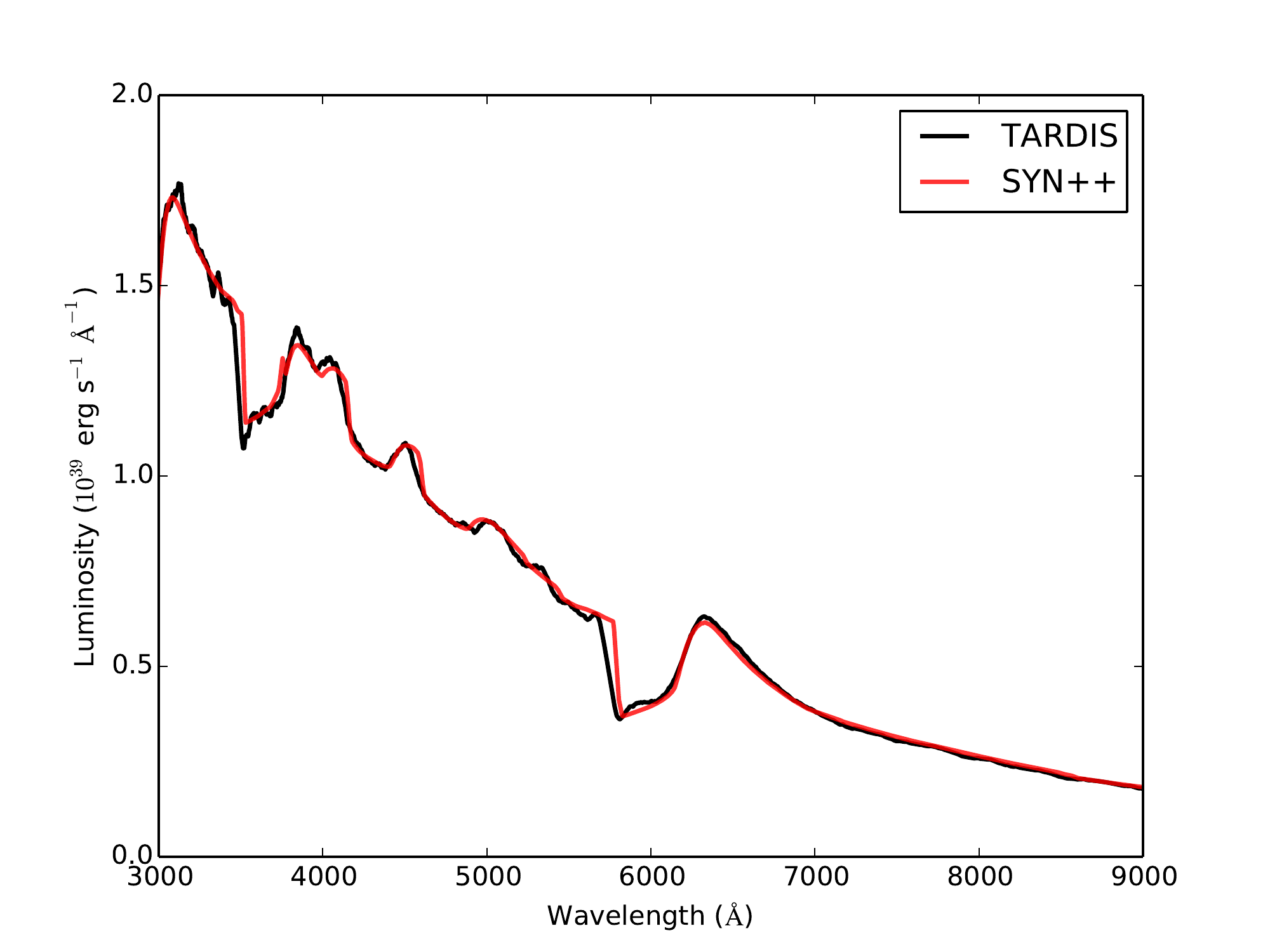} 

   \caption{Comparison of \gls{synpp} and \gls{tardis} synthetic spectra for
     single cell pure silicon atmospheres. For ease of comparison we have convolved the \gls{tardis} output using a Savitzky-Golay filter \citep{doi:10.1021/ac60214a047}.  }
   \label{fig:compare_synpp_tardis}
\end{figure}

For the test, we setup \gls{tardis} with a pure silicon one-zone model. Line interactions were treated in 
\texttt{scatter} mode and electron scattering was disabled for ease of comparison with \gls{synpp}. 
Ionization and excitation were both treated in \texttt{lte} modes and we used the atomic data set drawn only from the \gls{kuruczlinelist}.
The temperature of the inner boundary and of the temperature of the radiation field inside the only cell were set to 10000~K and $10^6$ \gls{montecarlo} quanta were used to generate the \gls{tardis} spectrum. The output \tausobolev\ for the reference lines of species \ion{Si}{1}-\textsc{iv} 
were taken from \gls{tardis} and used to calculate the spectrum with \gls{synpp}. In Fig.~\ref{fig:compare_synpp_tardis}, we illustrate the excellent agreement between both codes, validating the method of spectrum formation used by \gls{tardis}.

\subsection{Testing of NLTE level populations}
\label{sec:compare_chianti}
To test the accuracy of our NLTE level population scheme, we compare our results to those obtained using \gls{chiantipy}. For this test, we calculate the NLTE level populations for a plasma element of fixed electron density $\nelectron$ and kinetic temperature $\telectron$ irradiated by a diluted black body  ($W=0.5$) with radiation temperature $\trad$. Note that, as in \gls{tardis}, the current version of \gls{chiantipy} includes only bound-bound (radiative and electron collision) rates when solving for level populations (i.e. the influence of photo-ionization and recombination are not included).

\begin{figure}
   \centering
   \includegraphics[width=\columnwidth]{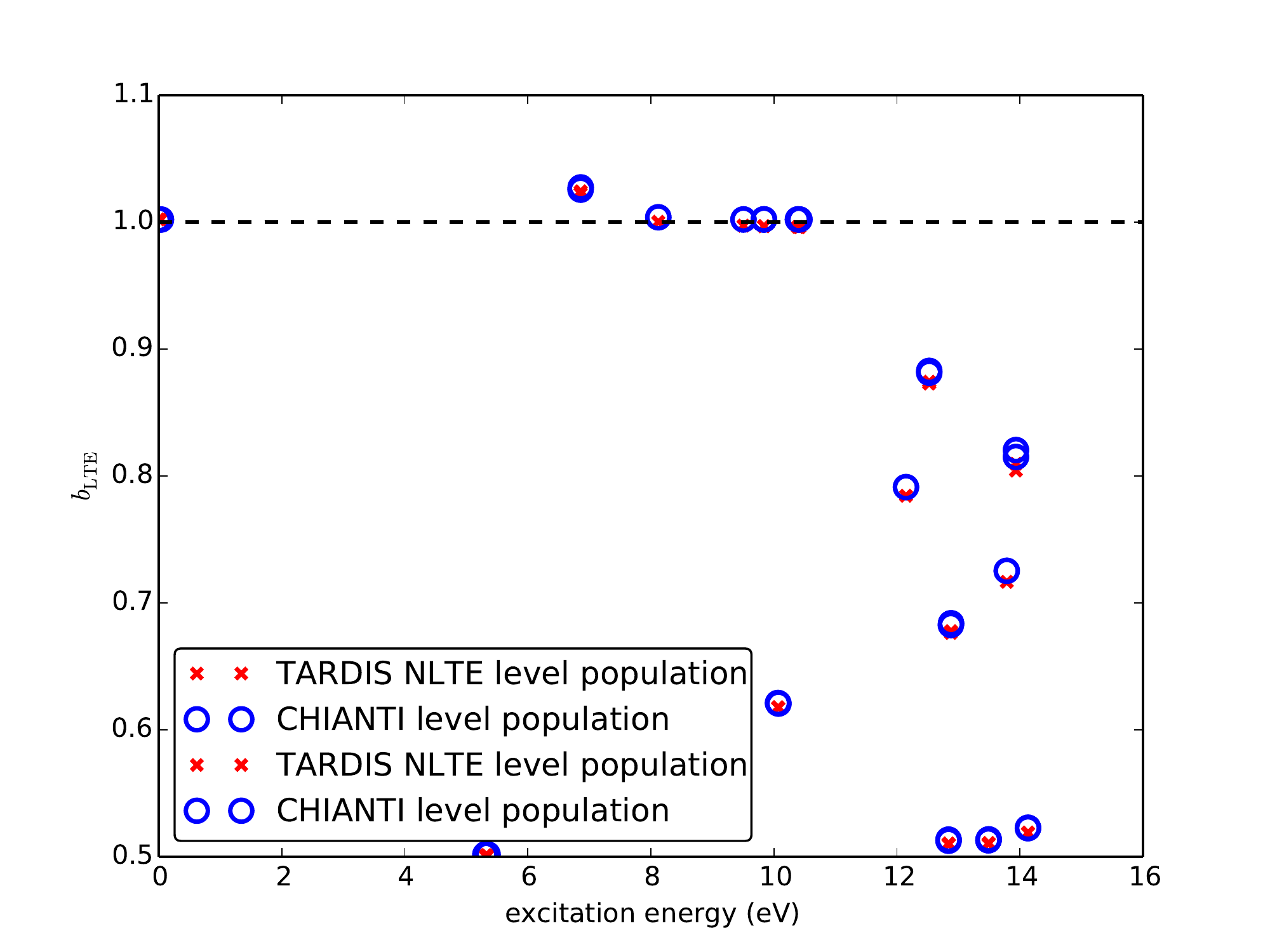} 
   \caption{Comparison of Si~{\sc ii} level populations calculated with
\gls{chianti} and \gls{tardis} as described in Section~\ref{sec:compare_chianti}.}
   \label{fig:tardis_chianti_classical_nebular}
\end{figure}

In general we found excellent agreement between the calculation performed by \gls{tardis} and \gls{chiantipy}. Fig.~\ref{fig:tardis_chianti_classical_nebular} shows the departure coefficient $b_\textrm{LTE}$ for \ion{Si}{2} computed for $\telectron=9000$~K, $\trad=10000$~K and $\nelectron=4.15\times10^8 \textrm{cm}^{-3}$. We find a bias of 0.006 and a standard deviation of 0.003 in the difference between the sets of departure coefficients obtained in the two calculations.


\subsection{Comparison to {\sc python}}
\label{sec:python_compare}

\gls{python_rt} is a 2D MC radiative transfer code, which has been used to model synthetic spectra for accretion disk winds in a variety of astrophysical systems: cataclysmic variables (LK02; \citealt{2012MNRAS.425.1430N}), massive young stellar objects \citep{2005MNRAS.363..615S} and active galactic nuclei \citep{2013MNRAS.436.1390H}. As in \gls{tardis}, an input radiation field is specified and synthetic spectra are computed by tracking the propagation of radiation through a computational domain. \gls{python_rt}  includes bound-bound, bound-free, free-free and electron scattering opacities. Although a \texttt{macroatom} treatment is implemented in \gls{python_rt}, to date this has only been used for modelling \ion{H}{1} recombination lines \citep{2005MNRAS.363..615S}: for metal lines, a two-level atom is used.

For our comparison calculation, we adopt a simple SN~Ia ejecta model, following \citet{1999A&A...345..211L}. Specifically, we choose a density profile based on a simple fit \citep[$\rho \propto v^{-7}$; cf.][]{1985ApJ...294..619B} to the W7 model of \citet{1984ApJ...286..644N}. We carry out the comparison for an epoch of 13~days post explosion, choosing an emergent bolometric luminosity of $\log_{10} \louter/L_{\odot} = 9.44$ and setting the inner boundary of the computational domain at $\vinner = 11000$~km~s$^{-1}$. We set the outer boundary at $\vouter = 20000$~\kms. Since \gls{python_rt} cannot currently handle position-dependent compositions, we adopt a uniform composition for the ejecta. At these epochs, the observed optical spectra of normal \sneia are dominated by features associated with \glspl{ime}. Therefore, we adopted an \gls{ime}-rich composition (see Table~\ref{tab:python_tardis_abund}) with relative abundances based on those found in modern SN~Ia explosion models \citep[specifically][]{2013MNRAS.429.1156S}. 
The \gls{tardis} calculation was carried out using \texttt{nebular} ionization and \texttt{dilute-lte} excitation modes
(see Section~\ref{sec:plasma_state}).
For ease of comparison, we used our \texttt{scatter} mode for line opacity, which is our closest equivalent 
to the two-level treatment of metal lines used in \gls{python_rt},
and our atomic data set drawn from \gls{kuruczlinelist} (see Section~\ref{sec:datasources}). 

\begin{table}
\caption{Elemental mass fractions adopted for the test model used in Section~\ref{sec:python_compare}.}
\label{tab:python_tardis_abund}
\begin{tabular}{llllll} \hline
Element & Mass fraction & Element & Mass fraction \\ \hline
O & 0.19 & Mg & 0.03\\
Si & 0.52 & S & 0.19\\
Ar & 0.04 & Ca & 0.03\\ \hline
\end{tabular}
\end{table}

For the \gls{python_rt} comparison run we set up the same model, using spherical geometry and imposing an homologous velocity law. The input radiation field for the \gls{python_rt} calculation was forced to be the same as in the \gls{tardis} calculation (i.e. we inject a black-body radiation field through the inner boundary with temperature set to be the same as that used by \gls{tardis}). The \gls{python_rt} calculations were also carried out using an atomic data from the \gls{kuruczlinelist}.

\begin{figure}
   \centering
   \includegraphics[width=\columnwidth]{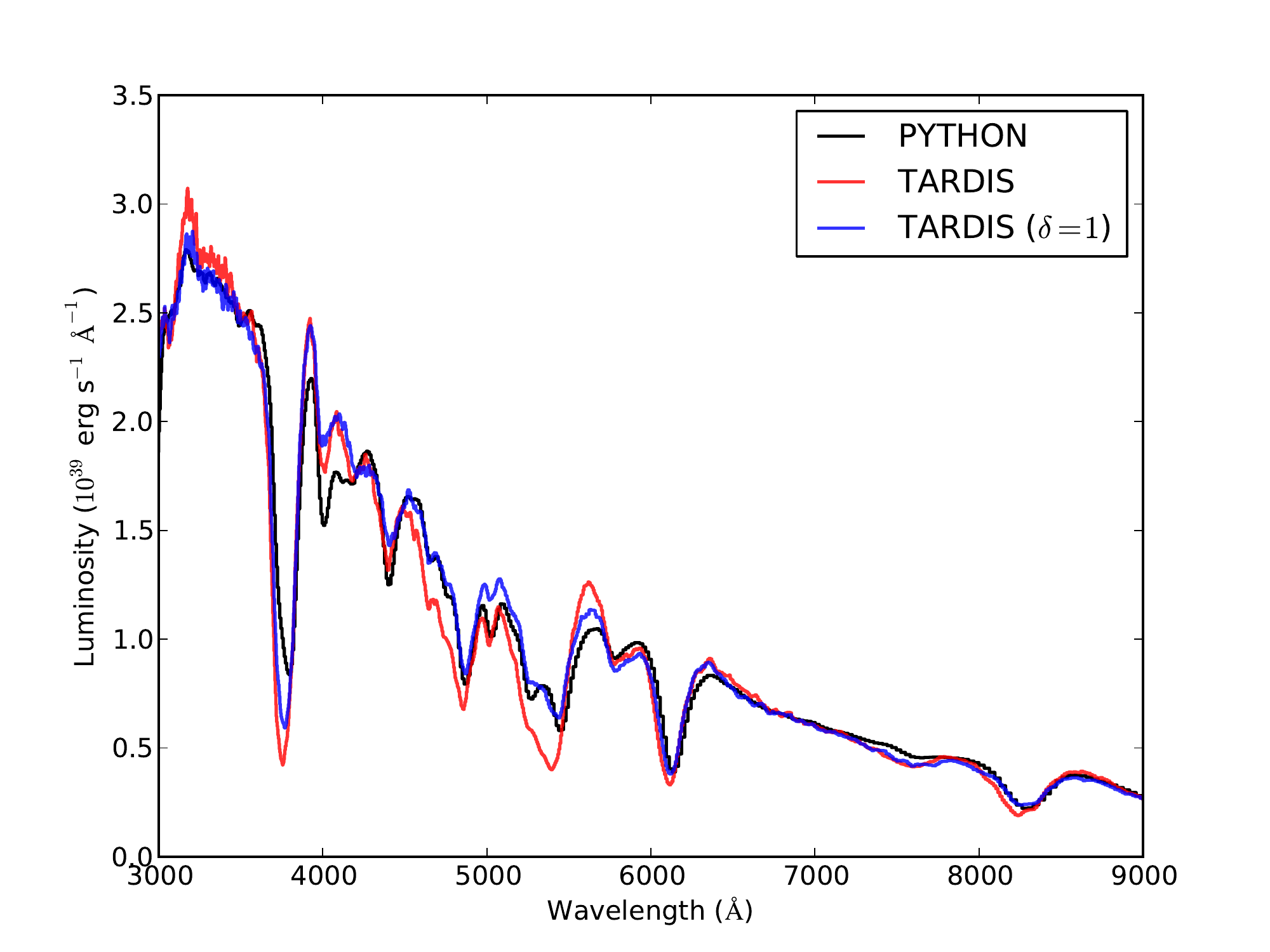} 

   \caption{Synthetic spectra computed with \gls{python_rt} and \gls{tardis} for the test model described in Section~\ref{sec:python_compare}. The \gls{tardis} calculations were carried out using our \texttt{scatter} and \texttt{nebular} ionization modes. In addition to our standard case, we also plot the spectrum obtained if we enforce $\delta = 1$ in \gls{tardis} (see Equation~\ref{eq:modified_nebular}).}
   \label{fig:python_compare}
\end{figure}

The synthetic spectra are compared in Fig.~\ref{fig:python_compare}. In general, the agreement is very good: the strengths and shapes of the line feature are well-matched, as is the total emergent intensity (in these test calculations, roughly 30 per cent of the radiation launched through the inner boundary is backscattered by the ejecta such that the total amplitude of the emergent spectrum is not trivially reproduced by imposing the same radiative flux at the inner boundary). 

\begin{figure*}
   \centering
   \includegraphics[width=\columnwidth]{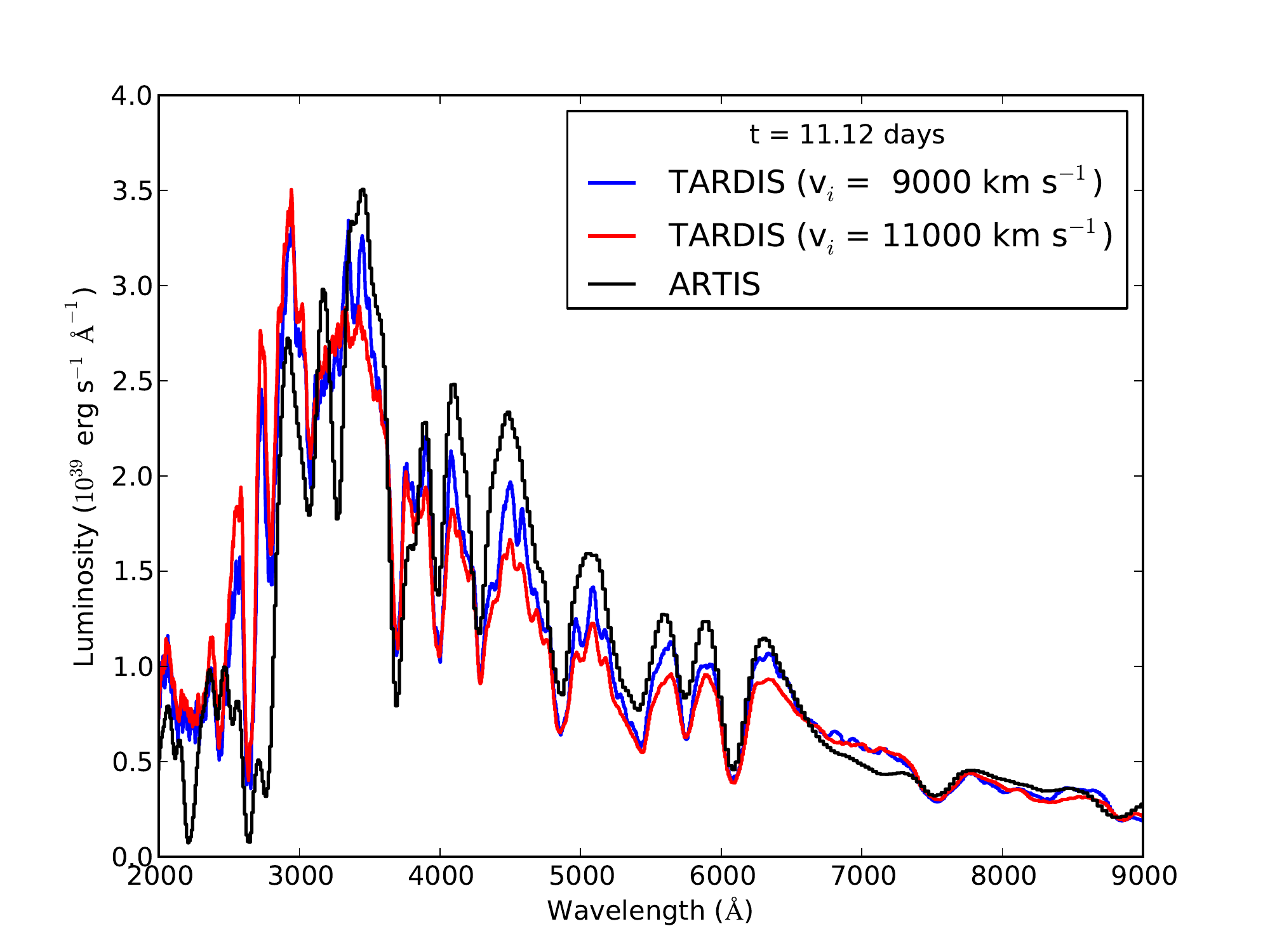} 
   \includegraphics[width=\columnwidth]{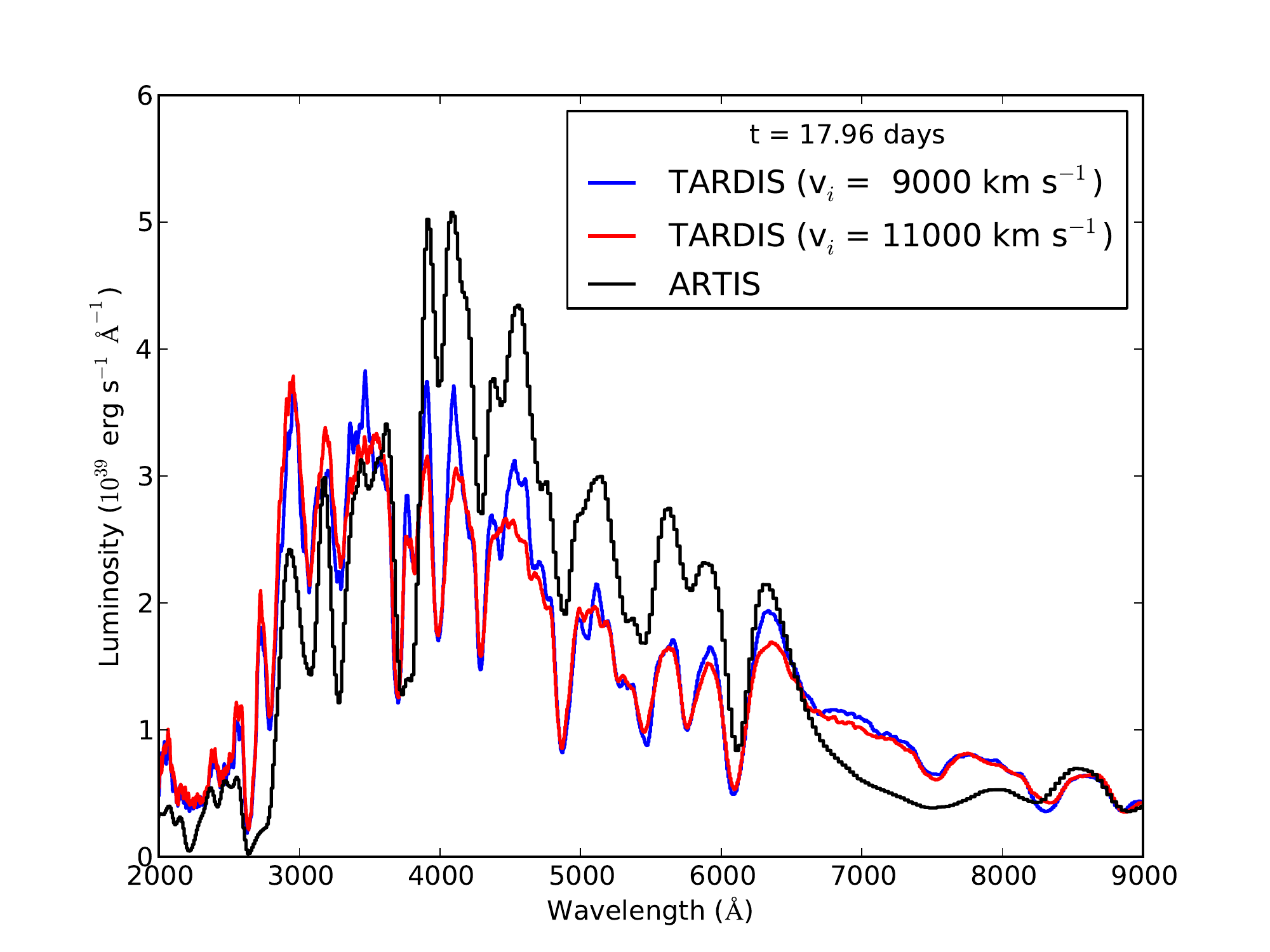} \\
   \includegraphics[width=\columnwidth]{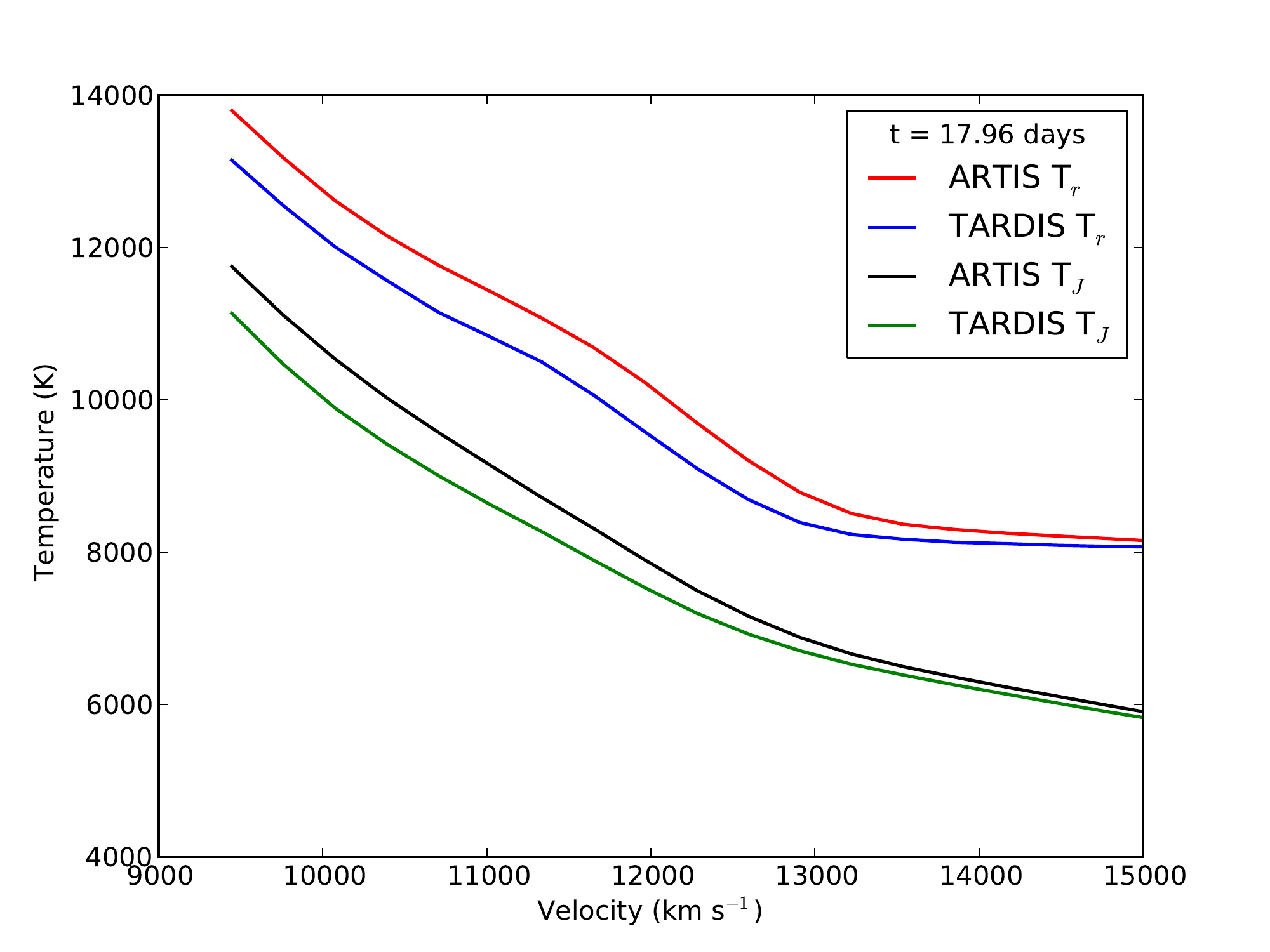} 
   \includegraphics[width=\columnwidth]{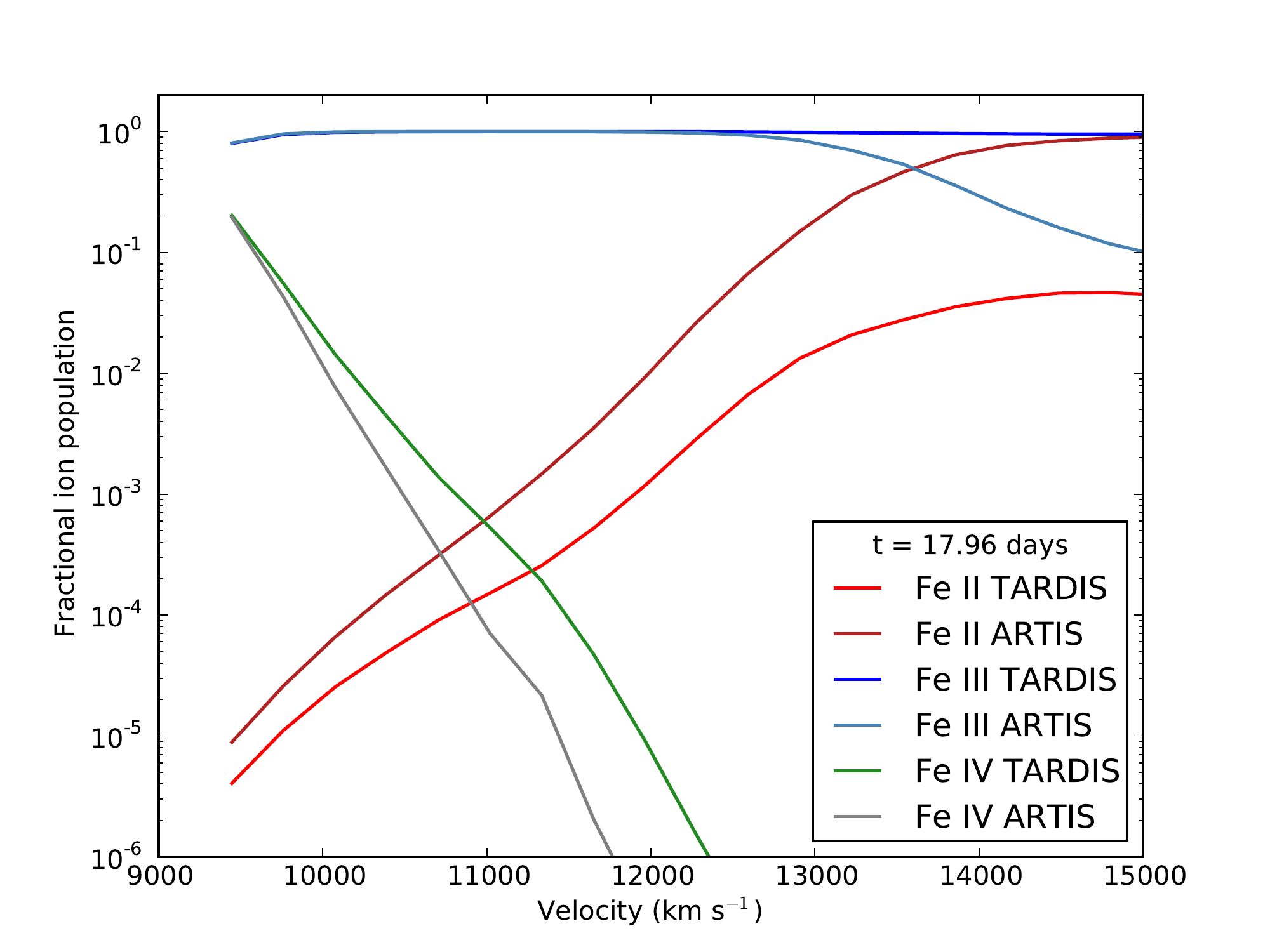} 
   \caption{Comparisons of \gls{artis} and \gls{tardis} calculations for the WD detonation model (see Section~\ref{sec:numerics}). The upper panels show synthetic spectra for 11.1 (left) and 18.0 (right) days after explosion. For \gls{tardis}, spectra are shown for two calculations with different choices of \vinner.
For the later epoch considered (18~days after explosion), the lower panels compare radiation field temperatures (left) and Fe ionization fractions (right) between \gls{artis} and the \gls{tardis} calculation with $v_{i} = 9000$~km~s$^{-1}$.}
   \label{fig:artis_model_spec}
\end{figure*}

There are, however, some discrepancies. For example, the strength of the \ion{S}{2} feature ($\sim 5300$~\AA) is noticeably different and there are small wavelength offsets: for example, \ion{Si}{2} $\lambda$ 6355, 5050 and the \ion{Ca}{2} H+K / IR triplet features have larger blueshifts in the \gls{tardis} spectrum. Many of these discrepancies can be attributed to subtle differences in the treatment of ionization. In particular, the modified nebular approximation used by \gls{python_rt} does not incorporate the correction factor $\delta$ proposed by ML93. If we do not include this correction (i.e. we force $\delta = 1$ in Equation 3), we find an even closer match between the two codes (see Fig.~\ref{fig:python_compare}). Although still imperfect, this level of agreement between two independent codes gives confidence that both are performing well. Our comparisons also confirm that, although not qualitatively critical, even relatively minor changes to the ionization approximation are relevant to high-precision modelling.

\subsection{Comparison to {\sc artis}}
\label{sec:artis_compare}

The \gls{artis} code is a purpose-build, multi-dimensional SN radiative transfer code. In contrast to \gls{tardis} and \gls{python_rt}, \gls{artis} carries out time-dependent calculations that produce time sequences of synthetic spectra for an input model. \gls{artis} calculations do not involve an inner boundary through which radiation is injected -- instead, the energy injection is followed in detail by simulating the emission of $\gamma$-rays origination from radioactive decays \citep[see][]{2005A&A...429...19L,2007MNRAS.375..154S,2009MNRAS.398.1809K}. Avoiding the need for an artificial inner boundary is the greatest advantage of codes such as \gls{artis}, but this comes at considerable computational expense: a single \gls{artis} calculation typically takes 10~--~100 kCPU hours, which is prohibitive for studies in which large parameter spaces are to be searched. 

Like \gls{python_rt}, \gls{artis} includes bound-bound, bound-free, free-free and electron scattering opacities. \gls{artis} also includes an implementation of the macro atom scheme for all ions and an ionization approximation that is based on detailed photo-ionization rate estimators \citep[see][for details]{2009MNRAS.398.1809K}. As in \gls{tardis}, the ejecta are assumed to be in homologous expansion.

For our comparisons between \gls{tardis} and \gls{artis}, we use the same detonation model described in Section~\ref{sec:numerics}.
The \gls{artis} simulations for the model yield a sequence of synthetic spectra covering the ultraviolet to infrared wavelength regions. We shall focus on comparisons at two epochs, 11.1 and 18~days after explosion (i.e. roughly at, and one week before, the time of maximum light in the optical bands).
For the \gls{tardis} comparison calculations, we use the luminosities calculated by \gls{artis} ($\log_{10} \louter / _{\odot} = 9.34$, 9.50 for $t = 11.1, 18.0$~days, respectively) and adopt $\vouter = 22000$~km~s$^{-1}$. We must also choose a location for the inner boundary of the \gls{tardis} computational domain. Based on the model structure (Fig.~{\ref{fig:artis_model}}), it seems inappropriate to consider $v_{i} \simlt 8000$~km~s$^{-1}$, since this would place the inner boundary within the $^{56}$Ni-rich layers. Consequently, we have carried out several \gls{tardis} runs for both epochs that explore the effect of choosing a range of values for $v_{i} \ge 9000$~km~s$^{-1}$. 


\begin{figure*}
   \centering
   \includegraphics[width=0.7\textwidth]{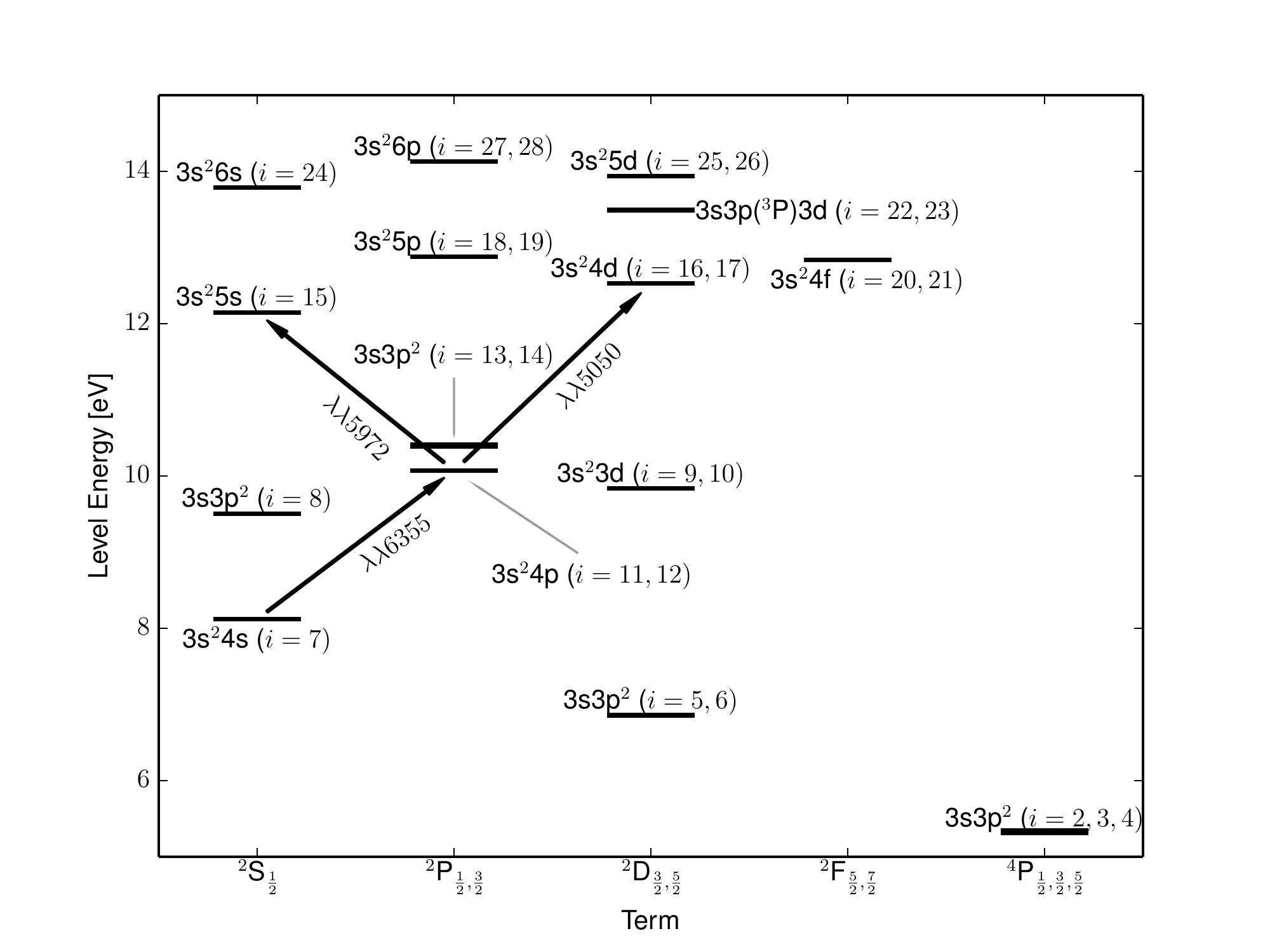} 

   \caption{A Grotrian diagram of the \ion{Si}{2} atomic model in the \gls{chianti} database.}
   \label{fig:grotrian_si2}
\end{figure*}

Synthetic optical and ultraviolet spectra from our \gls{tardis} calculations are compared to  \gls{artis} spectra in Fig.~\ref{fig:artis_model_spec}. First, we note that the choice of $v_{i}$ does not significantly affect the optical spectrum, even when varied over a fairly wide range ($9000 - 11000$~km~s$^{-1}$). This is reassuring since significant sensitivity to the choice of $v_{i}$ would pose a challenge for the rapid exploration of model parameter space for which \gls{tardis} is intended. If $v_{i}$ is made very large ($v_{i} \simgt 12000$~km~s$^{-1}$) consequences do appear, primarily in the ultraviolet and blue regions of the spectra, due to the lack of line-blocking by iron-group elements when only the high velocity ejecta are included (see Fig.~{\ref{fig:artis_model}}).

Second, for a reasonable choice of $v_i$, the shapes and strengths of the optical spectral features agree fairly well in the two codes, suggesting that the simplifications made in \gls{tardis} do not severely limit its ability to model optical lines, particularly for early epochs. Our comparison do, however, highlight important differences: 
in the \gls{tardis} calculations there is more ultraviolet emission (blueward of $\sim 3000$~\AA) and correspondingly less optical emission. This discrepancy is significantly stronger at the later epoch considered -- at $18$ days, the optical pseudo continuum in \gls{artis} is clearly brighter than predicted by \gls{tardis}. The origin of these differences is likely a combination of factors arising from the different approximations used. In Fig.~\ref{fig:artis_model_spec}, we show that both the radiation temperature (\trad) and the mean intensity\citep[described in terms of $T_{J} = W^{1/4} T_{r}$; see ][]{2009MNRAS.398.1809K} and fairly similar in both calculations. However, there is a distinct offset in the computed ionization state. For example, ionization fractions for Fe are shown in Fig.~\ref{fig:artis_model_spec}: while the overall pattern of the ionization is similar in both calculations, the ionization approximation used by \gls{artis} leads to systematically lower mean ionization. Since higher abundances of the singly ionized iron-group elements will lead to more effective line-blocking at blue wavelengths, this likely is responsible for much of the difference between the ultraviolet flux level in the two calculations.

Clearly, further study and improvement of the ionization approximations used is needed for quantitative modelling, and the limitations of these approximations must be borne in mind when comparing synthetic spectra from any of the codes discussed here to observations. Nevertheless, we conclude that, despite a considerable reduction in complexity, the \gls{tardis} code is already capable of providing a reasonable approximation to the \gls{artis} calculations (for suitably chosen luminosity), lending credence to its use for initial attempts to fit spectra and explore model parameter spaces.

\section{Results}
\label{sec:results}

Thanks to its modularity, \gls{tardis} is well suited to study the influences of different physical assumptions on the spectrum. 
Here we present results of simple calculations that investigate the differential effects of two important approximations that are sometimes adopted in the modelling of SN~Ia spectra.
For these tests we again use the 
sub-Chandrasekhar mass detonation model introduced in Section~\ref{sec:numerics}. We will show results computed for two epochs for which we adopt \louter and \vinner  as given in Table~\ref{tab:artis_epoch_models}.
\ctable[
width=\columnwidth,
caption = {Parameters used for the calculations in our comparisons
between different physics modules. The luminosities adopted are
extracted from the \gls{artis} simulations.},
label = {tab:artis_epoch_models}, nostar
]{lXXX}{}
{\FL
\texp & Luminosity & \vinner & \vouter \\
days & $\log{\louter/\lsun}$ & \kms &\kms
\ML
11.1 & 9.34 & 11000 & 22000\\
18.0 & 9.50 & 9000 &22000
\LL
}


\begin{figure*}
   \centering
   \includegraphics[width=\columnwidth, trim=50 0 20 0]{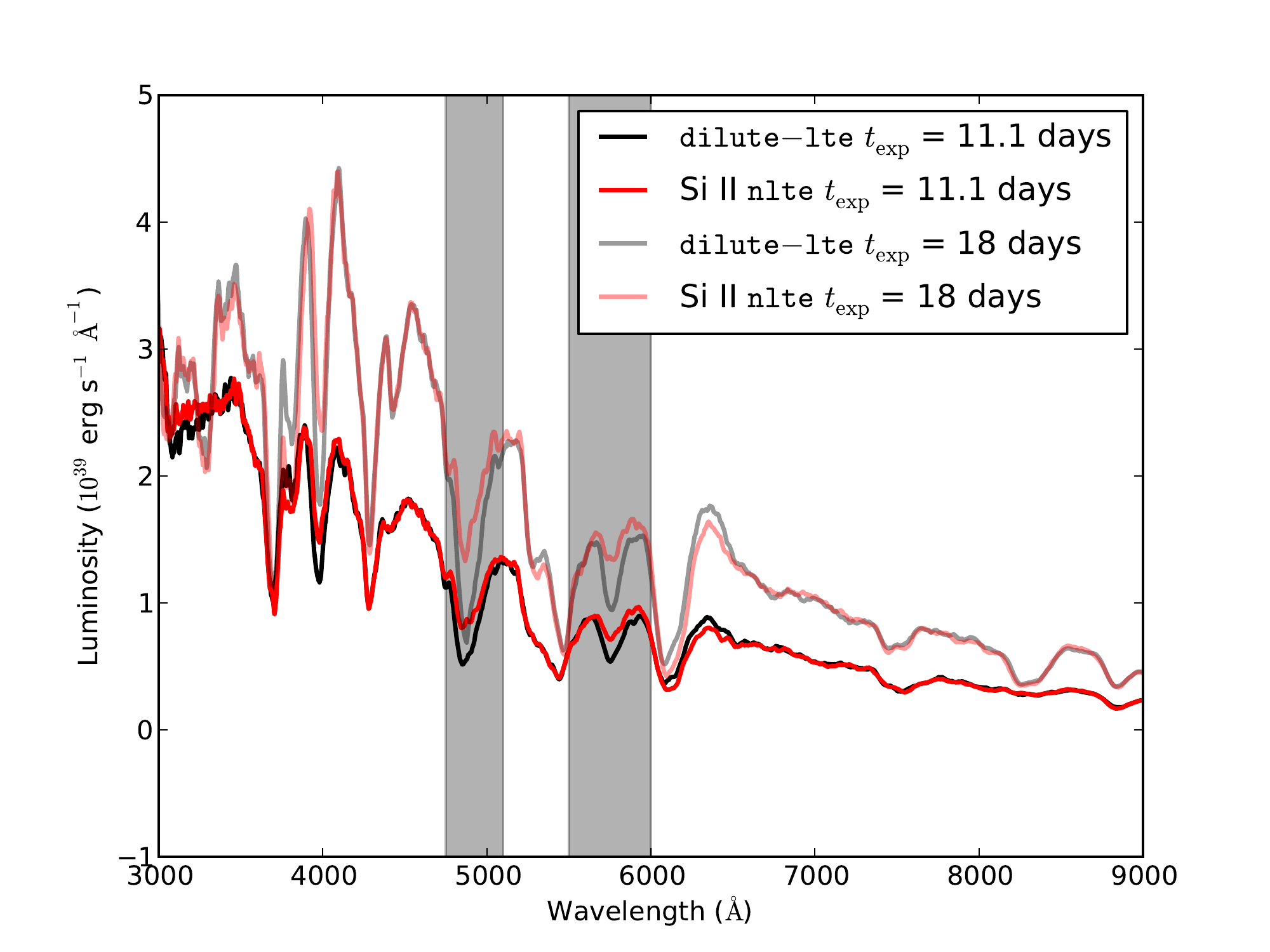} 
   \includegraphics[width=\columnwidth, trim=20 0 50 0]{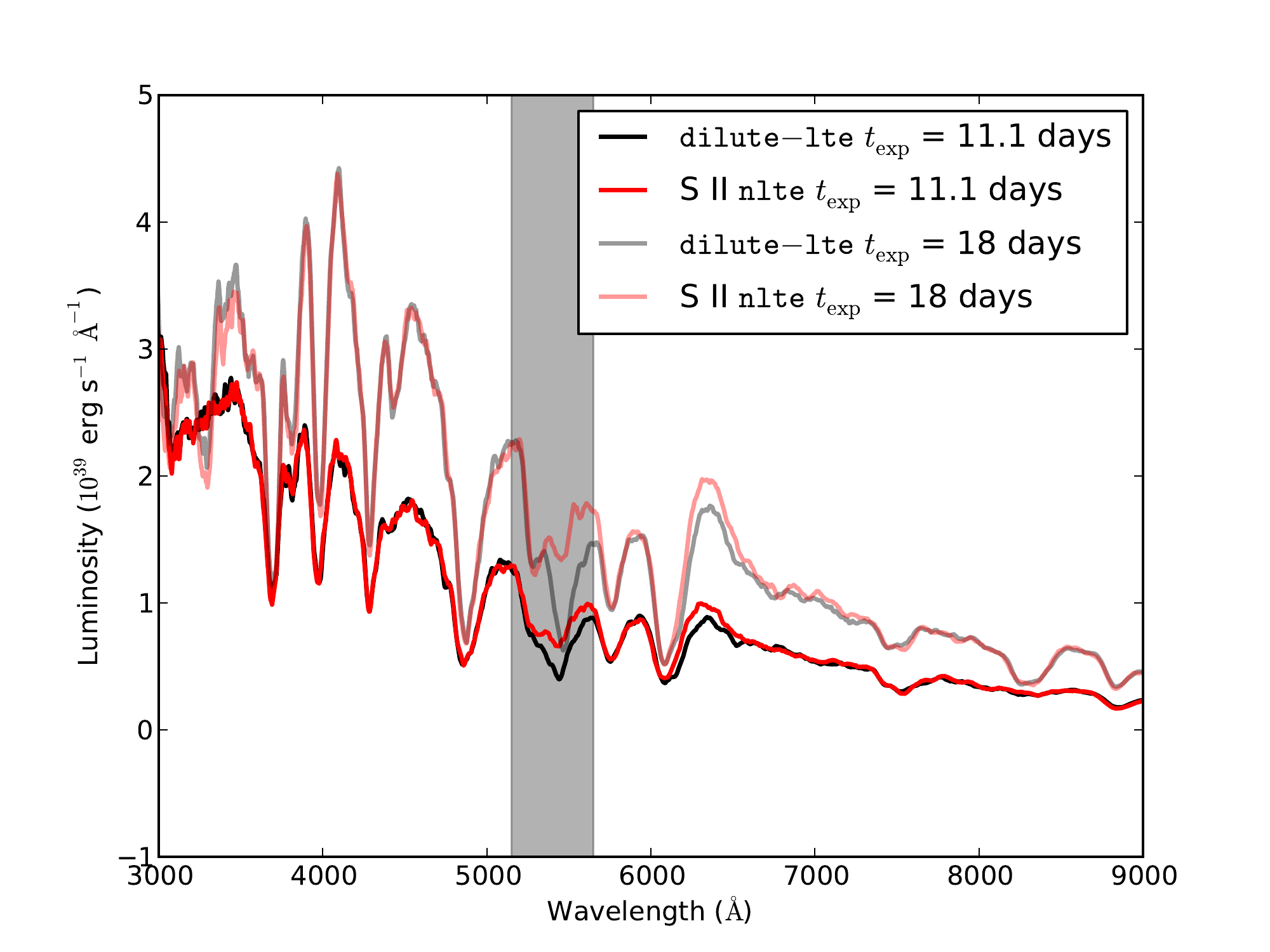}\\
   \includegraphics[width=\columnwidth, trim=50 0 20 0]{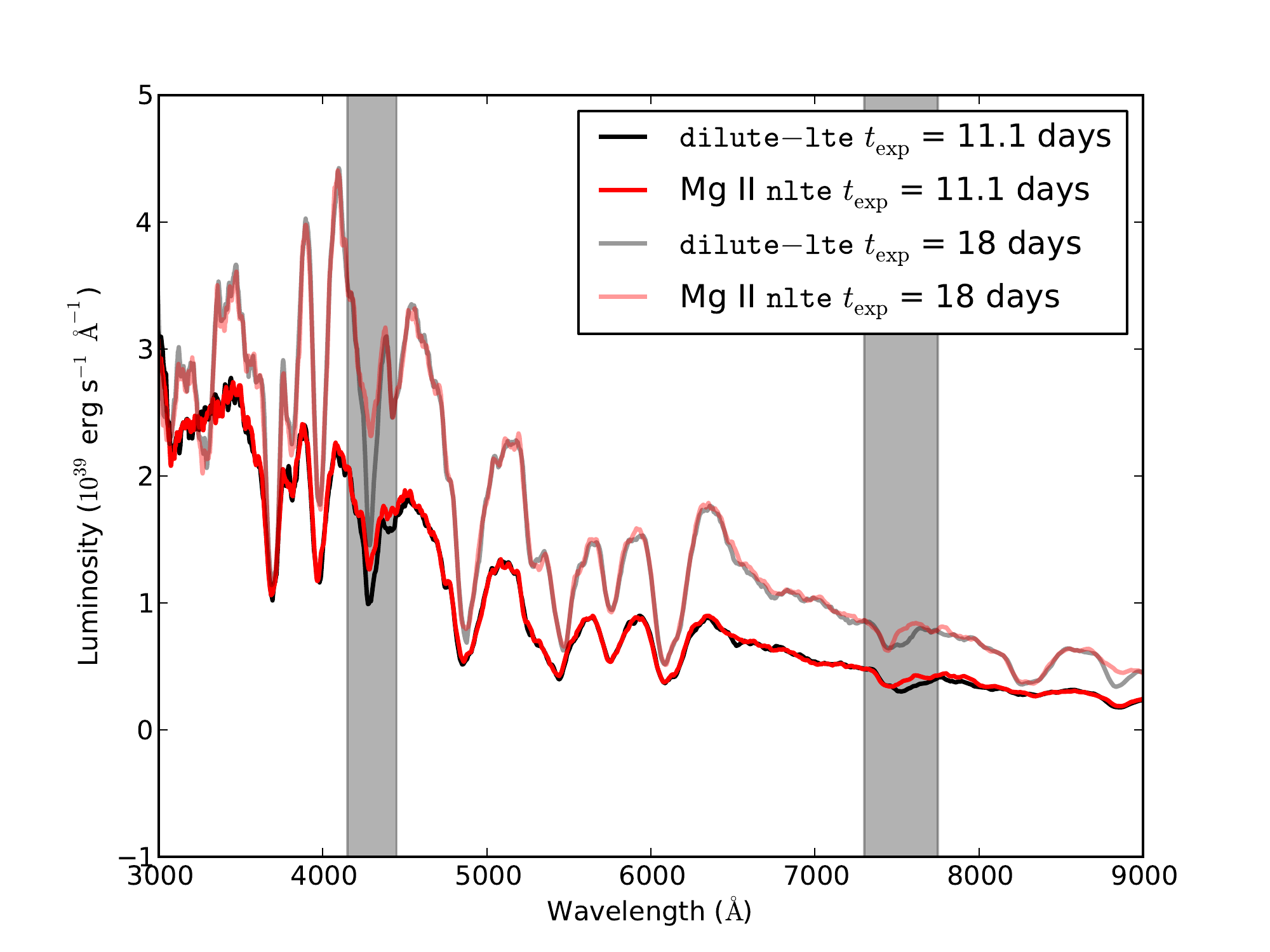} 
   \includegraphics[width=\columnwidth, trim=20 0 50 0]{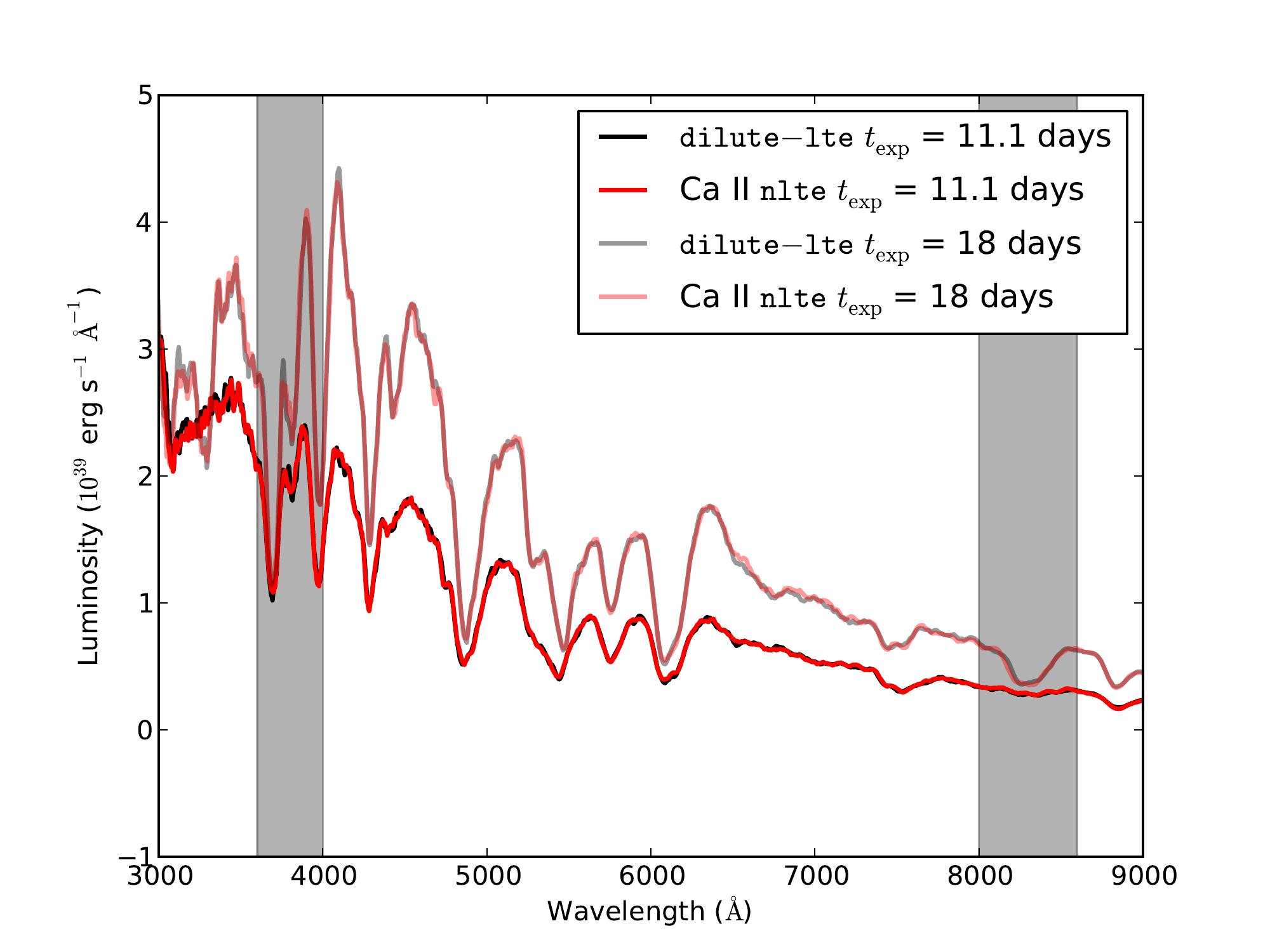} 

   \caption{Comparison of spectra computed using \texttt{dilute-lte}
     (black) and \texttt{nlte} (red)
     excitation modes for select species (see text).
In each plot the regions most strongly affected are marked by a grey shade.} 
   \label{fig:tardis_result_nlte}
\end{figure*}

\subsection{NLTE excitation}
\label{sec:results_nlte}

As noted in Section~\ref{sec:intro}, several of the existing \gls{montecarlo} radiative transfer codes (including \gls{mlmc}, \gls{artis} and \gls{python_rt}) make use of simple approximations for the treatment of excited level populations. This has advantages for computational expediency and, particularly for multi-dimensional simulations, for reducing memory requirements but it comes at the cost of reduced accuracy. In contrast, studies by \citet{1996A&A...312..525P}, \citet{1996MNRAS.283..297B} and \citet{2013MNRAS.429.2127B} have made use of the full equations of statistical equilibrium incorporating both ionization and excitation and shown that NLTE effects can be important. Here, we focus on exploring approximations for excitation (rather than ionization) with a view to quantifying the systematic errors introduced by the simple (and computationally cheap) \texttt{dilute-lte} formula to estimate excitation states. 

We focus on four important ions, namely \ion{Si}{2}, \ion{S}{2}, \ion{Mg}{2} and \ion{Ca}{2}. Each of these exhibit prominent features in the spectra of \sneia and understanding potential systematic uncertainties in their modelling is an important ingredient in attempts to make synthetic fits to observed data sets. 
For our comparisons, we 
adopted the \texttt{nebular} ionization, \texttt{macroatom} line interaction and \texttt{detailed} radiative rates modes (i.e. the most sophisticated set of assumptions currently implemented) and then compared results obtained with \texttt{dilute-lte} and \texttt{nlte} excitation modes. We considered each of our selected ions in turn: i.e. for both epochs considered, we have computed a total of five synthetic spectra -- one in which each of the four selected ions is treated in \texttt{nlte} excitation mode (all other ions in \texttt{dilute-lte} excitation mode) and one comparison calculation in which all ions are handled in \texttt{dilute-lte} excitation mode. This approach is convenient to assess the direct consequences of the level populations in key ions but we note that it does not capture any complex NLTE effects where one ion can influence the radiation field and therefore the excitation state of other ions. 
In all calculations, $10^7$ \gls{montecarlo} quanta were used and 30 iterations were carried out.
Synthetic spectra from these calculations are compared in Fig.~\ref{fig:tardis_result_nlte}.

To help quantify the effects of our different excitation treatment, we define a departure coefficient for multiplets 

\begin{equation}
b_\textrm{LTE}^{m}=\frac{\sum_i n_i/n_0}{\sum_i n_i^\textrm{LTE}/n_0^{\textrm{LTE}}},
\end{equation}
where $i$ runs over the $J$ sub-states of the term.

\subsubsection{Silicon}

The \ion{Si}{2} $\lambda 6355$ feature 
($3\textrm{s}^2 4\textrm{s}~^2\textrm{S} \rightarrow 3\textrm{s}^2 4\textrm{p}~^2\textrm{P}$; see Fig.~\ref{fig:grotrian_si2})
is characteristic of \snia spectra and one of the main identifiers for this class. 
We find that the population of the $3\textrm{s}^2 4\textrm{s}~^2\textrm{S}$ state is relatively unaffected by our choice of excitation mode, resulting in 
a very small difference of departure coefficients between the two excitation treatments (see Table~\ref{tab:departure_coefficients}) and little change in the strength or shape of the
spectral feature (Fig.~\ref{fig:tardis_result_nlte}). 

\begin{figure}
   \centering
   \includegraphics[width=\columnwidth]{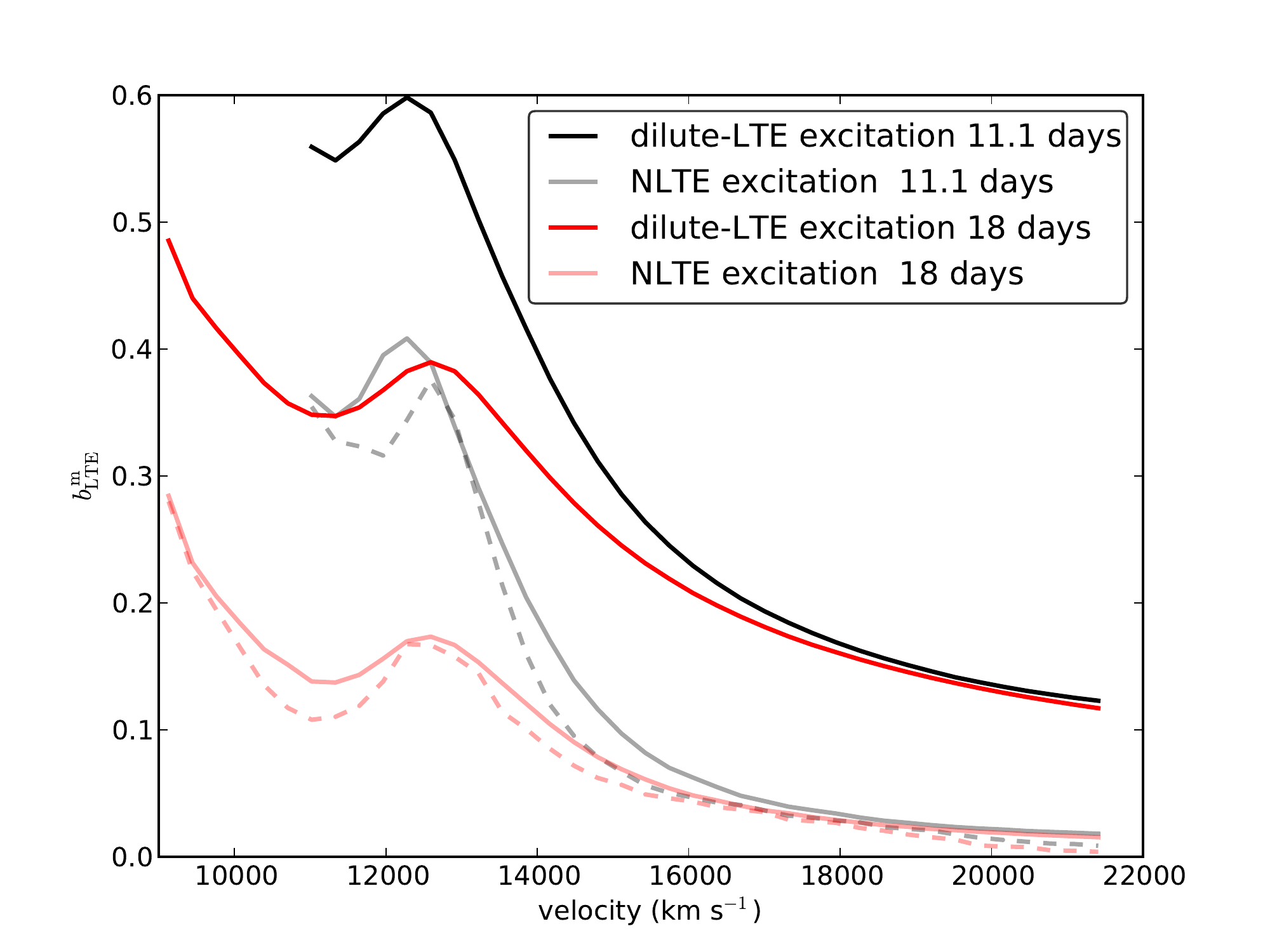} 

   \caption{Comparison between the departure coefficients for 
     Si~{\sc ii} $4\textrm{p}~^2\textrm{P}$ using \texttt{dilute-lte} and \texttt{nlte}
     excitation modes. For \texttt{nlte} mode we show results using
     both \texttt{dilute-blackbody} (solid lines) and \texttt{detailed}
     (dashed lines) treatments of the radiative rates.}
   \label{fig:tardis_result_departure_evolution}
\end{figure}

In contrast, both the $\lambda 5050$ 
($3\textrm{s}^2 4\textrm{p}~^2\textrm{P} \rightarrow 3\textrm{s}^2 4\textrm{d}~^2\textrm{D}$)
and the $\lambda 5972$
($3\textrm{s}^2 4\textrm{p}~^2\textrm{P} \rightarrow 3\textrm{s}^2 5\textrm{s}~^2\textrm{S}$) features
 are affected by the excitation mode:
departure coefficients for $3\textrm{s}^2 4\textrm{p}~^2\textrm{P}$ are illustrated in Fig.~\ref{fig:tardis_result_departure_evolution} (and also listed in Table~\ref{tab:departure_coefficients}). 
Clearly our \texttt{nlte} excitation mode leads to significantly reduced populations for this state and correspondingly weaker spectral features compared to the \texttt{dilute-lte} approximation.
E.g., in our $11.1$~d (18.0~d) spectra, the equivalent width (EW) of $\lambda 5972$ drops from 48 (38) to 24 (18)~\AA.

The different sensitivities of the departure coefficients for $3\textrm{s}^2 4\textrm{s}~^2\textrm{S}$  and $3\textrm{s}^2 4\textrm{p}~^2\textrm{P}$ mean that the choice of excitation treatment will systematically affect quantitative modelling of the equivalent width ratio $EW(\lambda 6355) / EW(\lambda 5972)$, which has been shown to act as a luminosity indicator 
\citep{1995ApJ...455L.147N, 2008MNRAS.389.1087H}. 
Thus our \gls{tardis} calculations verify that, as in \cite{2013MNRAS.429.2127B},  NLTE effects must be included if quantitative analysis (factor $\sim 2$ accuracy or better) of this line ratio is to be used to test models.

\ctable[
width=\textwidth,
caption = {Departure Coefficients},
label = {tab:departure_coefficients},
star
]{lllXXXXXXX}{\tnote[a]{The departure coefficient in
\texttt{dilute-lte} model is 1 for meta-stable levels and the dilution factor $W$ for other levels.}}{\FL
 &&& \multicolumn{2}{c}{\ion{Si}{2}} & \multicolumn{2}{c}{\ion{S}{2}} & \multicolumn{1}{c}{\ion{Ca}{2}} & \multicolumn{2}{c}{\ion{Mg}{2}} \NN
\cmidrule(r){4-5}\cmidrule(rl){6-7}\cmidrule(rl){8-8}\cmidrule(l){9-10}
\texp & shell velocity & excitation treatment  & $4\textrm{p}~^2\textrm{P}$%
& $4\textrm{s}~^4\textrm{P}$ & $3\textrm{d}~^4\textrm{F}$%
& $4\textrm{s}~^2\textrm{S}$ & $3\textrm{d}~^2\textrm{D}$%
& $3\textrm{d}~^2\textrm{D}$ & $4\textrm{p}~^2\textrm{P}$\\
day&\kms&&&&&&&&
\ML
11.1 &11000& \texttt{nlte}& 0.55 & 0.35&0.42&0.40&0.83&0.25& 0.21\\
 & & \texttt{dilute-lte}\tmark[a] & 0.56 & 0.56&0.56&1.0&1.0&0.56&0.56\\
 &17300& \texttt{nlte}& 0.16 & 0.03&0.20&0.15& 1.11&0.07& 0.02\\
 & & \texttt{dilute-lte}\tmark[a] & 0.19 & 0.19&0.19&1.0&1.0&0.19&0.19\\
18.0  & 9000&\texttt{nlte}& 0.52 &0.29&0.38&0.36&0.88&0.25&0.17\\
 & &\texttt{dilute-lte}\tmark[a]& 0.49 &0.49&0.49&1.00&1.0&0.49&0.49\\
 & 15500&\texttt{nlte}& 0.20 &0.05&0.24&0.21&1.74&0.02&0.01\\
 & &\texttt{dilute-lte}\tmark[a]& 0.23 &0.23&0.23&1.0&1.0&0.23&0.23\\

\LL}


\subsubsection{Sulphur}

\ion{S}{2} is identified by iconic features around $\lambda 5449$ and  $\lambda5623$ in the spectra of SNe~Ia (see Fig.~\ref{fig:tardis_result_nlte}): these are dominated by the $3\textrm{s}^2 3\textrm{p}^2 (3\textrm{P}) 4\textrm{s}~^4\textrm{P} \rightarrow 3\textrm{s}^2 3\textrm{p}^2 (3\textrm{P}) 4\textrm{p}~^4\textrm{D}$ and $3\textrm{s}^2 3\textrm{p}^2 (3\textrm{P}) 3\textrm{d}~^4\textrm{F} \rightarrow 3\textrm{s}^2 3\textrm{p}^2 (3\textrm{P}) 4\textrm{p}~^4\textrm{D}$ multiplets. As for the $3\textrm{s}^2 4\textrm{p}~^2\textrm{P}$ states in \ion{Si}{2}, we find that our \texttt{nlte} excitation mode results in smaller level populations than \texttt{dilute-lte}, particularly for $3\textrm{d}~^4\textrm{F}$, which is classified as meta-stable in the simpler scheme. This has quite noticeable ramifications for the shape of the S~{\sc ii} feature (Fig.~\ref{fig:tardis_result_nlte}), again highlighting the sensitivity to excitation treatment in high-precision modelling.

\subsubsection{Magnesium}

Two particular features associated with Mg~{\sc ii} could be of note in the modelling of optical SN~Ia spectra: the strong $\lambda 4481$ line ($3\textrm{d}~^2\textrm{D} \rightarrow 4 \textrm{f}~^2\textrm{F}$) and the $4 \textrm{p}~^2\textrm{P} \rightarrow 4 \textrm{d}~^2\textrm{D}$ ($\lambda 7896$) transition, which is a potential 
contaminant of the oxygen feature at $\lambda 7774$. 
Once again we find that our \texttt{nlte} excitation treatment noticeably depopulates the lower levels of these transitions (compared to \texttt{dilute-lte}). This causes the
$\lambda 4481$ to become weaker (particularly at the later epoch, when it is less saturated) and the contribution of \ion{Mg}{2} to the blend with $\lambda 7774$ to nearly vanish. This lends confidence to the use of $\lambda 7774$ as an oxygen abundance tracer \citep{2013MNRAS.429.2127B}.

\subsubsection{Calcium}

Ca~{\sc ii} imprints distinct spectral features at both ends of the optical spectrum:  
in the blue, the H and K lines ($\lambda 3950$~arising from $4\textrm{s}~^2\textrm{S} \rightarrow 4\textrm{p}~^2\textrm{P}$) and in the near-infrared at $\lambda 8498, 8542, 8662$ 
($3\textrm{d}~^2\textrm{D} \rightarrow 4\textrm{p}~^2\textrm{P}$). In both epochs considered, we find only a modest difference between departure coefficients in \texttt{nlte} and \texttt{dilute-lte} excitation modes for the $3\textrm{d}~^2\textrm{D}$ state (see Table~\ref{tab:departure_coefficients}), but this does not lead to a significant change in the shape or strength of the near-infrared feature (the absorption is already saturated). Thus, of those considered, the \ion{Ca}{2} features are the least sensitive to the choice of excitation mode and are well-represented by the simple excitation formula.

\subsection{Line interaction schemes}
\label{sec:results_line_interaction}
\ctable[
width=\textwidth,
caption = {Equivalent widths for spectral features measures from
spectra calculated with various line interaction schemes},
label = {tab:line_interaction_eqw}, star
]{lXXXXXX}{}
{\FL
&\multicolumn{3}{c}{\ion{Si}{2} $\lambda 5987$} &
\multicolumn{3}{c}{\ion{S}{2} $\lambda 5449 + \lambda 5623$}\NN
\cmidrule(r){2-4}\cmidrule(l){5-7}
\texp &  \texttt{scatter} & \texttt{downbranch} & \texttt{macroatom} & \texttt{scatter} & \texttt{downbranch} & \texttt{macroatom}\\
days & \AA & \AA & \AA & \AA & \AA & \AA
\ML
11 &  41 & 37 & 43 & 153 & 142 & 171\\
18 &  50 & 38 & 39 & 158 & 143 & 175
\LL
}


\begin{figure*}
   \centering
   \includegraphics[width=0.48\textwidth, trim=40 0 10 0]{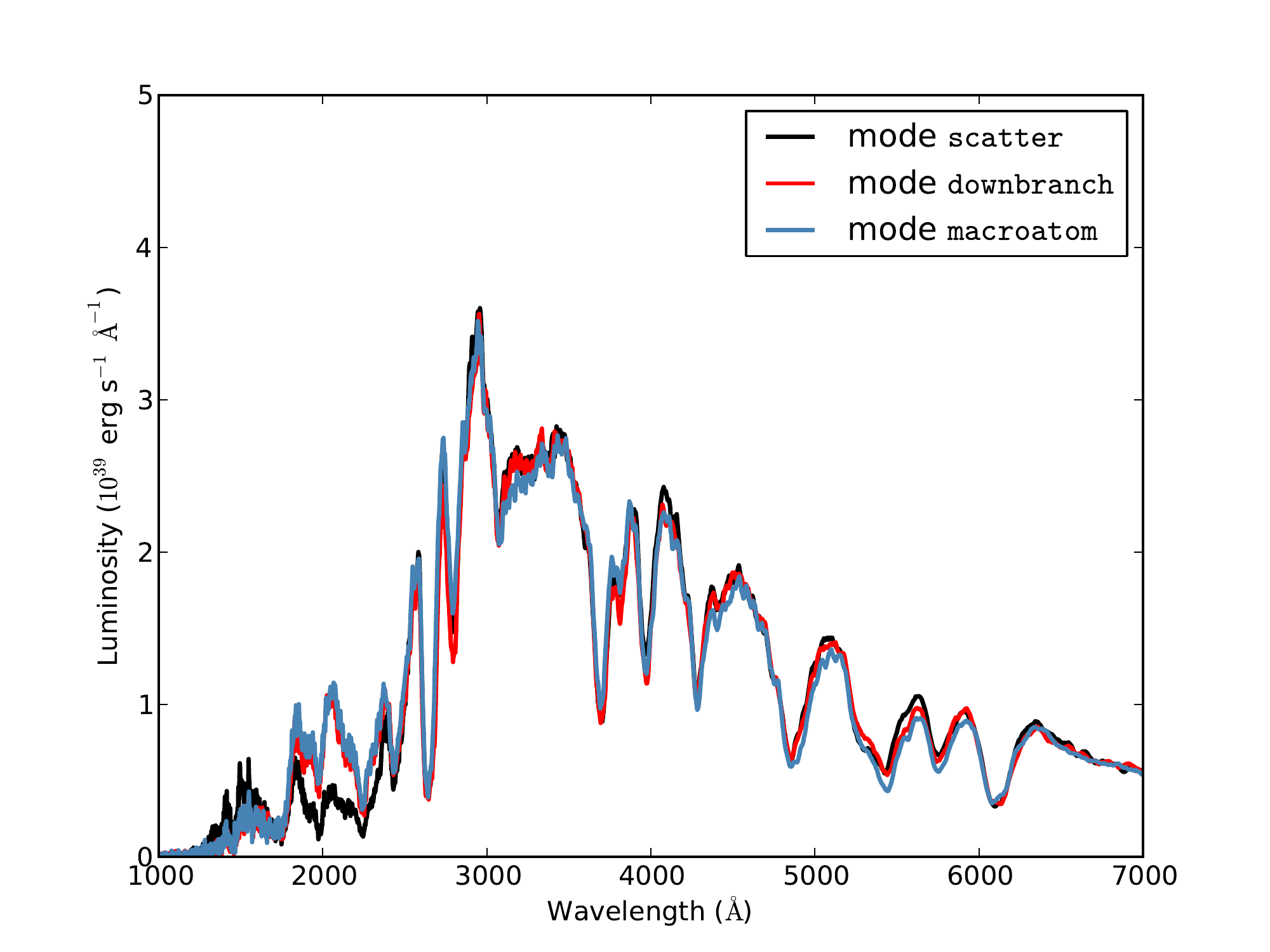} 
   \includegraphics[width=0.48\textwidth, trim=40 0 10 0]{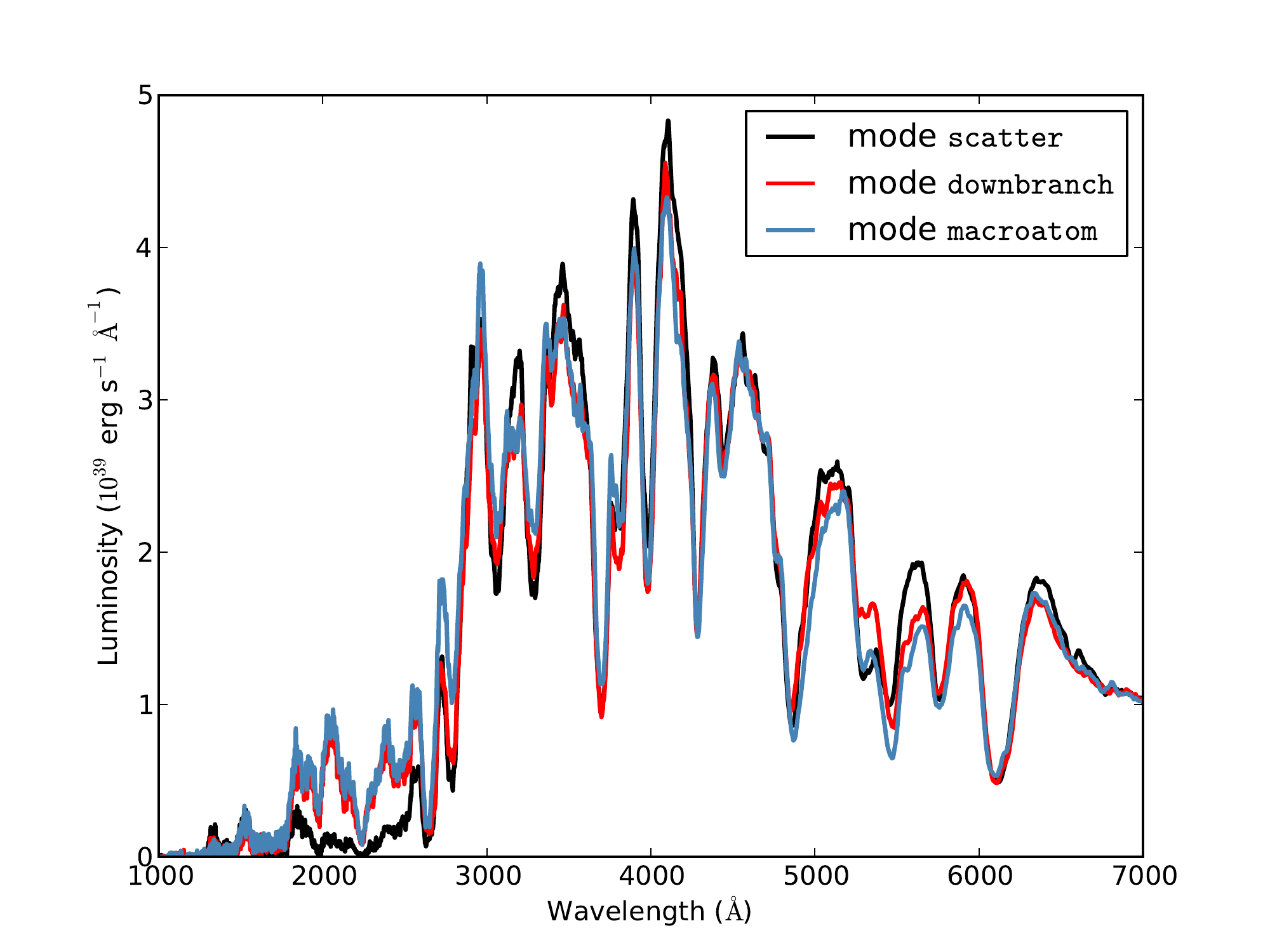} 

   \caption{Comparison of \gls{tardis} spectra computed for the
     detonation model with
     different line interaction schemes (\texttt{scatter},
     \texttt{downbranch} and \texttt{macroatom} modes). The left
   panel shows results for \texp = 11.2 days (with \vinner=11000~\kms) and the right panel for \texp = 18 days (with \vinner= 9000~\kms)}
   \label{fig:tardis_result_line_interaction}
\end{figure*}


\begin{figure*}
   \centering
   \includegraphics[width=0.33\textwidth, ]{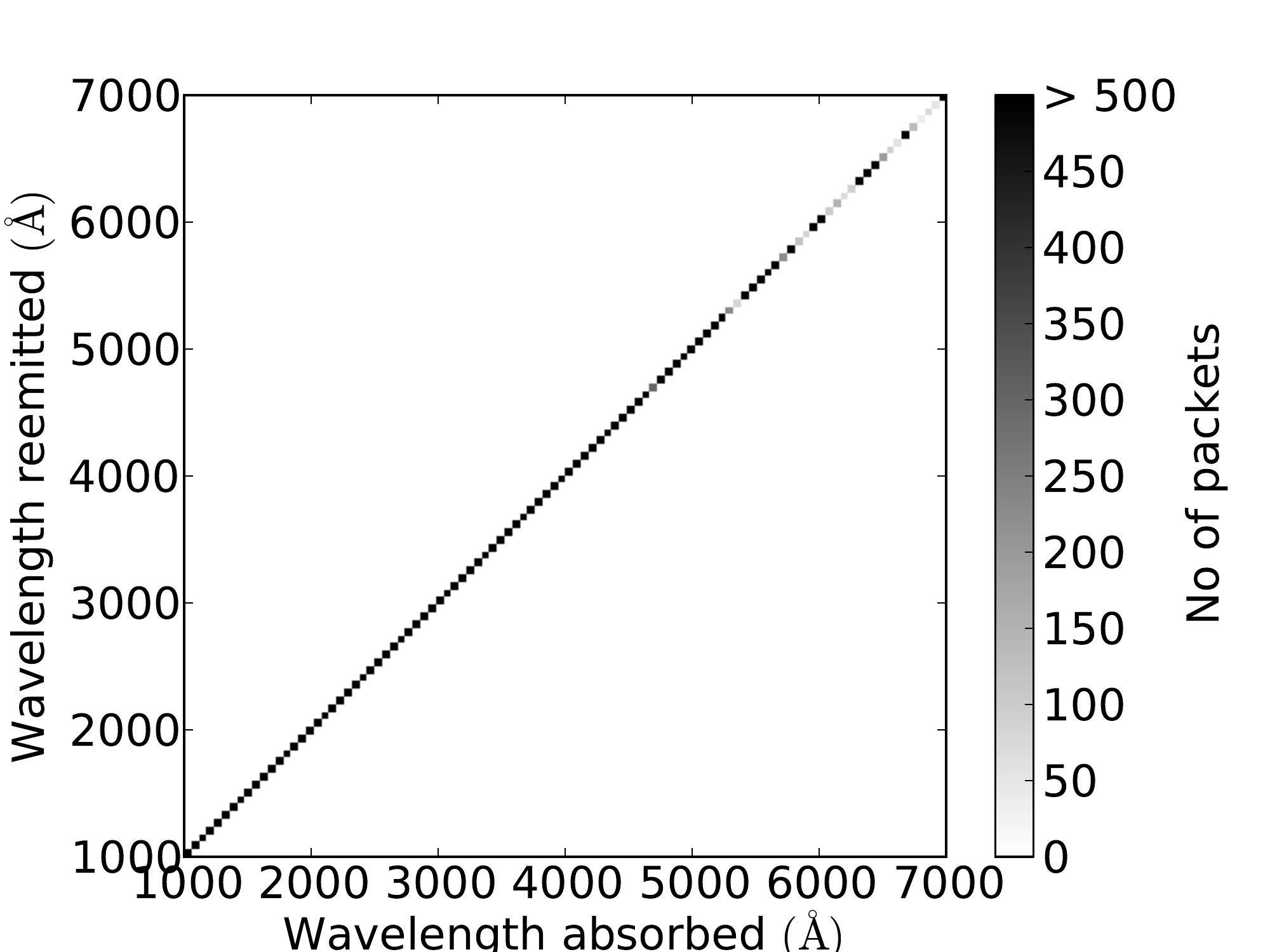} 
   \includegraphics[width=0.33\textwidth, ]{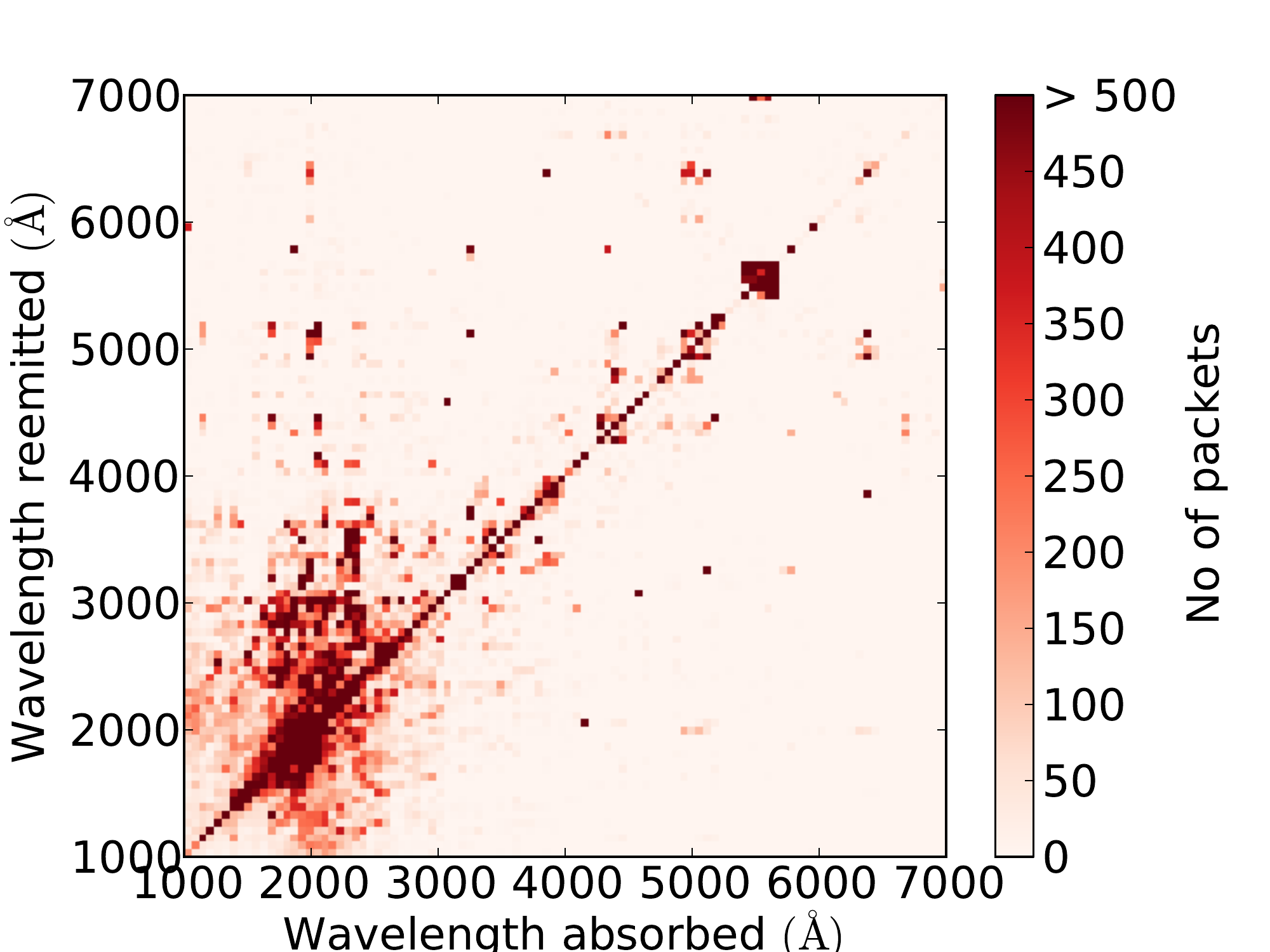} 
   \includegraphics[width=0.33\textwidth, ]{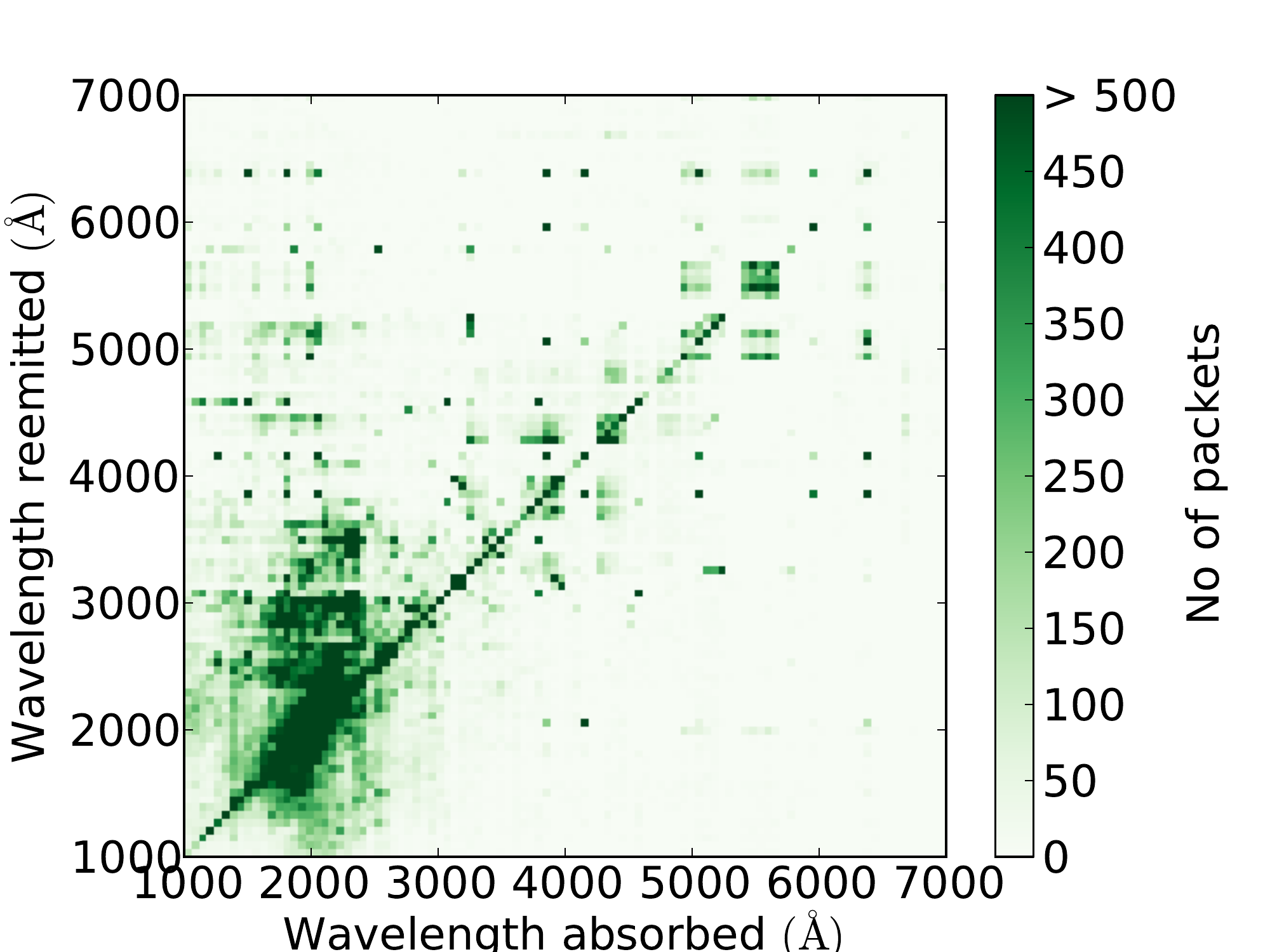} 

   \caption{Each panel illustrates the wavelength redistribution that occurred during the last interaction of escaping packets (co-moving frame wavelength prior to interaction versus wavelength after interaction) for calculations at $\texp=18~\textrm{days}$ (\texttt{scatter} mode left, \texttt{downbranch} mode centre, \texttt{macroatom} mode right).}
   \label{fig:tardis_result_last_line_interaction}
\end{figure*}

As described in Section~\ref{sec:rad_matter_interaction}, \gls{tardis} allows for three different bound-bound interaction schemes: \texttt{scatter}, \texttt{downbranch} and \texttt{macroatom} modes. In this section, we explore the influence of the interaction scheme on the spectrum. 

\citet{2000A&A...363..705M} showed that the ultraviolet flux is significantly affected if line branching is taken into account. When compared to calculations that include only resonance scattering, inverse fluorescence by iron group elements leads to enhanced emission in the ultraviolet. Our calculations also show this behaviour quite clearly -- 
spectra calculated using 
\texttt{scatter} and \texttt{downbranch} modes
are compared in Fig.~\ref{fig:tardis_result_line_interaction}.

With \gls{tardis} we can also test the extent to which the macro atom scheme (see Section~\ref{sec:rad_matter_interaction}) alters the spectrum.
In our calculations, the difference between \texttt{macroatom} (here combined with \texttt{detailed} mode for radiative rates)  and \texttt{downbranch} mode is most apparent near maximum light (see Fig.~\ref{fig:tardis_result_last_line_interaction}) in the region around the \ion{S}{2} features, $\sim$5500\AA. The differences arise from the more complex pattern of frequency redistribution afforded by the macro atom scheme -- 
this is illustrated in Fig.~\ref{fig:tardis_result_last_line_interaction} where we show the increasing complexity of the frequency redistribution associated with the last line-interaction events that occurred for \gls{montecarlo} quanta in simulations with each of our line interaction modes. 

To quantify the 
effects of the different line interaction methods we have measured equivalent widths of 
the \ion{S}{2} feature ($\lambda 5449$ and $\lambda5623$) and the  \ion{Si}{2} $\lambda 5987$ line (see Table~\ref{tab:line_interaction_eqw}). We find that the former is sensitive to the line interaction method on the level of $\sim 30$~\AA\ but the later by only $\sim 3$~\AA.

\subsection{Radiative rates estimation}
\label{sec:results_jblue}

Amongst the options currently implemented in \gls{tardis}, the treatment of the bound-bound radiative rates has the most significant implications for the computational cost of a calculation.
In \texttt{detailed} mode, it is necessary to have a sufficiently large number of \gls{montecarlo} quanta (many millions) to provide adequate statistics for the individual \jblue estimators (see Section~\ref{sec:montecarlo_estimators}) in every grid cell. Although more accurate, this is considerably more demanding than \texttt{dilute-blackbody} mode (Section~\ref{sec:plasma_state})
for which good convergence can be obtained with numbers of packets that are one to two orders of magnitude smaller.
In this section we compare results from our two modes for bound-bound radiative rates to investigate whether the extra computation cost of \texttt{detailed} mode is warranted.

\begin{figure}
   \centering
   \includegraphics[width=\columnwidth]{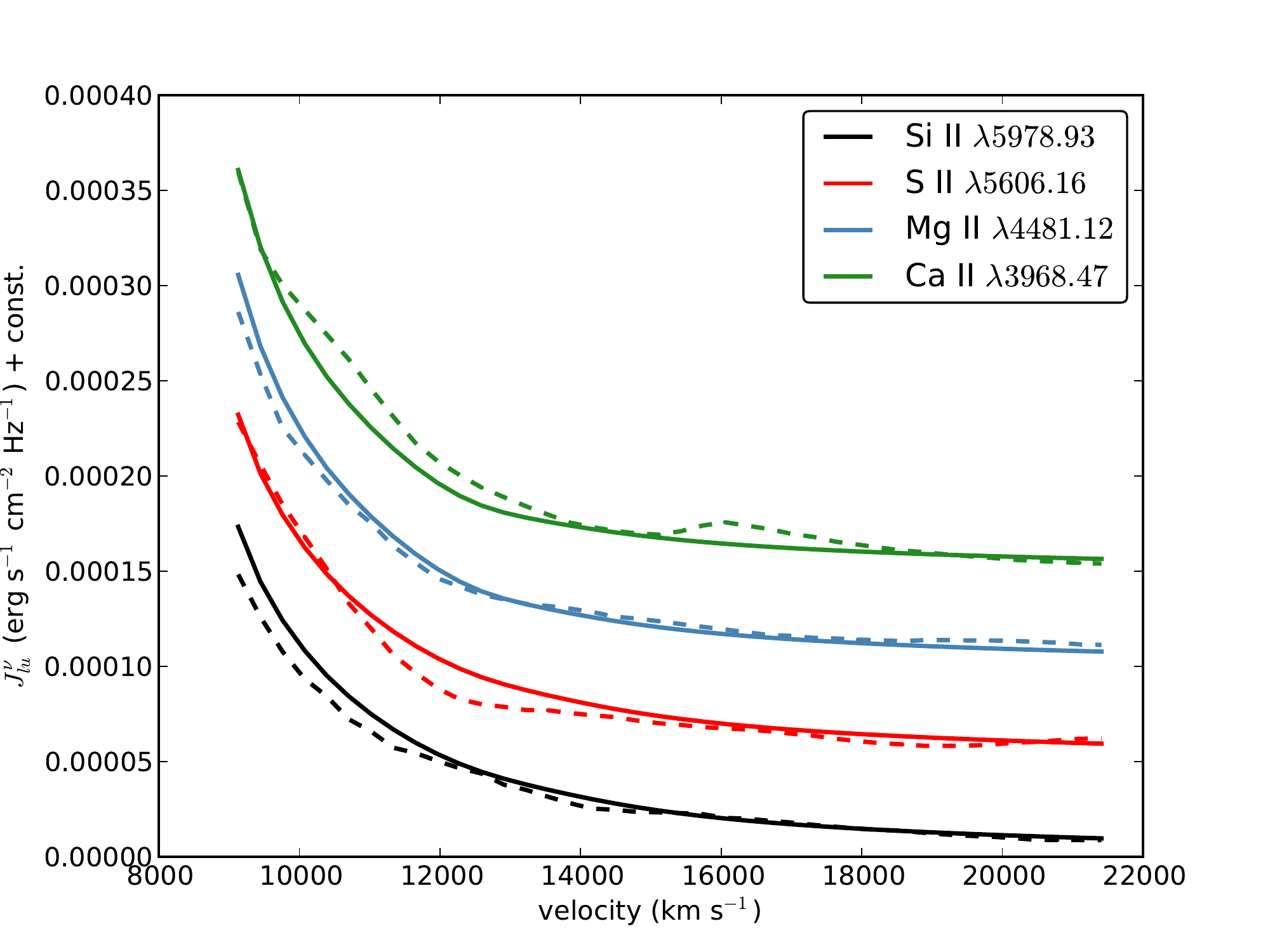} 

   \caption{Comparison between \jblue obtained using the estimators in
     Equation~\ref{eq:jblue} (dashed) and $\jblue = W B(\nu)$ (solid)
     for selected transitions at \texp=18~days. For clarity we have applied offsets to
     each curve (increment $5 \times 10^{-5}$).}
   \label{fig:tardis_result_jblue_evolution}
\end{figure}

For the comparison,  we repeat the \ion{Si}{2} calculation from Section~\ref{sec:results_nlte} adopting \texttt{dilute-blackbody} mode for the radiative rates.
In Fig.~\ref{fig:tardis_result_jblue_evolution}, we compare values of \jblue for a selection of line transitions between the calculations with \texttt{detailed} and \texttt{dilute-blackbody} mode. We find that, although rather simplistic, the \texttt{dilute-blackbody} assumption for \jblue performs rather well -- although there are modest deviations, the overall shape of the \texttt{detailed} calculation is well-matched, particularly in the inner regions where most of the spectrum formation occurs.

\begin{figure}
   \centering
   \includegraphics[width=\columnwidth]{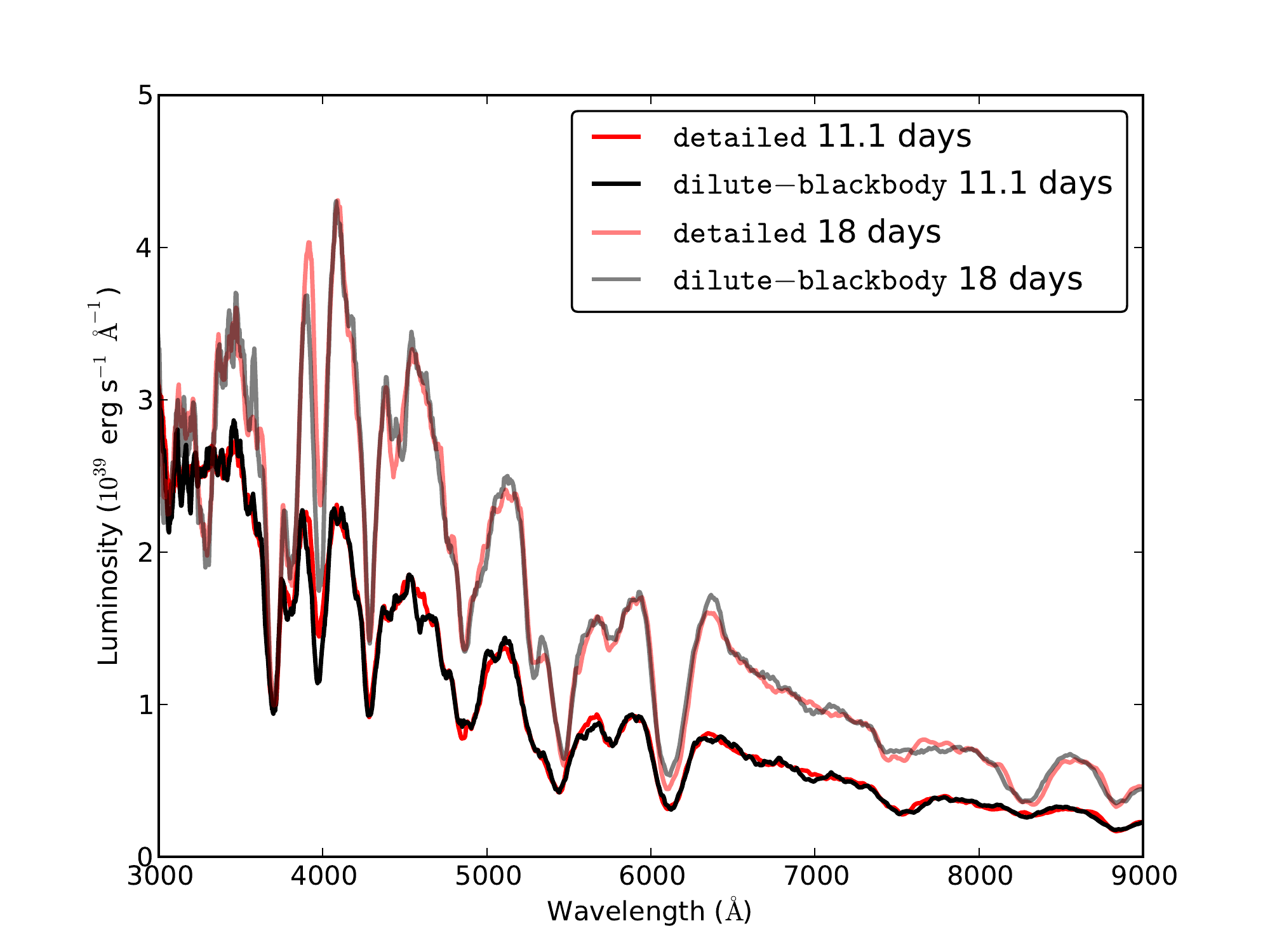} 

   \caption{Comparison of spectra obtained for \texttt{detailed} and
     \texttt{dilute-blackbody} radiative rates modes (\ion{Si}{2}
     treated in \texttt{nlte} excitatrion mode; all other ions in
     \texttt{dilute-lte} mode).}
   \label{fig:tardis_result_j_blue_spectral_comparison}
\end{figure}

Given that \jblue is well-represented by the \texttt{dilute-blackbody} assumption, it is to expected that the important quantities that depend on the bound-bound radiative rates will not be very adversely affected by this assumption. This is confirmed in Fig.~\ref{fig:tardis_result_departure_evolution}, which compares departure coefficients for one example, 
and Fig.~\ref{fig:tardis_result_j_blue_spectral_comparison}, in which the complete synthetic spectra computed using our two treatments of the radiative rates are shown.
Consequently, we conclude that the \texttt{dilute-blackbody} assumption is generally acceptable and is therefore recommended for rapid modelling owing the considerable reduction in computational cost that it brings.

\section{Conclusion \& Future work}
\label{sec:conclusions}

We have presented a new 1D radiative \gls{montecarlo} code (\gls{tardis}) for modelling of \gls{sn} spectra, which is based on the indivisible-packet methods developed by Lucy  (\citealt{1985ApJ...288..679A,1993ApJ...405..738L}; ML93; \citealt{1999A&A...345..211L,2002A&A...384..725L,2003A&A...403..261L}). The purpose of the code is to allow rapid but accurate synthesis of \gls{sn} spectra with only a few input parameters (see Appendix~\ref{sec:using_tardis}). It is built in a modular fashion, making it possible for the user to activate or deactivate different physics and approximations so that it can be tuned for use in different applications. 

Our successful comparisons with other codes  (\gls{synpp}, \gls{python_rt}, \gls{artis})
verify that \gls{tardis} is operating as expected and can accurately reproduce the shapes and strengths of complex spectral features obtained from independent calculations. Our results also highlight the importance of different ionization treatments in reconciling the results obtained with different codes and the ongoing need to pursue a good balance between accuracy and computational expediency in the treatment of ionization and excitation.

The modularity built into \gls{tardis} allows for the exploration of different treatments of microphysics. In this work, we first explored the effect of simple excitation approximations (applied to four ions of relevance to \sneia, namely \ion{Si}{2}, \ion{S}{2}, \ion{Mg}{2}, \ion{Ca}{2}). We conclude that, for all of these except \ion{Ca}{2}, 
the excitation treatment has a significant effect on the strength of line features which should be considered when attempting quantitative modelling of observations.
We defer to a future study an equivalent investigation of approximate NLTE ionization treatments, which are likely to have more significant ramifications for the overall spectral shape 
\citep[see][]{1996A&A...312..525P,1996MNRAS.283..297B,2013MNRAS.429.2127B}.

Secondly, we investigated the effect of different line interaction schemes on synthetic spectra. Similar to previous studies \citep{1999A&A...345..211L,2000A&A...363..705M} we found that the assumption of pure resonance scattering underestimates the ultraviolet luminosity of \sneia. 
Comparing the full macro atom approach to a simple down branching scheme shows that there is a minor but noticeable difference. As the computational overhead of using the macro atom scheme is only 10\% above the simpler down branching, we advocate for the use of the macro atom scheme in future studies.

We also explored the sensitivity of \gls{tardis} results to the different schemes for obtaining radiative rates for bound-bound transitions. Comparing results obtained using radiative rates recorded with line-by-line \gls{montecarlo} estimators to a simplified dilute-blackbody model (as adopted for the radiative rates in \gls{artis}), we see only minute changes in the synthetic spectrum. Given that the detailed estimators lead to an increase in computational time by orders of magnitude, our findings support the use of the simpler treatment.

\gls{tardis} is now publicly available for use. For approximate modelling of \snia\ observations, we would generally recommend combining the \texttt{nebular} ionization / \texttt{dilute-lte} excitation plasma modes with the \texttt{dilute-blackbody} and \texttt{macroatom} interaction modes as a viable (computationally manageable) mode of operation. To fit a particular observation, a user would need to supply the luminosity and then develop a model by choosing a density profile [which could be empirical or based on an explosion model such as W7 \citep{1984ApJ...286..644N}] and a set of ejecta abundances (which can be uniform or stratified). The density and abundances can then be modified to attempt to improve the fit (and therefore constrain the SN properties, as in e.g. \citealt{2005MNRAS.360.1231S,2008MNRAS.386.1897M,2009MNRAS.399.1238H,2011MNRAS.410.1725T}). However, we stress that this is only one potential use/mode of operation for \gls{tardis} and we encourage potential users to refer to the manual \url{http://tardis.rtfd.org} for further details.


In the near future, we plan to focus on two distinct \gls{tardis}  projects. First and foremost, the implementation of additional physics (bound-free/thermalization processes) with the goal of adding modules that include more sophisticated ionization approximations and allow for spectral synthesis for SNe~II (Klauser \& Kromer et al., in prep.). 
Secondly, since \gls{tardis} was mainly developed to provide a means to fit \sneia with an approach similar to \citet{2007Sci...315..825M}, we aim to couple \gls{tardis} with a suitable algorithm for automatic fitting of observations.  
We have explored this problem already, using the ML93 spectral synthesis code and genetic algorithms as the optimization algorithm and find that this is a promising approach \citep[see ][priv. comm. S. Hachinger, P. Mazzali]{2011PhDT.......324K}.

\section{Acknowledgements}

We gratefully acknowledge Markus Kromer and Michael Klauser for many useful discussions and suggestions during all stages in the development of \gls{tardis}. We thank Paolo Mazzali and Stephan Hachinger for helpful discussions of ionization/excitation treatments and formulation of the automatic spectral fitting project from which this work arose.
We thank Knox Long for making available the \gls{python_rt} code, the LK02 $\zeta$-values and for advice in making our comparisons with \gls{python_rt}.
We are grateful to Ken Dere for help with {\sc ChiantiPy} and to Brian Schmidt for suggestions and support throughout the development of this project. 
We thank Adam Suban-Loewen for help in developing the \gls{tardis} user interface and Aoife Boyle for participating in the testing of the code. We also would like to thank Erik Bray for helping with the installation frameworks implemented in \gls{tardis}.  We acknowledge the referee R. C. Thomas for very helpful suggestions and critical assessment of the paper. This research made use of Astropy, a community-developed core Python package for Astronomy \citep{2013A&A...558A..33A}, and of the \textsc{randomkit}\footnote{\nolinkurl{http://js2007.free.fr/code/index.html\#RandomKit}} library.

\bibliographystyle{mn2e}
\bibliography{tardis}

\begin{thebibliography}{}

\bibitem[\protect\citeauthoryear{{Abbott} \& {Lucy}}{{Abbott} \&
  {Lucy}}{1985}]{1985ApJ...288..679A}
{Abbott} D.~C.,  {Lucy} L.~B.,  1985, \apj, 288, 679

\bibitem[\protect\citeauthoryear{Anderson et~al.,}{Anderson
  et~al.}{1999}]{laug}
Anderson E.  et~al., 1999, {LAPACK} Users' Guide, third edn.
Society for Industrial and Applied Mathematics, Philadelphia, PA

\bibitem[\protect\citeauthoryear{{Astropy Collaboration} et~al.,}{{Astropy
  Collaboration} et~al.}{2013}]{2013A&A...558A..33A}
{Astropy Collaboration} et~al., 2013, \aap, 558, A33

\bibitem[\protect\citeauthoryear{{Baron}, {Hauschildt}, {Nugent} \&
  {Branch}}{{Baron} et~al.}{1996}]{1996MNRAS.283..297B}
{Baron} E.,  {Hauschildt} P.~H.,  {Nugent} P.,    {Branch} D.,  1996, \mnras,
  283, 297

\bibitem[\protect\citeauthoryear{{Blondin}, {Dessart}, {Hillier} \&
  {Khokhlov}}{{Blondin} et~al.}{2013}]{2013MNRAS.429.2127B}
{Blondin} S.,  {Dessart} L.,  {Hillier} D.~J.,    {Khokhlov} A.~M.,  2013,
  \mnras, 429, 2127

\bibitem[\protect\citeauthoryear{{Branch}, {Doggett}, {Nomoto} \&
  {Thielemann}}{{Branch} et~al.}{1985}]{1985ApJ...294..619B}
{Branch} D.,  {Doggett} J.~B.,  {Nomoto} K.,    {Thielemann} F.-K.,  1985,
  \apj, 294, 619

\bibitem[\protect\citeauthoryear{{Dere}, {Landi}, {Mason}, {Monsignori Fossi}
  \& {Young}}{{Dere} et~al.}{1997}]{1997A&AS..125..149D}
{Dere} K.~P.,  {Landi} E.,  {Mason} H.~E.,  {Monsignori Fossi} B.~C.,
  {Young} P.~R.,  1997, \aaps, 125, 149

\bibitem[\protect\citeauthoryear{{Dessart} \& {Hillier}}{{Dessart} \&
  {Hillier}}{2005}]{2005A&A...437..667D}
{Dessart} L.,  {Hillier} D.~J.,  2005, \aap, 437, 667

\bibitem[\protect\citeauthoryear{{Hachinger}, {Mazzali}, {Tanaka},
  {Hillebrandt} \& {Benetti}}{{Hachinger} et~al.}{2008}]{2008MNRAS.389.1087H}
{Hachinger} S.,  {Mazzali} P.~A.,  {Tanaka} M.,  {Hillebrandt} W.,    {Benetti}
  S.,  2008, \mnras, 389, 1087

\bibitem[\protect\citeauthoryear{{Hachinger}, {Mazzali}, {Taubenberger},
  {Pakmor} \& {Hillebrandt}}{{Hachinger} et~al.}{2009}]{2009MNRAS.399.1238H}
{Hachinger} S.,  {Mazzali} P.~A.,  {Taubenberger} S.,  {Pakmor} R.,
  {Hillebrandt} W.,  2009, \mnras, 399, 1238

\bibitem[\protect\citeauthoryear{{Higginbottom}, {Knigge}, {Long}, {Sim} \&
  {Matthews}}{{Higginbottom} et~al.}{2013}]{2013MNRAS.436.1390H}
{Higginbottom} N.,  {Knigge} C.,  {Long} K.~S.,  {Sim} S.~A.,    {Matthews}
  J.~H.,  2013, \mnras, 436, 1390

\bibitem[\protect\citeauthoryear{{H{\"o}flich}}{{H{\"o}flich}}{2003}]{2003ASPC..288..185H}
{H{\"o}flich} P.,  2003, in {Hubeny} I.,  {Mihalas} D.,   {Werner} K.,  eds,
  Astronomical Society of the Pacific Conference Series Vol. 288, Stellar
  Atmosphere Modeling. p.~185

\bibitem[\protect\citeauthoryear{{Kasen}, {Thomas} \& {Nugent}}{{Kasen}
  et~al.}{2006}]{2006ApJ...651..366K}
{Kasen} D.,  {Thomas} R.~C.,    {Nugent} P.,  2006, \apj, 651, 366

\bibitem[\protect\citeauthoryear{{Kerzendorf}}{{Kerzendorf}}{2011}]{2011PhDT.......324K}
{Kerzendorf} W.~E.,  2011, PhD thesis, Australian National University, Research
  School of Astronomy \& Astrophysics

\bibitem[\protect\citeauthoryear{Kramida, Ralchenko, Reader \& {NIST ASD
  Team}}{Kramida et~al.}{2012}]{kramida2012nist}
Kramida A.,  Ralchenko Y.,  Reader J.,    {NIST ASD Team} 2012, Atomic Spectra
  Database (v5.0), National Institute of Standards and Technology,
  Gaithersburg, MD.

\bibitem[\protect\citeauthoryear{{Kromer} \& {Sim}}{{Kromer} \&
  {Sim}}{2009}]{2009MNRAS.398.1809K}
{Kromer} M.,  {Sim} S.~A.,  2009, \mnras, 398, 1809

\bibitem[\protect\citeauthoryear{{Kurucz} \& {Bell}}{{Kurucz} \&
  {Bell}}{1995}]{1995KurCD..23.....K}
{Kurucz} R.,  {Bell} B.,  1995, Atomic Line Data Kurucz CD-ROM
  No.~23.~Cambridge, Mass.: Smithsonian Astrophysical Observatory

\bibitem[\protect\citeauthoryear{{Lamers} \& {Cassinelli}}{{Lamers} \&
  {Cassinelli}}{1999}]{1999isw..book.....L}
{Lamers} H.~J.~G.~L.~M.,  {Cassinelli} J.~P.,  1999, {Introduction to Stellar
  Winds}

\bibitem[\protect\citeauthoryear{{Landi}, {Del Zanna}, {Young}, {Dere} \&
  {Mason}}{{Landi} et~al.}{2012}]{2012ApJ...744...99L}
{Landi} E.,  {Del Zanna} G.,  {Young} P.~R.,  {Dere} K.~P.,    {Mason} H.~E.,
  2012, \apj, 744, 99

\bibitem[\protect\citeauthoryear{{Lentz}, {Baron}, {Branch} \&
  {Hauschildt}}{{Lentz} et~al.}{2001}]{2001ApJ...557..266L}
{Lentz} E.~J.,  {Baron} E.,  {Branch} D.,    {Hauschildt} P.~H.,  2001, \apj,
  557, 266

\bibitem[\protect\citeauthoryear{{Long} \& {Knigge}}{{Long} \&
  {Knigge}}{2002}]{2002ApJ...579..725L}
{Long} K.~S.,  {Knigge} C.,  2002, \apj, 579, 725 (LK02)

\bibitem[\protect\citeauthoryear{{Lucy}}{{Lucy}}{1999}]{1999A&A...345..211L}
{Lucy} L.~B.,  1999, \aap, 345, 211

\bibitem[\protect\citeauthoryear{{Lucy}}{{Lucy}}{2002}]{2002A&A...384..725L}
{Lucy} L.~B.,  2002, \aap, 384, 725

\bibitem[\protect\citeauthoryear{{Lucy}}{{Lucy}}{2003}]{2003A&A...403..261L}
{Lucy} L.~B.,  2003, \aap, 403, 261

\bibitem[\protect\citeauthoryear{{Lucy}}{{Lucy}}{2005}]{2005A&A...429...19L}
{Lucy} L.~B.,  2005, \aap, 429, 19

\bibitem[\protect\citeauthoryear{{Lucy} \& {Abbott}}{{Lucy} \&
  {Abbott}}{1993}]{1993ApJ...405..738L}
{Lucy} L.~B.,  {Abbott} D.~C.,  1993, \apj, 405, 738

\bibitem[\protect\citeauthoryear{Matsumoto \& Nishimura}{Matsumoto \&
  Nishimura}{1998}]{Matsumoto:1998:MTE:272991.272995}
Matsumoto M.,  Nishimura T.,  1998, ACM Trans. Model. Comput. Simul., 8, 3

\bibitem[\protect\citeauthoryear{{Mazzali}}{{Mazzali}}{2000}]{2000A&A...363..705M}
{Mazzali} P.~A.,  2000, \aap, 363, 705

\bibitem[\protect\citeauthoryear{{Mazzali} \& {Lucy}}{{Mazzali} \&
  {Lucy}}{1993}]{1993A&A...279..447M}
{Mazzali} P.~A.,  {Lucy} L.~B.,  1993, \aap, 279, 447 (ML93)

\bibitem[\protect\citeauthoryear{{Mazzali}, {R{\"o}pke}, {Benetti} \&
  {Hillebrandt}}{{Mazzali} et~al.}{2007}]{2007Sci...315..825M}
{Mazzali} P.~A.,  {R{\"o}pke} F.~K.,  {Benetti} S.,    {Hillebrandt} W.,  2007,
  Science, 315, 825

\bibitem[\protect\citeauthoryear{{Mazzali}, {Sauer}, {Pastorello}, {Benetti} \&
  {Hillebrandt}}{{Mazzali} et~al.}{2008}]{2008MNRAS.386.1897M}
{Mazzali} P.~A.,  {Sauer} D.~N.,  {Pastorello} A.,  {Benetti} S.,
  {Hillebrandt} W.,  2008, \mnras, 386, 1897

\bibitem[\protect\citeauthoryear{{Noebauer}, {Sim}, {Kromer}, {R{\"o}pke} \&
  {Hillebrandt}}{{Noebauer} et~al.}{2012}]{2012MNRAS.425.1430N}
{Noebauer} U.~M.,  {Sim} S.~A.,  {Kromer} M.,  {R{\"o}pke} F.~K.,
  {Hillebrandt} W.,  2012, \mnras, 425, 1430

\bibitem[\protect\citeauthoryear{{Nomoto}, {Thielemann} \& {Yokoi}}{{Nomoto}
  et~al.}{1984}]{1984ApJ...286..644N}
{Nomoto} K.,  {Thielemann} F.-K.,    {Yokoi} K.,  1984, \apj, 286, 644

\bibitem[\protect\citeauthoryear{{Nugent}, {Phillips}, {Baron}, {Branch} \&
  {Hauschildt}}{{Nugent} et~al.}{1995}]{1995ApJ...455L.147N}
{Nugent} P.,  {Phillips} M.,  {Baron} E.,  {Branch} D.,    {Hauschildt} P.,
  1995, \apjl, 455, L147

\bibitem[\protect\citeauthoryear{{Pauldrach}, {Duschinger}, {Mazzali}, {Puls},
  {Lennon} \& {Miller}}{{Pauldrach} et~al.}{1996}]{1996A&A...312..525P}
{Pauldrach} A.~W.~A.,  {Duschinger} M.,  {Mazzali} P.~A.,  {Puls} J.,  {Lennon}
  M.,    {Miller} D.~L.,  1996, \aap, 312, 525

\bibitem[\protect\citeauthoryear{{Pinto} \& {Eastman}}{{Pinto} \&
  {Eastman}}{2000}]{2000ApJ...530..757P}
{Pinto} P.~A.,  {Eastman} R.~G.,  2000, \apj, 530, 757

\bibitem[\protect\citeauthoryear{{R{\"o}pke}}{{R{\"o}pke}}{2005}]{2005A&A...432..969R}
{R{\"o}pke} F.~K.,  2005, \aap, 432, 969

\bibitem[\protect\citeauthoryear{{Sauer}, {Hoffmann} \& {Pauldrach}}{{Sauer}
  et~al.}{2006}]{2006A&A...459..229S}
{Sauer} D.~N.,  {Hoffmann} T.~L.,    {Pauldrach} A.~W.~A.,  2006, \aap, 459,
  229

\bibitem[\protect\citeauthoryear{Savitzky \& Golay}{Savitzky \&
  Golay}{1964}]{doi:10.1021/ac60214a047}
Savitzky A.,  Golay M. J.~E.,  1964, Analytical Chemistry, 36, 1627

\bibitem[\protect\citeauthoryear{{Seitenzahl} et~al.,}{{Seitenzahl}
  et~al.}{2013}]{2013MNRAS.429.1156S}
{Seitenzahl} I.~R.  et~al., 2013, \mnras, 429, 1156

\bibitem[\protect\citeauthoryear{{Sim}}{{Sim}}{2007}]{2007MNRAS.375..154S}
{Sim} S.~A.,  2007, \mnras, 375, 154

\bibitem[\protect\citeauthoryear{{Sim}, {Drew} \& {Long}}{{Sim}
  et~al.}{2005}]{2005MNRAS.363..615S}
{Sim} S.~A.,  {Drew} J.~E.,    {Long} K.~S.,  2005, \mnras, 363, 615

\bibitem[\protect\citeauthoryear{{Sim} \& {Mazzali}}{{Sim} \&
  {Mazzali}}{2008}]{2008MNRAS.385.1681S}
{Sim} S.~A.,  {Mazzali} P.~A.,  2008, \mnras, 385, 1681

\bibitem[\protect\citeauthoryear{{Sim}, {Miller}, {Long}, {Turner} \&
  {Reeves}}{{Sim} et~al.}{2010}]{2010MNRAS.404.1369S}
{Sim} S.~A.,  {Miller} L.,  {Long} K.~S.,  {Turner} T.~J.,    {Reeves} J.~N.,
  2010, \mnras, 404, 1369

\bibitem[\protect\citeauthoryear{{Sim}, {R{\"o}pke}, {Hillebrandt}, {Kromer},
  {Pakmor}, {Fink}, {Ruiter} \& {Seitenzahl}}{{Sim}
  et~al.}{2010}]{2010ApJ...714L..52S}
{Sim} S.~A.,  {R{\"o}pke} F.~K.,  {Hillebrandt} W.,  {Kromer} M.,  {Pakmor} R.,
   {Fink} M.,  {Ruiter} A.~J.,    {Seitenzahl} I.~R.,  2010, \apjl, 714, L52

\bibitem[\protect\citeauthoryear{{Stehle}, {Mazzali}, {Benetti} \&
  {Hillebrandt}}{{Stehle} et~al.}{2005}]{2005MNRAS.360.1231S}
{Stehle} M.,  {Mazzali} P.~A.,  {Benetti} S.,    {Hillebrandt} W.,  2005,
  \mnras, 360, 1231

\bibitem[\protect\citeauthoryear{{Tanaka}, {Mazzali}, {Stanishev}, {Maurer},
  {Kerzendorf} \& {Nomoto}}{{Tanaka} et~al.}{2011}]{2011MNRAS.410.1725T}
{Tanaka} M.,  {Mazzali} P.~A.,  {Stanishev} V.,  {Maurer} I.,  {Kerzendorf}
  W.~E.,    {Nomoto} K.,  2011, \mnras, 410, 1725

\bibitem[\protect\citeauthoryear{{Thomas}, {Nugent} \& {Meza}}{{Thomas}
  et~al.}{2011}]{2011PASP..123..237T}
{Thomas} R.~C.,  {Nugent} P.~E.,    {Meza} J.~C.,  2011, \pasp, 123, 237

\bibitem[\protect\citeauthoryear{Wieser \& Coplen}{Wieser \&
  Coplen}{2011}]{wieser2011atomic}
Wieser M.~E.,  Coplen T.~B.,  2011, Pure and Applied Chemistry, 83, 359

\end{thebibliography}

\appendix


\section{Using {\sc tardis}}
\label{sec:using_tardis}

At runtime \gls{tardis} requires (1) an atomic database and (2) a user specified configuration file.

The atomic database is supplied in HDF5\footnote{\nolinkurl{http://www.hdfgroup.org/HDF5/}} format and consists of tables containing  the data described in Section~\ref{sec:datasources}. A relatively simple atomic database is downloadable from the online manual (\url{http://tardis.rtfd.org} - in the ``Running \gls{tardis}'' section). For custom versions of the atomic database users are currently encouraged to contact the authors.

The configuration file  is supplied in YAML\footnote{\nolinkurl{http://www.yaml.org}} markup language and defines both the physical and numerical parameters of the calculation and selects which of the available modes of operation are to be used. A simple example of the input file is shown and explained in Section~\ref{sec:example_input} (further details in the code manual - \url{http://tardis.rtfd.org}).

In the standard mode of operation, the major physical parameters that can be varied are the output luminosity, the time since explosion and the density/composition profile of the model (including the locations of the inner and outer boundaries). The density profile can be chosen from preset standard options [e.g. constant density or a W7-like \citep{1984ApJ...286..644N} density, as in the example given below] or it can be supplied via an addition ASCII input table (density tabulated as a function of velocity; see manual). The abundance distribution can be uniform (see example given below) or stratified (again using an ASCII input table; see manual).

\subsection{Example input file}
\label{sec:example_input}

\begin{figure*}

\includegraphics{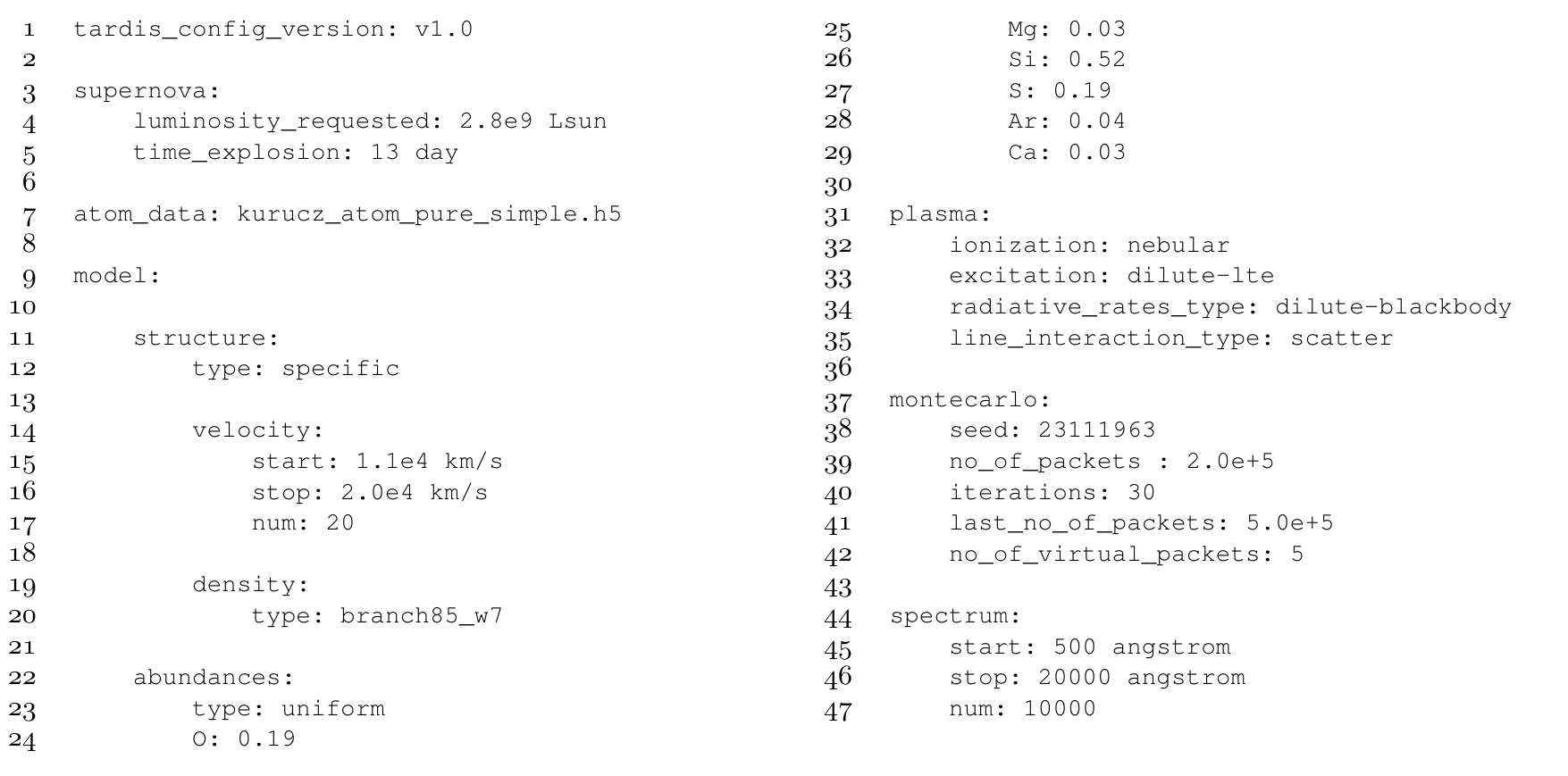} 
\label{fig:tardis_yaml_input}
\caption{YAML input file for \gls{tardis}.}
\end{figure*}

Fig.~\ref{fig:tardis_yaml_input} shows a configuration file for \gls{tardis}.
This file initiates a
calculation equivalent to that used in Section~\ref{sec:python_compare} (i.e. see Fig.~\ref{fig:python_compare}) and runs successfully with version 0.9 of \gls{tardis}. The sections of the file are described in detail below (with reference to line numbers). We stress that this is only one simple example and refer users to the manual (\url{http://tardis.rtfd.org}) for further information.

All dimensional quantities in the configuration file require a unit [these are parsed using the \gls{astropy} \texttt{Quantity} framework].

The first line of the configuration file identifies the \gls{tardis} version. Currently only \texttt{v1.0} is supported but future versions of \gls{tardis} may involve a restructure of the configuration file and will recognise the layout according to the version number.

The section (\texttt{supernova}) defines the requested emergent luminosity (\louter; see Section~\ref{sec:montecarlo_iteration}) and the time since explosion (\texp). The standard \gls{astropy} units are available. 

The atomic data file (HDF5 format) is specified at line 7 (absolute or relative path can be given).

The \texttt{model} section controls the setup of the radial density and abundance profiles. In the example, we specify 20 shells (line 17) equally spaced in the velocity range 11000~\kms to 20000~\kms (lines 15--16). We adopt a W7-like density profile (specifically, a polynomial fit to the W7 structure similar to \citealt{1985ApJ...294..619B}; lines 19--20). The example calculation adopts uniform (line 23) composition of all zones by specifying mass fractions for six elements in lines 24--29. Abundances of elements not specified are set to zero. More complex radial stratification of density and/or composition can be specified (see user manual).

The settings to be used in determining the ionization/excitation/line-interactions modes are provided in the \texttt{plasma} section (lines 31--35) using the keywords described in Table~\ref{tab:modes}. 

In the \texttt{montecarlo} section, \gls{tardis} is supplied with the seed value for the random number generator (line 38), the requested number of packets to be used in each MC simulation (line 39) and the number of iterations to be performed in determining the plasma state (line 40). In addition (line 41), one can specify a different number of packets for the last MC simulation (making it possible to increase the signal-to-noise of the final spectrum). If the keyword \texttt{no\_of\_virtual\_packets} is greater than zero (line 42), the virtual packet mode will be activated in the last MC simulation with $N_{v}$ set to the value given (see Section~\ref{sec:synthetic_spectrum}). 

In the last section (\texttt{spectrum}), the spectral range for the synthetic spectrum is set by specifying start and end wavelengths (lines 45 and 46). The spectral resolution is determined by choosing the total number of wavelength bins (line 47).

Running the model defined by the example configuration file on a mid 2012 MacBook Air (Intel Core i7 2.00 GHz, 8 GB RAM) takes approximately $\sim 3$ minutes to complete all 30 iterations. The runtime scales roughly linearly with the number of MC packets.

\end{document}